\RequirePackage[displaymath, mathlines,running]{lineno}
\documentclass[floatfix,aps,prd,twocolumn,letterpaper,lengthcheck,superscriptaddress,amssymb,amsmath,amsfonts,nofootinbib,altaffilletter]{revtex4-2}
\usepackage{color}
\usepackage[dvipsnames]{xcolor}
\usepackage{calc}
\usepackage{mathtools,graphicx}
\usepackage{pifont}
\usepackage{bm}
\usepackage{microtype}
\usepackage{booktabs}
\usepackage{times}
\usepackage[varg]{txfonts}
\usepackage[colorlinks, pdfborder={0 0 0}]{hyperref}
\usepackage[utf8]{inputenc}
\usepackage[caption=false]{subfig}
\usepackage{paralist}
\usepackage[normalem]{ulem}
\usepackage{multirow}
\usepackage{etoolbox}
\usepackage{xstring}
\usepackage{xparse}
\usepackage{dcolumn}
\usepackage{hyperref}
\usepackage{orcidlink}
\usepackage{pgf,interval}

\definecolor{LinkColor}{rgb}{0.75, 0, 0}
\definecolor{CiteColor}{rgb}{0, 0.5, 0.5}
\definecolor{UrlColor}{rgb}{0, 0, 0.75}
\hypersetup{linkcolor=LinkColor}
\hypersetup{citecolor=CiteColor}
\hypersetup{urlcolor=UrlColor}
\allowdisplaybreaks
\DeclareFontFamily{OT1}{pzc}{}
\DeclareFontShape{OT1}{pzc}{m}{it}{<-> s * [1.10] pzcmi7t}{}
\DeclareMathAlphabet{\mathpzc}{OT1}{pzc}{m}{it}

\AtBeginDocument{%
    \newwrite\bibnotes
    \def\bibnotesext{Notes.bib}
    \immediate\openout\bibnotes=\jobname\bibnotesext
    \immediate\write\bibnotes{@CONTROL{REVTEX41Control}}
    \immediate\write\bibnotes{@CONTROL{%
    apsrev41Control,author="08",editor="1",pages="0",title="0",year="1"}}
     \if@filesw
     \immediate\write\@auxout{\string\citation{apsrev41Control}}%
    \fi
}%

\newtoggle{fullauthorlist}
\toggletrue{fullauthorlist}

\newtoggle{endauthorlist}
\togglefalse{endauthorlist}

\graphicspath{{./fig/}}

\usepackage{scrextend}
\usepackage{textgreek}

\usepackage{graphicx}

\usepackage{makerobust}
\makeatletter
\@ifundefined{caption@xref}{}{%
  \MakeRobustCommand\caption@xref
}
\makeatother

\def\mnras{\ref@jnl{MNRAS}}             

\renewcommand{\today}{\number\day\space\ifcase\month\or
  January\or February\or March\or April\or May\or June\or
  July\or August\or September\or October\or November\or December\fi
  \space\number\year}

\newcommand{\QGR}{\ensuremath{\mathcal{Q}_{\rm GR}^{\rm{2D}}}}
\newcommand{\dMf}{\ensuremath{\Delta M_{\rm f} / \bar{M}_{\rm f}}}
\newcommand{\dchif}{\ensuremath{\Delta \chi_{\rm f} / \bar{\chi}_{\rm f}}}

\input{macros/PE_macros.tex}
\input{macros/event_macros.tex}
\newcommand{\OBSERVINGINSTRUMENTS}[1]{\IfEqCase{#1}{{GW191105E}{HLV}}}

\newcommand{\PEINSTRUMENTS}[1]{\IfEqCase{#1}{{GW191105E}{HLV}}}

\newcommand{\MITIGATIONMETHOD}[1]{\IfEqCase{#1}{{GW191105E}{None}}}

\input{macros/imrp_macros.tex}

\newtoggle{cleancomments}
\togglefalse{cleancomments}

\newcommand{\blue}[1]{{\color{Blue} #1}}

\iftoggle{cleancomments}{

}{

}

\newcommand{\IMRP}{\textsc{IMRPhenomXP}}
\newcommand{\IMRPHM}{\textsc{IMRPhenomXPHM}}

\newcommand{\SEOBNR}{\textsc{SEOBNRv4}}
\newcommand{\SEOBNRHM}{\textsc{SEOBNRv4HM}}
\newcommand{\SEOBROM}{\textsc{SEOBNRv4\_ROM}}

\newcommand{\cmark}{\ding{51}}

\newcommand{\linf}{\textsc{LALInference}}

\newcommand{\gstlal}{\textsc{GstLAL}}
\newcommand{\pycbc}{\textsc{PyCBC}}
\newcommand{\cwb}{\textsc{cWB}}

\newcommand{\SimBestPosUL}[1]{\IfEqCase{#1}{{NAME0}{GW191216ap}{VALUE0}{10.65}{NAME1}{GW191216ap}{VALUE1}{11.33}}}
\newcommand{\SimCombinedCI}[1]{\IfEqCase{#1}{{HIER_POP}{\ensuremath{-26.3^{+45.8}_{-52.9}}}{SIMPLE_POP}{\ensuremath{-16.0^{+13.6}_{-16.7}}}{HIER_POP_NEG}{66.20}{HIER_POP_POS}{51.85}{SIMPLE_POP_NEG}{28.68}{SIMPLE_POP_POS}{6.66}{HIER_MU}{\ensuremath{-26.8^{+26.3}_{-34.1}}}{HIER_SIGMA}{41.8}}}
\newcommand{\SimBF}[1]{\IfEqCase{#1}{{SYM}{0.9}{POS}{2.2}}}
\newcommand{\LivMgUL}{\ensuremath{2.42 \times 10^{-23}}}

\newcommand{\LivImprov}[1]{\IfEqCase{#1}{{AMP}{1.3}{AMP_min}{0.8}{AMP_max}{2.1}{MG}{1.3}{EXP}{1.2}{MG_SOLAR}{1.3}}}
\newcommand{\LivImprovFoutyoneEvents}[1]{\IfEqCase{#1}{{AMP}{1.3}{AMP_min}{1.0}{AMP_max}{1.8}{MG}{1.1}{EXP}{1.2}{MG_SOLAR}{1.2}}}
\newcommand{\LivEvents}[1]{\IfEqCase{#1}{{GWTC-3 (43 events)}{43}{GWTC-3 (41 events)}{41}{GWTC-2}{31}}}
\newcommand{\ParBestSimplePopS}[1]{\IfEqCase{#1}{{NAME}{\ensuremath{\delta\hat{\varphi}_{-2}}}{VALUE}{\ensuremath{0.10^{+0.60}_{-0.40}\times 10^{-3}}}{QGR}{40\%}}}
\newcommand{\ParWorstSimplePopS}[1]{\IfEqCase{#1}{{NAME}{\ensuremath{\delta\hat{\varphi}_{6l}}}{VALUE}{\ensuremath{-0.89^{+1.05}_{-1.02}}}{QGR}{91\%}}}
\newcommand{\ParBestHierPopS}[1]{\IfEqCase{#1}{{NAME}{\ensuremath{\delta\hat{\varphi}_{-2}}}{VALUE}{\ensuremath{-0.07^{+1.58}_{-1.90}\times 10^{-3}}}{QGR}{54\%}}}
\newcommand{\ParWorstHierPopS}[1]{\IfEqCase{#1}{{NAME}{\ensuremath{\delta\hat{\varphi}_{6l}}}{VALUE}{\ensuremath{-0.61^{+1.52}_{-1.48}}}{QGR}{78\%}}}
\newcommand{\ParBestHierZgrS}[1]{\IfEqCase{#1}{{NAME}{\ensuremath{\delta\hat{\varphi}_{-2}}}{VALUE}{\ensuremath{0.10}}}}
\newcommand{\ParWorstHierZgrS}[1]{\IfEqCase{#1}{{NAME}{\ensuremath{\delta\hat{\varphi}_{3}}}{VALUE}{\ensuremath{1.08}}}}
\newcommand{\ParAlphaBetaBest}[1]{\IfEqCase{#1}{{Zh_NAME}{\ensuremath{\delta\hathyper}}{Zh_VALUE}{0.1}{Qh_NAME}{\ensuremath{\delta\hathyper}}{Qh_VALUE}{0.0}{Zs_NAME}{\ensuremath{\delta\hathyper}}{Zs_VALUE}{0.3}}}
\newcommand{\ImrMfHierMu}[1]{\IfEqCase{#1}{{GWTC3}{\ensuremath{0.04^{+0.08}_{-0.07}}}{GWTC2}{\ensuremath{0.03^{+0.14}_{-0.09}}}{GWTC1}{\ensuremath{0.14^{+0.34}_{-0.23}}}}}
\newcommand{\ImrChifHierMu}[1]{\IfEqCase{#1}{{GWTC3}{\ensuremath{-0.04^{+0.12}_{-0.12}}}{GWTC2}{\ensuremath{-0.07^{+0.16}_{-0.17}}}{GWTC1}{\ensuremath{-0.01^{+0.21}_{-0.20}}}}}
\newcommand{\ImrMfHierSigma}[1]{\IfEqCase{#1}{{GWTC3}{\ensuremath{0.05^{+0.10}_{-0.04}}}{GWTC2}{\ensuremath{0.07^{+0.24}_{-0.06}}}{GWTC1}{\ensuremath{0.25^{+0.46}_{-0.23}}}}}
\newcommand{\ImrChifHierSigma}[1]{\IfEqCase{#1}{{GWTC3}{\ensuremath{0.19^{+0.17}_{-0.13}}}{GWTC2}{\ensuremath{0.26^{+0.20}_{-0.19}}}{GWTC1}{\ensuremath{0.11^{+0.28}_{-0.10}}}}}
\newcommand{\ImrMfHierPop}[1]{\IfEqCase{#1}{{GWTC3}{\ensuremath{0.03^{+0.14}_{-0.13}}}{GWTC2}{\ensuremath{0.03^{+0.26}_{-0.20}}}{GWTC1}{\ensuremath{0.13^{+0.73}_{-0.53}}}}}
\newcommand{\ImrChifHierPop}[1]{\IfEqCase{#1}{{GWTC3}{\ensuremath{-0.05^{+0.37}_{-0.38}}}{GWTC2}{\ensuremath{-0.07^{+0.50}_{-0.51}}}{GWTC1}{\ensuremath{-0.01^{+0.37}_{-0.32}}}}}

\newcommand{\ImrGWTCTHREE}[1]{\IfEqCase{#1}{{DMFGWTC3PHENOM}{\ensuremath{-0.02^{+0.07}_{-0.06}}}{DCHIFGWTC3PHENOM}{\ensuremath{-0.06^{+0.10}_{-0.07}}}{GRQUANTGWTC3}{\ensuremath{79.6}}}}

\newcommand{\ImrGWTCTWO}[1]{\IfEqCase{#1}{{DMFGWTC2PHENOM}{\ensuremath{-0.04^{+0.09}_{-0.06}}}{DCHIFGWTC2PHENOM}{\ensuremath{-0.09^{+0.11}_{-0.08}}}{GRQUANTGWTC2}{\ensuremath{78.7}}}}
\newcommand{\ImrEVENTSTATS}[1]{\IfEqCase{#1}{{S190503bfDMFGWTC3PHENOM}{\ensuremath{0.90^{+0.29}_{-0.59}}}{S190503bfDCHIFGWTC3PHENOM}{\ensuremath{0.40^{+0.89}_{-0.30}}}{S190503bfGRQUANTGWTC3}{\ensuremath{94.3}}{S190828jDMFGWTC3PHENOM}{\ensuremath{0.03^{+0.36}_{-0.25}}}{S190828jDCHIFGWTC3PHENOM}{\ensuremath{-0.07^{+0.44}_{-0.34}}}{S190828jGRQUANTGWTC3}{\ensuremath{21.0}}{GW150914DMFGWTC3PHENOM}{\ensuremath{0.22^{+0.32}_{-0.22}}}{GW150914DCHIFGWTC3PHENOM}{\ensuremath{0.16^{+0.61}_{-0.41}}}{GW150914GRQUANTGWTC3}{\ensuremath{54.3}}{S200129mDMFGWTC3PHENOM}{\ensuremath{0.04^{+0.26}_{-0.19}}}{S200129mDCHIFGWTC3PHENOM}{\ensuremath{-0.03^{+0.36}_{-0.30}}}{S200129mGRQUANTGWTC3}{\ensuremath{1.5}}{S190408anDMFGWTC3PHENOM}{\ensuremath{0.02^{+0.40}_{-0.27}}}{S190408anDCHIFGWTC3PHENOM}{\ensuremath{-0.01^{+0.48}_{-0.46}}}{S190408anGRQUANTGWTC3}{\ensuremath{11.5}}{S190521rDMFGWTC3PHENOM}{\ensuremath{0.10^{+0.36}_{-0.22}}}{S190521rDCHIFGWTC3PHENOM}{\ensuremath{0.07^{+0.30}_{-0.28}}}{S190521rGRQUANTGWTC3}{\ensuremath{0.4}}{S200224caDMFGWTC3PHENOM}{\ensuremath{0.30^{+0.62}_{-0.37}}}{S200224caDCHIFGWTC3PHENOM}{\ensuremath{0.18^{+0.30}_{-0.33}}}{S200224caGRQUANTGWTC3}{\ensuremath{20.7}}{GW170818DMFGWTC3PHENOM}{\ensuremath{0.18^{+0.49}_{-0.32}}}{GW170818DCHIFGWTC3PHENOM}{\ensuremath{-0.03^{+0.60}_{-0.47}}}{GW170818GRQUANTGWTC3}{\ensuremath{24.5}}{S200225qDMFGWTC3PHENOM}{\ensuremath{0.09^{+0.43}_{-0.59}}}{S200225qDCHIFGWTC3PHENOM}{\ensuremath{0.05^{+0.61}_{-0.38}}}{S200225qGRQUANTGWTC3}{\ensuremath{1.3}}{GW170104DMFGWTC3PHENOM}{\ensuremath{-0.08^{+0.34}_{-0.23}}}{GW170104DCHIFGWTC3PHENOM}{\ensuremath{-0.08^{+0.41}_{-0.34}}}{GW170104GRQUANTGWTC3}{\ensuremath{28.9}}{GW170809DMFGWTC3PHENOM}{\ensuremath{-0.05^{+0.38}_{-0.29}}}{GW170809DCHIFGWTC3PHENOM}{\ensuremath{-0.19^{+0.51}_{-0.46}}}{GW170809GRQUANTGWTC3}{\ensuremath{24.7}}{S190513bmDMFGWTC3PHENOM}{\ensuremath{-0.03^{+0.33}_{-0.27}}}{S190513bmDCHIFGWTC3PHENOM}{\ensuremath{-0.21^{+0.56}_{-0.42}}}{S190513bmGRQUANTGWTC3}{\ensuremath{34.6}}{S190630agDMFGWTC3PHENOM}{\ensuremath{-0.13^{+0.27}_{-0.20}}}{S190630agDCHIFGWTC3PHENOM}{\ensuremath{-0.15^{+0.45}_{-0.28}}}{S190630agGRQUANTGWTC3}{\ensuremath{58.5}}{S200208qDMFGWTC3PHENOM}{\ensuremath{0.17^{+1.46}_{-0.50}}}{S200208qDCHIFGWTC3PHENOM}{\ensuremath{0.03^{+1.66}_{-0.66}}}{S200208qGRQUANTGWTC3}{\ensuremath{10.5}}{GW170823DMFGWTC3PHENOM}{\ensuremath{0.82^{+0.21}_{-0.38}}}{GW170823DCHIFGWTC3PHENOM}{\ensuremath{0.20^{+0.63}_{-0.43}}}{GW170823GRQUANTGWTC3}{\ensuremath{95.1}}{S200311bgDMFGWTC3PHENOM}{\ensuremath{0.06^{+0.38}_{-0.25}}}{S200311bgDCHIFGWTC3PHENOM}{\ensuremath{-0.24^{+0.49}_{-0.63}}}{S200311bgGRQUANTGWTC3}{\ensuremath{15.2}}{GW170814DMFGWTC3PHENOM}{\ensuremath{0.50^{+0.68}_{-0.70}}}{GW170814DCHIFGWTC3PHENOM}{\ensuremath{-0.01^{+0.55}_{-0.24}}}{GW170814GRQUANTGWTC3}{\ensuremath{9.9}}{S190814bvDMFGWTC3PHENOM}{\ensuremath{-0.34^{+0.48}_{-1.11}}}{S190814bvDCHIFGWTC3PHENOM}{\ensuremath{-0.91^{+0.27}_{-0.19}}}{S190814bvGRQUANTGWTC3}{\ensuremath{99.9}}}}
\newcommand{\EVENTSELECTION}[1]{\IfEqCase{#1}{{S200316bjFCIMR}{\ensuremath{376}}{S200316bjFCFTA}{\ensuremath{153}}{S200316bjOPTSNRPREIMR}{\ensuremath{11.4}}{S200316bjOPTSNRPOSTIMR}{\ensuremath{1.7}}{S200316bjOPTSNRPREFTA}{\ensuremath{10.7}}{S200316bjOPTSNR}{\ensuremath{11.5}}{S200311bgFCIMR}{\ensuremath{122}}{S200311bgFCFTA}{\ensuremath{54}}{S200311bgOPTSNRPREIMR}{\ensuremath{13.5}}{S200311bgOPTSNRPOSTIMR}{\ensuremath{11.0}}{S200311bgOPTSNRPREFTA}{\ensuremath{6.5}}{S200311bgOPTSNR}{\ensuremath{17.5}}{S200225qFCIMR}{\ensuremath{213}}{S200225qFCFTA}{\ensuremath{91}}{S200225qOPTSNRPREIMR}{\ensuremath{11.1}}{S200225qOPTSNRPOSTIMR}{\ensuremath{6.6}}{S200225qOPTSNRPREFTA}{\ensuremath{6.8}}{S200225qOPTSNR}{\ensuremath{12.9}}{S200224caFCIMR}{\ensuremath{107}}{S200224caFCFTA}{\ensuremath{42}}{S200224caOPTSNRPREIMR}{\ensuremath{14.3}}{S200224caOPTSNRPOSTIMR}{\ensuremath{13.1}}{S200224caOPTSNRPREFTA}{\ensuremath{4.7}}{S200224caOPTSNR}{\ensuremath{19.4}}{S200219acFCIMR}{\ensuremath{86}}{S200219acFCFTA}{\ensuremath{39}}{S200219acOPTSNRPREIMR}{\ensuremath{7.4}}{S200219acOPTSNRPOSTIMR}{\ensuremath{8.4}}{S200219acOPTSNRPREFTA}{\ensuremath{2.8}}{S200219acOPTSNR}{\ensuremath{11.2}}{S200208qFCIMR}{\ensuremath{98}}{S200208qFCFTA}{\ensuremath{42}}{S200208qOPTSNRPREIMR}{\ensuremath{7.2}}{S200208qOPTSNRPOSTIMR}{\ensuremath{6.8}}{S200208qOPTSNRPREFTA}{\ensuremath{3.0}}{S200208qOPTSNR}{\ensuremath{9.9}}{S200202acFCIMR}{\ensuremath{502}}{S200202acFCFTA}{\ensuremath{216}}{S200202acOPTSNRPREIMR}{\ensuremath{10.9}}{S200202acOPTSNRPOSTIMR}{\ensuremath{1.6}}{S200202acOPTSNRPREFTA}{\ensuremath{10.5}}{S200202acOPTSNR}{\ensuremath{11.1}}{S200129mFCIMR}{\ensuremath{136}}{S200129mFCFTA}{\ensuremath{57}}{S200129mOPTSNRPREIMR}{\ensuremath{20.1}}{S200129mOPTSNRPOSTIMR}{\ensuremath{16.0}}{S200129mOPTSNRPREFTA}{\ensuremath{10.4}}{S200129mOPTSNR}{\ensuremath{25.7}}{S200115jFCIMR}{\ensuremath{806}}{S200115jFCFTA}{\ensuremath{364}}{S200115jOPTSNRPREIMR}{\ensuremath{12.3}}{S200115jOPTSNRPOSTIMR}{\ensuremath{0.3}}{S200115jOPTSNRPREFTA}{\ensuremath{12.2}}{S200115jOPTSNR}{\ensuremath{12.3}}{S191222nFCIMR}{\ensuremath{76}}{S191222nFCFTA}{\ensuremath{35}}{S191222nOPTSNRPREIMR}{\ensuremath{8.6}}{S191222nOPTSNRPOSTIMR}{\ensuremath{10.0}}{S191222nOPTSNRPREFTA}{\ensuremath{3.1}}{S191222nOPTSNR}{\ensuremath{13.1}}{S191216apFCIMR}{\ensuremath{456}}{S191216apFCFTA}{\ensuremath{151}}{S191216apOPTSNRPREIMR}{\ensuremath{17.7}}{S191216apOPTSNRPOSTIMR}{\ensuremath{2.4}}{S191216apOPTSNRPREFTA}{\ensuremath{15.6}}{S191216apOPTSNR}{\ensuremath{17.9}}{S191215wFCIMR}{\ensuremath{158}}{S191215wFCFTA}{\ensuremath{68}}{S191215wOPTSNRPREIMR}{\ensuremath{9.0}}{S191215wOPTSNRPOSTIMR}{\ensuremath{5.6}}{S191215wOPTSNRPREFTA}{\ensuremath{5.5}}{S191215wOPTSNR}{\ensuremath{10.6}}{S191204rFCIMR}{\ensuremath{451}}{S191204rFCFTA}{\ensuremath{183}}{S191204rOPTSNRPREIMR}{\ensuremath{17.7}}{S191204rOPTSNRPOSTIMR}{\ensuremath{3.2}}{S191204rOPTSNRPREFTA}{\ensuremath{16.3}}{S191204rOPTSNR}{\ensuremath{18.0}}{S191129uFCIMR}{\ensuremath{464}}{S191129uFCFTA}{\ensuremath{174}}{S191129uOPTSNRPREIMR}{\ensuremath{14.0}}{S191129uOPTSNRPOSTIMR}{\ensuremath{2.1}}{S191129uOPTSNRPREFTA}{\ensuremath{12.8}}{S191129uOPTSNR}{\ensuremath{14.1}}{S191109dFCIMR}{\ensuremath{55}}{S191109dFCFTA}{\ensuremath{27}}{S191109dOPTSNRPREIMR}{\ensuremath{9.3}}{S191109dOPTSNRPOSTIMR}{\ensuremath{17.9}}{S191109dOPTSNRPREFTA}{\ensuremath{0.8}}{S191109dOPTSNR}{\ensuremath{20.2}}}}

\newcommand{\twoc}[1]{\multicolumn{2}{c}{#1}}

\newcommand{\threec}[1]{\multicolumn{3}{c}{#1}}

\begin{document}

\title{Tests of General Relativity with GWTC-3}

\iftoggle{endauthorlist}{
 %
 %
 \let\mymaketitle\maketitle
 \let\myauthor\author
 \let\myaffiliation\affiliation
 \author{The LIGO Scientific Collaboration, Virgo Collaboration, and KAGRA Collaboration}
}{
 %
 %
 \iftoggle{fullauthorlist}{


\author{R.~Abbott}
\affiliation{LIGO Laboratory, California Institute of Technology, Pasadena, CA 91125, USA}
\author{H.~Abe}
\affiliation{Graduate School of Science, Tokyo Institute of Technology, Meguro-ku, Tokyo 152-8551, Japan  }
\author{F.~Acernese}
\affiliation{Dipartimento di Farmacia, Universit\`a di Salerno, I-84084 Fisciano, Salerno, Italy  }
\affiliation{INFN, Sezione di Napoli, Complesso Universitario di Monte S. Angelo, I-80126 Napoli, Italy  }
\author{K.~Ackley\,\orcidlink{0000-0002-8648-0767}}
\affiliation{OzGrav, School of Physics \& Astronomy, Monash University, Clayton 3800, Victoria, Australia}
\author{N.~Adhikari\,\orcidlink{0000-0002-4559-8427}}
\affiliation{University of Wisconsin-Milwaukee, Milwaukee, WI 53201, USA}
\author{R.~X.~Adhikari\,\orcidlink{0000-0002-5731-5076}}
\affiliation{LIGO Laboratory, California Institute of Technology, Pasadena, CA 91125, USA}
\author{V.~K.~Adkins}
\affiliation{Louisiana State University, Baton Rouge, LA 70803, USA}
\author{V.~B.~Adya}
\affiliation{OzGrav, Australian National University, Canberra, Australian Capital Territory 0200, Australia}
\author{C.~Affeldt}
\affiliation{Max Planck Institute for Gravitational Physics (Albert Einstein Institute), D-30167 Hannover, Germany}
\affiliation{Leibniz Universit\"at Hannover, D-30167 Hannover, Germany}
\author{D.~Agarwal}
\affiliation{Inter-University Centre for Astronomy and Astrophysics, Pune 411007, India}
% ...




 }{
 \author{The LIGO Scientific Collaboration, Virgo Collaboration, and KAGRA Collaboration}
}
}

\date[\relax]{compiled \today}

\begin{abstract}
The ever-increasing number of detections of gravitational waves from compact binaries by the Advanced LIGO and Advanced Virgo detectors allows us to perform ever-more sensitive tests of general relativity (GR) in the dynamical and strong-field regime of gravity.
We perform a suite of tests of GR using the compact binary signals observed during the second half of the third observing run of those detectors. We restrict our analysis to the 15 confident signals that have false alarm rates $\leq 10^{-3}\, {\rm yr}^{-1}$.
In addition to signals consistent with binary black hole mergers, the new events include \NNAME{GW200115A}, a signal consistent with a neutron star--black hole merger. We find the residual power,  after subtracting the best fit waveform from the data for each event, to be consistent with the detector noise. Additionally, we find all the post-Newtonian deformation coefficients to be consistent with the predictions from GR, with an improvement by a factor of $\sim 2$ in the $-1$PN parameter.
We also find that the spin-induced quadrupole moments of the binary black hole constituents are consistent with those of Kerr black holes in GR. 
We find no evidence for dispersion of gravitational waves, non-GR modes of polarization, or post-merger echoes in the events that were analyzed. We update the bound on the mass of the graviton, at 90\% credibility, to   $m_g \leq \LivMgUL \mathrm{eV}/c^2$.
The final mass and final spin as inferred from the pre-merger and post-merger parts of the waveform are consistent with each other. 
The studies of the properties of the remnant black holes, including deviations of the quasi-normal mode frequencies and damping times, show consistency with the predictions of GR.
In addition to considering signals individually, we also combine results from the catalog of gravitational waves signals to calculate more precise population constraints. We find no evidence in support of physics beyond general relativity.

\end{abstract}

\maketitle

\section{Introduction}
\label{sec:intro}

The first three observing runs of Advanced LIGO~\cite{TheLIGOScientific:2014jea} and Advanced Virgo~\cite{TheVirgo:2014hva} have led to 
detections of signals consistent with coming from the three canonical classes of compact binary systems: binary black holes~(BBH)~\cite{GW150914_paper}, binary neutron stars~(BNS)~\cite{DetectionPaper}, and neutron star--black holes~(NSBH)~\cite{LIGOScientific:2021qlt}. These observations had, in particular, a profound impact on fundamental physics as they allowed us to probe the properties of gravity in the highly nonlinear and dynamical regime. These detections subjected Einstein's general relativity (GR), which had passed all previous experimental tests to date with flying colors, to scrutiny in an entirely new regime. 
These new gravitational-wave tests~\cite{PhysRevLett.116.221101,Abbott:2017oio,Abbott:2017vtc,Abbott:2018lct,LIGOScientific:2019fpa,LIGOScientific:2020tif} complement existing laboratory and astrophysical tests of GR~\cite{Will:2014kxa,Berti:2015itd}.

GR has a well posed initial value formulation, making it possible to calculate the two-body evolution.
Despite the progress made on the analytical fronts (see for example \cite{Will:1994fb,Lang:2013fna,Lang:2014osa,Sennett:2016klh,Khalil:2018aaj,Bernard:2018hta,Bernard:2018ivi,Tahura:2018zuq,Sotiriou:2006pq,Shiralilou:2021mfl}) and numerical fronts (see for example \cite{Barausse:2012da,Shibata:2013pra,Okounkova:2019zjf,Okounkova:2020rqw,Cayuso:2020lca,East:2020hgw,East:2021bqk}), modelling compact binaries in modified gravity theories is still in its infancy.
We are thus not yet able to carry out tests of GR that rely on high-accuracy waveforms in alternative theories. Instead, it is possible to devise a strategy based on the currently best-understood theory of gravity (GR) and look for possible departures from its predictions~\cite{YunesPretorius:2009,Agathos:2013upa}.  This approach enables constraints to be placed on potential deviations from GR, although it has been argued that without reference to specific alternatives, it is difficult to assess the ability of these methods to detect GR violations~\cite{Chua:2020oxn}.

Given the major advances in the compact binary dynamics modeling using analytical (see for instance \cite{Blanchet:2013haa,Barack:2018yvs} for reviews) and numerical relativity (NR) techniques within GR, e.g.,~\cite{Lehner:2014asa,Jani:2016wkt,Boyle:2019kee,Healy:2020vre},
 such \emph{null tests of GR} have been successful so far.
However, the null tests are also sensitive to the physics \emph{within GR} that is not accounted for in the waveform model that is used (such as eccentricity, presence of exotic compact objects, etc) in addition to beyond-GR physics. For example, some of the tests performed here, which involve the dynamics of the merger remnant, may be interpreted as tests of GR or as a test of the black hole nature of the remnant.  However, for uniformity, they will be referred to as tests of GR in this paper. The tests are not completely independent of each other and vary in sensitivity to different types of deviations~\cite{Johnson-McDaniel:2021yge}, and by considering them together we obtain the best overview of how well predictions and observations agree.

Theoretically, gravitational waves in a modified theory of gravity may differ from GR broadly in three different ways: generation, propagation, and polarization. Gravitational-wave generation relates the outgoing radiation to the properties of the source, a hard problem even in GR. Propagation of gravitational waves in a modified theory of gravity can differ from that in GR via effects such as dispersion~\cite{Mirshekari:2011yq}, birefringence~\cite{Okounkova:2021xjv}, and amplitude damping~\cite{Nishizawa:2017nef,Belgacem:2017ihm}.  An effective-field theoretic approach to the problem can be found in \cite{Kostelecky:2016kfm,Wang:2020pgu,Shao:2020shv,Mewes:2019dhj,ONeal-Ault:2021uwu}. Tests based on propagation effects target those modified theories which predict generation of gravitational waves to be very close to that of GR, but differ in the way the waves propogate. This can happen in theories like massive graviton theories where the generation effects are suppressed by powers of $r/\lambda_g \ll 1$ where $\lambda_g$ is the Compton wavelength of the graviton and $r$ the size the binary~\cite{Will:1997bb}.
 
A general metric theory of gravity can allow up to six modes of polarization: two tensor, two vector, and two scalar modes~\cite{Eardley:1973br,Eardley:1974nw}. In GR, one only has the two tensor modes referred to as \emph{plus} and \emph{cross}. Hence, searching for non-GR modes of polarization is also an effective method to search for violation of GR. 
Nevertheless, there are alternative theories which predict modified gravitational-wave generation or propagation but do not predict additional non-GR modes of polarization, e.g.~\cite{Iyonaga:2018vnu,Wagle:2019mdq}. Hence, confidently detecting signatures of one or more of these three effects would strongly suggest a possible GR violation.

While these theoretical insights get reflected in our analysis strategies while searching for possible departures from GR,  the organization of this paper instead classifies the tests of GR as \emph{consistency tests} and \emph{parameterized tests}. Consistency tests, as the name indicates, search for possible violations of GR by asking how consistent the observed signal is with that of GR and do not invoke any parametrization of the deviation from GR. Consistency tests may be tests of {\emph {self-consistency}} of the signal or {\emph {overall consistency of the signal with the data}}. We consider one of each type of test: The inspiral--merger--ringdown (IMR) consistency test checks for consistency between the low- and high-frequency parts of the signal, while the residuals test subtracts the best-fit GR waveform from the data and asks whether there is any statistically significant residual power.
Parameterized tests, on the other hand, invoke a specific parametrization  that is appropriate for searching for possible deviations from GR in terms of certain physical effects. For example, the parameterized tests of post-Newtonian (PN) coefficients are sensitive to the physical effects that appear at different PN orders. Similarly, by parameterizing the dispersion relation, one can look for possible imprints of non-GR propagation effects in the gravitational waveforms. A variant of the null-stream based method  allows us to probe non-GR modes of gravitational-wave polarization.

%

\newcommand{\ImrMfImprovement}{\ensuremath{1.15}}
\newcommand{\ImrChifImprovement}{\ensuremath{1.12}}
\newcommand{\ImrMfHierImprovement}{\ensuremath{1.81}}
\newcommand{\ImrChifHierImprovement}{\ensuremath{1.36}}
\newcommand{\ImrMinImprovement}{\ensuremath{1.1}}
\newcommand{\ImrMaxImprovement}{\ensuremath{1.8}}
\begin{table*}
\caption{\label{tab:summary}
Summary of methods and results. 
This table summarizes the names of the tests performed, the corresponding sections, the parameters involved in the test, and the improvement with regard to our previous analysis. The analyses performed are: RT = residuals test; IMR = inspiral--merger--ringdown consistency test; PAR = parametrized tests of gravitational-wave generation; SIM = spin-induced moments; MDR = modified gravitational-wave dispersion relation; POL = polarization content; RD = ringdown; ECH = echoes searches.  
The last column provides the \emph{approximate} improvement in the bounds over the previous analyses reported in~\cite{LIGOScientific:2020tif}. This is defined as $X_{\rm GWTC-2}/X_{\rm GWTC-3}$, where $X$ denotes the width of the 90\% credible interval for the parameters for each test, using the combined results on all events considered.
For the MDR test, some of the bounds have worsened in comparison to GWTC-2. See the corresponding section for details.
The high improvement factor for $\textsc{pSEOB}$ is due to the larger number of events from GWTC-2 analysed here compared to~\cite{LIGOScientific:2020tif}.
} 

\begin{tabular}{c c c c c c c}
\toprule
\multirow{1}{*}{Test}&\multirow{1}{*}{Section } &  \multirow{1}{*}{Quantity } &  \multirow{1}{*}{Parameter } & \multicolumn{1}{c}{Improvement w.r.t.\ GWTC-2}\\
\\
\midrule
RT &\ref{sec:res} &  $p$-value & $p$-value & Not applicable
\\
IMR & \ref{sec:imr} & Fractional deviation in remnant mass and spin& $\biggl\{\dfrac{\Delta M_{\rm f}}{\bar{M}_{\rm f}}$, $\dfrac{\Delta \chi_{\rm f}}{\bar{\chi}_{\rm f}}\biggr\}$ & \ImrMinImprovement{}--\ImrMaxImprovement{}
 \\
PAR& \ref{sec:par} & PN deformation parameter & $\delta{\hat{\phi}_k}$ & 1.2--3.1
 \\
SIM & \ref{sec:sim} & Deformation in spin-induced multipole parameter & $\delta \kappa_s$ & 1.1--1.2 \\
MDR & \ref{sec:liv}& Magnitude of dispersion& $|A_{\alpha}|$ & $\LivImprov{AMP_min}$--$\LivImprov{AMP_max}$
 \\
POL& \ref{sec:pol}& Bayes Factors between different polarization hypotheses & $\log_{10}{\cal B}^{\rm X}_{\rm T}$ & New Test 
\\
RD & \ref{subsec:rin1}&Fractional deviations in frequency (\textsc{pyRing}) &
 $\delta \hat{f}_{221}$ & 1.1\\
 & \ref{subsec:rin3}&Fractional deviations in frequency and damping time (\textsc{pSEOB}) &$\bigl\{\delta \hat{f}_{220},\delta \hat{\tau}_{220}\bigr\}$ & 1.7--5.5
\\

ECH & \ref{sec:ech}&  Signal-to-noise Bayes Factor & $\log_{10}{\rm {\cal B}}_{\rm S/N}$ & New Test
\\
\bottomrule
\end{tabular}
\end{table*}

The detected coalescences of massive compact objects may involve not only black holes of classical GR, but also different compact objects described by exotic physics, commonly referred to as exotic compact objects (ECOs). They include objects like firewalls \cite{Almheiri:2012rt}, fuzzballs \cite{Lunin:2001jy}, gravastars \cite{Mazur:2004fk}, boson stars \cite{Liebling:2012fv}, AdS black bubbles~\cite{Danielsson:2021ykm} and dark matter stars \cite{Giudice_2016}. What these objects have in common is the absence of a horizon, causing ingoing gravitational waves (e.g., resulting from merger) to reflect multiple times off effective radial potential barriers, with wave packets leaking  out to infinity, potentially, at regular times; these are called echoes \cite{Cardoso:2016rao,Cardoso:2017cqb,Cardoso:2016oxy}.

We also perform a  series of tests that search for possible GR violation or non-Kerr nature of the merger remnant, specifically in the postinspiral part of the waveform. Ringdown tests~\cite{Dreyer:2003bv,Berti:2005ys,Gossan:2011ha,PhysRevD.90.064009} probe the consistency of the post-merger dynamics with the predictions for Kerr black holes in GR, while searches for \emph{echoes} constrain the presence of repeating ringdown signals~\cite{Cardoso:2016rao,Cardoso:2017cqb,Abedi:2016hgu,Ashton:2016xff,Westerweck:2017hus,Uchikata:2019frs} expected in certain classes of exotic compact objects (ECOs).

Tests of GR performed on the data of the previous observing runs have set increasingly stringent limits~\cite{LIGOScientific:2019fpa,LIGOScientific:2020tif}. The bounds on the parameterized post-Newtonian deformation coefficients have been used to constrain the parameter space of alternatives to GR~\cite{Yunes:2016jcc,Nair:2019iur} relying on several assumptions about the underlying GR violation (critiqued in~\cite{Gupta:2020lxa,Johnson-McDaniel:2021yge}).
The joint detection of gravitational waves and the gamma rays from the binary neutron star merger GW170817 has placed extremely stringent bounds on the speed of gravitational waves~\cite{Monitor:2017mdv}, which in turn has had a profound impact on constraining certain classes of modified theories from a cosmological standpoint~\cite{Baker:2017hug,Creminelli:2017sry,Sakstein:2017xjx,Ezquiaga:2017ekz}. This joint detection also has been used to constrain the number of space-time dimensions~\cite{Pardo:2018ipy,Abbott:2018lct}. 
Results on the measurement of the properties of the merger remnant from ringdown, bounds on spin-induced quadrupole moment of compact binary constituents and the outcome of the search for echoes have implications for models of black hole mimickers~\cite{Cardoso:2019rvt}.

This paper analyzes events reported during the second half of the third observing run of LIGO and Virgo (O3b)~\cite{GWTC3}, extending our previous analysis~\cite{LIGOScientific:2020tif} which reported the status of the bounds up to and including the first half of the third observing run (O3a). 
This paper also provides joint bounds combining the events that occurred during the first three observing runs whenever possible, in addition to deriving limits on possible departures from GR for individual events. 

The analysis methods used here are largely similar to those used in our O3a analysis~\cite{LIGOScientific:2020tif}, and there are only significant differences for the polarization-based test of GR and the search for post-merger echoes. The polarization test was updated to probe mixed polarization content using a framework based upon the null stream.~\cite{Pang:2020pfz}. The morphology-dependent search for echoes is replaced by a wavelet-based~\cite{Cornish:2014kda} morphology-independent search~\cite{Tsang:2018uie,Tsang:2019zra}. Besides these changes, the waveform models employed in most analyses have been upgraded to more accurate and complete ones, accounting for more physics, the details of which are discussed in Sec.~\ref{sec:inference}. 
Table~\ref{tab:summary} summarizes the tests that are performed, the quantities that are used for the test, the fractional changes with regard to the previous analyses, and the section where details about each test can be found.

The paper is organized as follows. Section~\ref{sec:events} discusses the data used in this paper while Sec.~\ref{sec:inference} describes the method of extracting astrophysical information about events from the data. Section~\ref{sec:con}  discusses two tests of consistency with GR:  examining the residuals left in the data after subtracting the best-fit GR waveform (Sec.~\ref{sec:res}) and looking at consistency of the inspiral and postinspiral portions of the waveform with GR (Sec.~\ref{sec:imr}). Section \ref{sec:gen} discusses two tests of gravitational-wave generation, the parameterized tests of GR (Sec.~\ref{sec:par}) and the test for BBH nature using the spin-induced quadrupole moment (Sec.\ref{sec:sim}). Section \ref{sec:liv} discusses tests of gravitational-wave propagation looking for non-GR dispersion of gravitational waves. Section~\ref{sec:pol} reports results  from the searches for non-GR polarization. Section \ref{sec:rem} discusses various tests using the merger remnants, specifically two analyses of ringdown (Sec.~\ref{sec:rin}) and a search for the signatures of echoes (Sec.~\ref{sec:ech}). Section \ref{sec:conclusion} discusses the conclusions.

\section{Data, events, and significance}
\label{sec:events}

The global network of gravitational wave detectors completed their third observing run in March 2020.  O3b adds \NUMEVENTS{} candidate events with probability of being of astrophysical origin better than 0.5, including the first confident observations of NSBH systems \cite{GWTC3, LIGOScientific:2021qlt}.  The analyses presented here are focused on events from O3b, though the joint bounds that are reported also include events from previous observing runs.  
Following our O3a analysis \cite{LIGOScientific:2020tif}, we consider only those events with false alarm rates lower than $10^{-3}$ per year that were confidently observed in two or more detectors as determined by any search pipeline used in the catalog of O3b events \cite{GWTC3}.\footnote{Though some events included in the O3a analysis were assigned a false alarm rate above this threshold in the GWTC-2.1 catalog~\cite{LIGOScientific:2021usb}, we retain them here for consistency with \cite{LIGOScientific:2020tif}.} Of the 14 BBH mergers and the one NSBH merger (\NNAME{GW200115A}) that pass this threshold, nine events are observed with three detectors and six are observed with only two detectors. The median total masses in the detector frame of these analysed events range from $\sim$ 8--140 $M_{\odot}.$

The LIGO interferometers maintained sensitivities comparable to that in O3a~\cite{Buikema:2020dlj}, and the Virgo interferometer achieved a $\sim20\%$ improvement during O3b.  As with previous results, noise subtraction methods~\cite{Driggers:2018gii,Davis:2018yrz,Vajente:2019ycy,Cornish:2014kda} were applied to selected events in order to improve parameter estimation.  The third Gravitational-Wave Transient Catalog (GWTC-3)~\cite{GWTC3} includes details on instrument performance. 
Three events analysed here were identified in as requiring additional data quality mitigation.  One, \NNAME{GW191109A} \ had data quality issues in both LIGO detectors, while the other two (\NNAME{GW200115A} \  and \NNAME{GW200129D}) were each only affected by noise transients in one interferometer.  Details on the noise transient removal techniques can be found in GWTC-3's Table XIV.  Appendix~\ref{app:sys} below discusses how these data quality issues can affect analyses in this article.

Table~\ref{tab:events} shows selected source properties of events from the O3b observing run that are included in this paper.  Similar tables in O2~\cite{LIGOScientific:2019fpa} and O3a~\cite{LIGOScientific:2020tif} analyses provide selected source properties for the other events included in the analyses presented here.  
Every test detailed in the following sections has its own sub-selection criteria of events from this table based on the physics it explores; these criteria are detailed in the respective sections for each analysis.

Detection significance is given by four search pipelines, three of which rely on GR-based templates (\pycbc{}~\cite{pycbc-github, Canton:2014ena, Usman:2015kfa}, MBTA~\cite{Aubin:2020goo}, and \gstlal{}~\cite{Sachdev:2019vvd,Messick:2016aqy})
  and one that does not (\textsc{coherent WaveBurst}; \cwb{}~\cite{Klimenko:2008fu, Klimenko:2015ypf, TheLIGOScientific:2016uux}).  Details on these pipelines, including the two different \pycbc{} configurations used, is included in Appendix D of GWTC-3~\cite{GWTC3}.
While making significance determinations using searches based on GR could potentially lead to selection biases disfavoring events that deviate significantly from GR, the use of the minimally modeled search \cwb{} helps alleviate this concern: \cwb{} is sensitive to at least some of the potential chirp-like signals that deviate enough from GR that they would not be detected with high significance by the other three pipelines.  However, we are unable to fully exclude the possibility of a hidden population of signals that show significant departures from GR thus evading detection by all of our pipelines.

%
%
\begin{table*}
\caption{\label{tab:events}
List of O3b events considered in this paper.
The first block of columns gives the names of the events and lists the instruments (LIGO {\bf H}anford, LIGO {\bf L}ivingston, {\bf V}irgo) involved in each detection, as well as some relevant properties obtained assuming GR: luminosity distance $D_\text{L}$, redshifted total mass $(1+z)M$, redshifted chirp mass $(1+z)\mathcal{M}$, redshifted final mass $(1+z)M_\text{f}$, dimensionless final spin $\chi_\text{f} = c |\vec{S}_{\rm f}| / (G M^2_{\rm f})$, and network signal-to-noise ratio SNR.
Reported quantities correspond to the median and 90\% symmetric credible intervals, as computed in Table IV in GWTC-3 \cite{GWTC3}.  The final mass and final spin quantities are inferred from analysis of the entire signal and are for the remnant long after the coalescence and ringdown are complete, as described in~\cite{GW170104}.
The last block of columns indicates which analyses are performed on a given event according to the selection criteria in Sec.~\ref{sec:events}: RT = residuals test (Sec.~\ref{sec:res}); IMR = inspiral--merger--ringdown consistency test (Sec.~\ref{sec:imr}); PAR = parametrized tests of gravitational-wave generation (Sec.~\ref{sec:par}); SIM = spin-induced moments (Sec.~\ref{sec:sim}); MDR = modified gravitational-wave dispersion relation (Sec.~\ref{sec:liv}); POL = polarization content (Sec.~\ref{sec:pol}); RD = ringdown (Sec.~\ref{sec:rin}); ECH = echoes searches (Sec.~\ref{sec:ech}).   
}  
\begin{tabular}{ l l l l l l@{\;} l@{\;} l@{\;} l l c l c c c c c c c c}
\toprule
\multirow{2}{*}{Event} & \hphantom{X} & \multirow{2}{*}{Inst.} & \hphantom{X} & \multicolumn{5}{c}{Properties} & \hphantom{} & \multirow{2}{*}{SNR} & \hphantom{X} & \multicolumn{8}{c}{Tests performed}\\
\cline{5-9}
\cline{13-20}
& & & &\multicolumn{1}{c}{$D_\text{L}$} & \multicolumn{1}{c}{$(1+z)M$} & \multicolumn{1}{c}{$(1+z)\mathcal{M}$} &\multicolumn{1}{c}{$(1+z)M_\text{f}$} & \multicolumn{1}{c}{$\chi_\text{f}$} & & & & RT & IMR & PAR \ & SIM & MDR & POL & RD & ECH\\
& & & &\multicolumn{1}{c}{ [Gpc]} &\multicolumn{1}{c}{ [$M_\odot$] }&\multicolumn{1}{c}{ [$M_\odot$] }&\multicolumn{1}{c}{ [$M_\odot$]} & & & & & & & & & & &\\
\midrule
\NNAME{GW191109A} & &\INSTRUMENTS{GW191109A} & & $\luminositydistancemed{GW191109A}^{+\luminositydistanceplus{GW191109A}}_{-\luminositydistanceminus{GW191109A}}$ & $\totalmassdetmed{GW191109A}^{+\totalmassdetplus{GW191109A}}_{-\totalmassdetminus{GW191109A}}$ & $\chirpmassdetmed{GW191109A}^{+\chirpmassdetplus{GW191109A}}_{-\chirpmassdetminus{GW191109A}}$ & $\finalmassdetmed{GW191109A}^{+\finalmassdetplus{GW191109A}}_{-\finalmassdetminus{GW191109A}}$ & $\finalspinmed{GW191109A}^{+\finalspinplus{GW191109A}}_{-\finalspinminus{GW191109A}}$ & & $\networkmatchedfiltersnrIMRPmed{GW191109A}^{+\networkmatchedfiltersnrIMRPplus{GW191109A}}_{-\networkmatchedfiltersnrIMRPminus{GW191109A}}$ & & \cmark & -- & -- & --  & --    & \cmark & \cmark & \cmark \\[0.075cm]

\NNAME{GW191129G} & &\INSTRUMENTS{GW191129G} & & $\luminositydistancemed{GW191129G}^{+\luminositydistanceplus{GW191129G}}_{-\luminositydistanceminus{GW191129G}}$ & $\totalmassdetmed{GW191129G}^{+\totalmassdetplus{GW191129G}}_{-\totalmassdetminus{GW191129G}}$ & $\chirpmassdetmed{GW191129G}^{+\chirpmassdetplus{GW191129G}}_{-\chirpmassdetminus{GW191129G}}$ & $\finalmassdetmed{GW191129G}^{+\finalmassdetplus{GW191129G}}_{-\finalmassdetminus{GW191129G}}$ & $\finalspinmed{GW191129G}^{+\finalspinplus{GW191129G}}_{-\finalspinminus{GW191129G}}$ & & $\networkmatchedfiltersnrIMRPmed{GW191129G}^{+\networkmatchedfiltersnrIMRPplus{GW191129G}}_{-\networkmatchedfiltersnrIMRPminus{GW191129G}}$ & & \cmark & -- & \cmark & \cmark  & \cmark  & --  & -- & \cmark\\[0.075cm]

\NNAME{GW191204G} & &\INSTRUMENTS{GW191204G} & & $\luminositydistancemed{GW191204G}^{+\luminositydistanceplus{GW191204G}}_{-\luminositydistanceminus{GW191204G}}$ & $\totalmassdetmed{GW191204G}^{+\totalmassdetplus{GW191204G}}_{-\totalmassdetminus{GW191204G}}$ & $\chirpmassdetmed{GW191204G}^{+\chirpmassdetplus{GW191204G}}_{-\chirpmassdetminus{GW191204G}}$ & $\finalmassdetmed{GW191204G}^{+\finalmassdetplus{GW191204G}}_{-\finalmassdetminus{GW191204G}}$ & $\finalspinmed{GW191204G}^{+\finalspinplus{GW191204G}}_{-\finalspinminus{GW191204G}}$ & & $\networkmatchedfiltersnrIMRPmed{GW191204G}^{+\networkmatchedfiltersnrIMRPplus{GW191204G}}_{-\networkmatchedfiltersnrIMRPminus{GW191204G}}$ & & \cmark & -- & \cmark & \cmark  & \cmark  & \cmark  & -- & \cmark \\[0.075cm]

\NNAME{GW191215G} & &\INSTRUMENTS{GW191215G} & & $\luminositydistancemed{GW191215G}^{+\luminositydistanceplus{GW191215G}}_{-\luminositydistanceminus{GW191215G}}$ & $\totalmassdetmed{GW191215G}^{+\totalmassdetplus{GW191215G}}_{-\totalmassdetminus{GW191215G}}$ & $\chirpmassdetmed{GW191215G}^{+\chirpmassdetplus{GW191215G}}_{-\chirpmassdetminus{GW191215G}}$ & $\finalmassdetmed{GW191215G}^{+\finalmassdetplus{GW191215G}}_{-\finalmassdetminus{GW191215G}}$ & $\finalspinmed{GW191215G}^{+\finalspinplus{GW191215G}}_{-\finalspinminus{GW191215G}}$ & & $\networkmatchedfiltersnrIMRPmed{GW191215G}^{+\networkmatchedfiltersnrIMRPplus{GW191215G}}_{-\networkmatchedfiltersnrIMRPminus{GW191215G}}$ & & \cmark & -- & -- & --  & \cmark  & \cmark    & -- & \cmark \\[0.075cm]

\NNAME{GW191216G} & &\INSTRUMENTS{GW191216G} & & $\luminositydistancemed{GW191216G}^{+\luminositydistanceplus{GW191216G}}_{-\luminositydistanceminus{GW191216G}}$ & $\totalmassdetmed{GW191216G}^{+\totalmassdetplus{GW191216G}}_{-\totalmassdetminus{GW191216G}}$ & $\chirpmassdetmed{GW191216G}^{+\chirpmassdetplus{GW191216G}}_{-\chirpmassdetminus{GW191216G}}$ & $\finalmassdetmed{GW191216G}^{+\finalmassdetplus{GW191216G}}_{-\finalmassdetminus{GW191216G}}$ & $\finalspinmed{GW191216G}^{+\finalspinplus{GW191216G}}_{-\finalspinminus{GW191216G}}$ & & $\networkmatchedfiltersnrIMRPmed{GW191216G}^{+\networkmatchedfiltersnrIMRPplus{GW191216G}}_{-\networkmatchedfiltersnrIMRPminus{GW191216G}}$ & & \cmark & -- & \cmark & \cmark  &\cmark  & \cmark  & -- & \cmark \\[0.075cm]

\NNAME{GW191222A} & &\INSTRUMENTS{GW191222A} & & $\luminositydistancemed{GW191222A}^{+\luminositydistanceplus{GW191222A}}_{-\luminositydistanceminus{GW191222A}}$ & $\totalmassdetmed{GW191222A}^{+\totalmassdetplus{GW191222A}}_{-\totalmassdetminus{GW191222A}}$ & $\chirpmassdetmed{GW191222A}^{+\chirpmassdetplus{GW191222A}}_{-\chirpmassdetminus{GW191222A}}$ & $\finalmassdetmed{GW191222A}^{+\finalmassdetplus{GW191222A}}_{-\finalmassdetminus{GW191222A}}$ & $\finalspinmed{GW191222A}^{+\finalspinplus{GW191222A}}_{-\finalspinminus{GW191222A}}$ & & $\networkmatchedfiltersnrIMRPmed{GW191222A}^{+\networkmatchedfiltersnrIMRPplus{GW191222A}}_{-\networkmatchedfiltersnrIMRPminus{GW191222A}}$ & & \cmark & -- & -- & -- & \cmark   & \cmark & \cmark & \cmark \\[0.075cm]

\NNAME{GW200115A} & &\INSTRUMENTS{GW200115A} & & $\luminositydistancemed{GW200115A}^{+\luminositydistanceplus{GW200115A}}_{-\luminositydistanceminus{GW200115A}}$ & $\totalmassdetmed{GW200115A}^{+\totalmassdetplus{GW200115A}}_{-\totalmassdetminus{GW200115A}}$ & $\chirpmassdetmed{GW200115A}^{+\chirpmassdetplus{GW200115A}}_{-\chirpmassdetminus{GW200115A}}$ & $\finalmassdetmed{GW200115A}^{+\finalmassdetplus{GW200115A}}_{-\finalmassdetminus{GW200115A}}$ & $\finalspinmed{GW200115A}^{+\finalspinplus{GW200115A}}_{-\finalspinminus{GW200115A}}$ & & $\networkmatchedfiltersnrIMRPmed{GW200115A}^{+\networkmatchedfiltersnrIMRPplus{GW200115A}}_{-\networkmatchedfiltersnrIMRPminus{GW200115A}}$ & & \cmark & -- & \cmark & --  & --  & --  & -- & \cmark \\[0.075cm]

\NNAME{GW200129D} & &\INSTRUMENTS{GW200129D} & & $\luminositydistancemed{GW200129D}^{+\luminositydistanceplus{GW200129D}}_{-\luminositydistanceminus{GW200129D}}$ & $\totalmassdetmed{GW200129D}^{+\totalmassdetplus{GW200129D}}_{-\totalmassdetminus{GW200129D}}$ & $\chirpmassdetmed{GW200129D}^{+\chirpmassdetplus{GW200129D}}_{-\chirpmassdetminus{GW200129D}}$ & $\finalmassdetmed{GW200129D}^{+\finalmassdetplus{GW200129D}}_{-\finalmassdetminus{GW200129D}}$ & $\finalspinmed{GW200129D}^{+\finalspinplus{GW200129D}}_{-\finalspinminus{GW200129D}}$ & & $\networkmatchedfiltersnrIMRPmed{GW200129D}^{+\networkmatchedfiltersnrIMRPplus{GW200129D}}_{-\networkmatchedfiltersnrIMRPminus{GW200129D}}$& & \cmark & \cmark & \cmark & \cmark  & \cmark   & \cmark & \cmark & \cmark   \\[0.075cm]

\NNAME{GW200202F} & &\INSTRUMENTS{GW200202F} & & $\luminositydistancemed{GW200202F}^{+\luminositydistanceplus{GW200202F}}_{-\luminositydistanceminus{GW200202F}}$ & $\totalmassdetmed{GW200202F}^{+\totalmassdetplus{GW200202F}}_{-\totalmassdetminus{GW200202F}}$ & $\chirpmassdetmed{GW200202F}^{+\chirpmassdetplus{GW200202F}}_{-\chirpmassdetminus{GW200202F}}$ & $\finalmassdetmed{GW200202F}^{+\finalmassdetplus{GW200202F}}_{-\finalmassdetminus{GW200202F}}$ & $\finalspinmed{GW200202F}^{+\finalspinplus{GW200202F}}_{-\finalspinminus{GW200202F}}$ & & $\networkmatchedfiltersnrIMRPmed{GW200202F}^{+\networkmatchedfiltersnrIMRPplus{GW200202F}}_{-\networkmatchedfiltersnrIMRPminus{GW200202F}}$ & & \cmark & -- & \cmark & --  & \cmark & --  & -- & \cmark  \\[0.075cm]

\NNAME{GW200208G} & &\INSTRUMENTS{GW200208G} & & $\luminositydistancemed{GW200208G}^{+\luminositydistanceplus{GW200208G}}_{-\luminositydistanceminus{GW200208G}}$ & $\totalmassdetmed{GW200208G}^{+\totalmassdetplus{GW200208G}}_{-\totalmassdetminus{GW200208G}}$ & $\chirpmassdetmed{GW200208G}^{+\chirpmassdetplus{GW200208G}}_{-\chirpmassdetminus{GW200208G}}$ & $\finalmassdetmed{GW200208G}^{+\finalmassdetplus{GW200208G}}_{-\finalmassdetminus{GW200208G}}$ & $\finalspinmed{GW200208G}^{+\finalspinplus{GW200208G}}_{-\finalspinminus{GW200208G}}$ & & $\networkmatchedfiltersnrIMRPmed{GW200208G}^{+\networkmatchedfiltersnrIMRPplus{GW200208G}}_{-\networkmatchedfiltersnrIMRPminus{GW200208G}}$ & & \cmark & \cmark & -- & -- & \cmark  & \cmark  & --  &\cmark  \\[0.075cm]

\NNAME{GW200219D} & &\INSTRUMENTS{GW200219D} & & $\luminositydistancemed{GW200219D}^{+\luminositydistanceplus{GW200219D}}_{-\luminositydistanceminus{GW200219D}}$ & $\totalmassdetmed{GW200219D}^{+\totalmassdetplus{GW200219D}}_{-\totalmassdetminus{GW200219D}}$ & $\chirpmassdetmed{GW200219D}^{+\chirpmassdetplus{GW200219D}}_{-\chirpmassdetminus{GW200219D}}$ & $\finalmassdetmed{GW200219D}^{+\finalmassdetplus{GW200219D}}_{-\finalmassdetminus{GW200219D}}$ & $\finalspinmed{GW200219D}^{+\finalspinplus{GW200219D}}_{-\finalspinminus{GW200219D}}$ & & $\networkmatchedfiltersnrIMRPmed{GW200219D}^{+\networkmatchedfiltersnrIMRPplus{GW200219D}}_{-\networkmatchedfiltersnrIMRPminus{GW200219D}}$& & \cmark & -- & -- & -- & \cmark & \cmark   & --  & \cmark \\[0.075cm]

\NNAME{GW200224H} & &\INSTRUMENTS{GW200224H} & & $\luminositydistancemed{GW200224H}^{+\luminositydistanceplus{GW200224H}}_{-\luminositydistanceminus{GW200224H}}$ & $\totalmassdetmed{GW200224H}^{+\totalmassdetplus{GW200224H}}_{-\totalmassdetminus{GW200224H}}$ & $\chirpmassdetmed{GW200224H}^{+\chirpmassdetplus{GW200224H}}_{-\chirpmassdetminus{GW200224H}}$ & $\finalmassdetmed{GW200224H}^{+\finalmassdetplus{GW200224H}}_{-\finalmassdetminus{GW200224H}}$ & $\finalspinmed{GW200224H}^{+\finalspinplus{GW200224H}}_{-\finalspinminus{GW200224H}}$ & & $\networkmatchedfiltersnrIMRPmed{GW200224H}^{+\networkmatchedfiltersnrIMRPplus{GW200224H}}_{-\networkmatchedfiltersnrIMRPminus{GW200224H}}$ & & \cmark & \cmark & -- & -- & \cmark & \cmark  & \cmark & \cmark  \\[0.075cm]

\NNAME{GW200225B} & &\INSTRUMENTS{GW200225B} & & $\luminositydistancemed{GW200225B}^{+\luminositydistanceplus{GW200225B}}_{-\luminositydistanceminus{GW200225B}}$ & $\totalmassdetmed{GW200225B}^{+\totalmassdetplus{GW200225B}}_{-\totalmassdetminus{GW200225B}}$ & $\chirpmassdetmed{GW200225B}^{+\chirpmassdetplus{GW200225B}}_{-\chirpmassdetminus{GW200225B}}$ & $\finalmassdetmed{GW200225B}^{+\finalmassdetplus{GW200225B}}_{-\finalmassdetminus{GW200225B}}$ & $\finalspinmed{GW200225B}^{+\finalspinplus{GW200225B}}_{-\finalspinminus{GW200225B}}$ & & $\networkmatchedfiltersnrIMRPmed{GW200225B}^{+\networkmatchedfiltersnrIMRPplus{GW200225B}}_{-\networkmatchedfiltersnrIMRPminus{GW200225B}}$ & & \cmark & \cmark & \cmark & \cmark  & \cmark  & \cmark & -- & \cmark  \\[0.075cm]

\NNAME{GW200311L} & &\INSTRUMENTS{GW200311L} & & $\luminositydistancemed{GW200311L}^{+\luminositydistanceplus{GW200311L}}_{-\luminositydistanceminus{GW200311L}}$ & $\totalmassdetmed{GW200311L}^{+\totalmassdetplus{GW200311L}}_{-\totalmassdetminus{GW200311L}}$ & $\chirpmassdetmed{GW200311L}^{+\chirpmassdetplus{GW200311L}}_{-\chirpmassdetminus{GW200311L}}$ & $\finalmassdetmed{GW200311L}^{+\finalmassdetplus{GW200311L}}_{-\finalmassdetminus{GW200311L}}$ & $\finalspinmed{GW200311L}^{+\finalspinplus{GW200311L}}_{-\finalspinminus{GW200311L}}$ & & $\networkmatchedfiltersnrIMRPmed{GW200311L}^{+\networkmatchedfiltersnrIMRPplus{GW200311L}}_{-\networkmatchedfiltersnrIMRPminus{GW200311L}}$& & \cmark & \cmark & \cmark & --  & \cmark  & \cmark  & \cmark & \cmark  \\[0.075cm]

\NNAME{GW200316I} & &\INSTRUMENTS{GW200316I} & & $\luminositydistancemed{GW200316I}^{+\luminositydistanceplus{GW200316I}}_{-\luminositydistanceminus{GW200316I}}$ & $\totalmassdetmed{GW200316I}^{+\totalmassdetplus{GW200316I}}_{-\totalmassdetminus{GW200316I}}$ & $\chirpmassdetmed{GW200316I}^{+\chirpmassdetplus{GW200316I}}_{-\chirpmassdetminus{GW200316I}}$ & $\finalmassdetmed{GW200316I}^{+\finalmassdetplus{GW200316I}}_{-\finalmassdetminus{GW200316I}}$ & $\finalspinmed{GW200316I}^{+\finalspinplus{GW200316I}}_{-\finalspinminus{GW200316I}}$ & & $\networkmatchedfiltersnrIMRPmed{GW200316I}^{+\networkmatchedfiltersnrIMRPplus{GW200316I}}_{-\networkmatchedfiltersnrIMRPminus{GW200316I}}$ & & \cmark & -- & \cmark & \cmark   & -- & -- & -- & \cmark  \\[0.075cm]

\bottomrule
\end{tabular}
\end{table*}

\section{Parameter inference}
\label{sec:inference}
Many of the tests performed here build upon two BBH waveform families. Consistency tests, as well as tests of gravitational-wave generation and propagation are based on precessing phenomenological (\texttt{Phenom}) models. Phenomenological models are currently constructed following the twisting-up technique \cite{Schmidt:2010it,Schmidt:2012rh,Hannam:2013oca}, where the signal in a co-precessing frame is approximated by an aligned-spin model and is then transformed to the detector's inertial frame by means of frequency (or time) dependent transfer functions. Previous works made use of the frequency-domain \textsc{PhenomP} family, which includes \textsc{PhenomPv2} and \textsc{PhenomPv3HM} \cite{Khan:2019kot}, where version numbers distinguish between different PN-based approximations of the precession dynamics, with v2 (v3) adopting a single (double) spin prescription. The suffix ``HM'' is appended to indicate that the model includes some higher-order multipole moments.  We will employ these models only for spin-induced quadrupole moment tests and otherwise upgrade to the most recent frequency-domain \textsc{PhenomX} family \cite{Pratten:2020fqn,Garcia-Quiros:2020qpx}, which includes \IMRP{} and \IMRPHM{} \cite{Pratten:2020ceb}. While the precession prescriptions adopted by \textsc{PhenomPv3HM} and \IMRPHM{} are similar, the aligned-spin baseline of \textsc{PhenomX}, including the higher-order multipoles $(2,1),(3,3),(3,2)$ and $(4,4)$, is calibrated to a larger set of numerical simulations with respect to its predecessors. 

For parameterized tests of PN theory (Sec.\,\ref{sec:par}), as well as for the ringdown test of Sec.\,\ref{subsec:rin3}, we also employ the spinning effective one-body family (SEOB) \cite{Buonanno:1998gg,BuonannoDamour:2000,DIN,Barausse:2009xi}. In the EOB framework, the two-body dynamics is mapped onto that of an effective body moving in a deformed black-hole spacetime, the deformation being the symmetric mass ratio. The dynamics and gravitational radiation are obtained solving numerically the Hamilton equations, which are built from the EOB Hamiltonian and radiation-reaction force. The latter combines information from PN theory, gravitational self-force, perturbation theory, and NR simulations \cite{buonanno:124018,Barack:2009ey,Blanchet:2013haa,Pan:2013rra,Taracchini:2013}. In this work, we will employ approximants belonging to the latest generation of SEOB models, \SEOBNR{} \cite{Bohe:2016gbl}, specifically the aligned-spin model \SEOBNRHM{} \cite{Cotesta:2018fcv} and the frequency-domain model \SEOBROM, a reduced-order-model version of \SEOBNR{}, which allows a significant speed up with respect to the original time-domain version \cite{Bohe:2016gbl,Cotesta:2020qhw}. On top of the leading $(2,2)$ multipole, the \SEOBNRHM{} model also includes the spherical harmonic multipoles $(2,1),(3,3),(4,4),(5,5)$. 

For most tests, our choice of waveform model is dictated by the need for computational efficiency. However, for some tests, this is dictated by technical and physical considerations. For instance, we use time-domain waveform models for the ringdown tests, since such tests are most conveniently formulated in the time domain \cite{Isi:2021iql,Carullo_02,Ghosh:2021mrv}.

Most template-based tests presented here rely on the assumption that the signal was generated in a quasi-circular BBH coalescence. It has been shown that signatures of eccentricity may lead to biases in the estimated source parameters \cite{Klein:2018ybm,Lower:2018seu,Favata:2021vhw,Romero-Shaw:2019itr}. The issue could especially affect the analysis of short-duration signals such as GW190521, where the complementarity between inspiral and merger--ringdown cannot be efficiently leveraged to break degeneracies in current templates \cite{Romero-Shaw:2020thy,Gayathri:2020coq,Bustillo:2020ukp,Gamba:2021gap,Estelles:2021jnz}.  

Comprehensive BBH waveform models including precession, higher-order multipole moments \emph{and} eccentricity are not yet available, and a thorough assessment of the impact of waveform systematics is beyond the scope of this analysis. Work in this direction is crucial, as inaccuracies in waveform templates could lead to detecting false violations of GR \cite{Pang:2018hjb}. As for higher-order multipole moments, their relative contribution can be estimated by computing their SNR distribution on a reference set of posterior samples \cite{Mills:2020thr,GW190814,GW190412,LIGOScientific:2021qlt}. Based on the \IMRPHM{} parameter estimation runs presented in GWTC-3 \cite{GWTC3}, we find that the SNR distributions for individual higher-order multipoles is consistent with Gaussian noise for all the events reported in Table \ref{tab:events}, though the sum of several subdominant harmonics does contribute a non-negligible amount of SNR for \NNAME{GW191109A}{}. Therefore, we choose to neglect them in the most computationally expensive tests, restricting their inclusion to the residuals, IMR consistency, and ringdown tests. 

Our event list includes the NSBH candidate \NNAME{GW200115A}. We do not consider matter effects in its analysis and expect them to be negligible: tidal imprints are suppressed by the relatively extreme mass ratio of the system and a higher SNR would be needed for them to be confidently detected \cite{Kumar:2016zlj,Huang:2020pba,LIGOScientific:2021qlt}.  

We perform Bayesian parameter estimation using either the \texttt{Python}-based package \texttt{bilby} \cite{Ashton:2018jfp,Romero-Shaw:2020owr} or the software library \texttt{LALInference} \cite{Veitch:2014wba}, which belongs to the LIGO Scientific Collaboration Algorithm Library Suite \cite{lalsuite}. Runs performed with  \texttt{bilby} are automated through the package \texttt{bilby pipe}\,\cite{Romero-Shaw:2020owr} and rely on static nested sampling \cite{Skilling:2004ns} as available through the package \texttt{dynesty} \cite{Speagle:2020spe}, while \texttt{LALInference} runs use either nested or Markov-chain Monte Carlo sampling \cite{MCMC}. One of the ringdown tests presented here makes use of \texttt{pyRing} \cite{pyRing_soft}, which relies on \texttt{cpnest} \cite{CPNest}. Power spectral densities are generated through the software \texttt{BAYESWAVE} \cite{Cornish:2014kda,Littenberg:2014oda} and match those used in GWTC-3 \cite{GWTC3}, unless otherwise stated.

When combining results obtained from multiple events, we employ two methods. The first method relies on the multiplication of the individual likelihoods corresponding to the deformation parameters that are inferred from the data. This method assumes that the deformation parameters take the same value across events~\cite{Zimmerman:2019wzo}. This is a restrictive assumption for all the tests we consider except for the modified dispersion test. In order to address this, wherever possible, we combine the information from the tests for different events hierarchically using a model that does not make this restrictive assumption and hence provides bounds which are qualitatively more robust~\cite{Isi:2019asy}. 

We quantify the agreement of our results with GR using several statistical indicators. In Section \ref{sec:imr} we present GR quantiles $\QGR$ for joint distributions, which denote the fraction of the likelihood contained within the isoprobability contour passing through the GR value, with a smaller GR quantile indicating a better agreement with GR. In Sections \ref{sec:par} and \ref{sec:liv}, we report instead quantiles on one-dimensional distributions $Q_{\rm{GR}}$. When error bars are reported, they denote $90\%$ confidence intervals and, likewise, we show $90\%$ credible regions (intervals) when presenting joint (individual) posterior distributions.

\section{Consistency tests}
\label{sec:con}

    \subsection{Residuals test}
    \label{sec:res}
    \newcommand{\rhores}{\mathrm{SNR}_{90}}
\newcommand{\rhosig}{\mathrm{SNR}_\mathrm{GR}}
\newcommand{\ff}{\mathrm{FF}_{90}}

Measuring the  remnant coherent power  in the network data after the subtraction of the best-fit GR template 
can be used to quantify consistency of GR waveform model with the data. The random noise in different detectors can be taken to be incoherent. The presence of consistent noise in the network after removing the gravitational-wave signal from the data indicates an inconsistency between the signal present in the data and the GR template used. The residual analysis is designed to detect such  discrepancies of the data with GR~\cite{PhysRevLett.116.221101,LIGOScientific:2019fpa,GW190521g:imp,GWTC2:TGR:release}.

A residual data  set is obtained by subtracting the waveform corresponding to maximum-likelihood parameters  from individual detector data with a window size of one second around the trigger time. The window size of one second is used due to the relatively short length of the signal.  Then the  residual SNR, or $\rhores$, is computed as the $90\%$ credible upper limit on the remnant coherent network SNR in the residual data using the \texttt{BAYESWAVE} pipeline~\cite{bayeswave,Cornish:2014kda,Littenberg:2014oda}. \texttt{BAYESWAVE} uses a  template-independent model to characterize any excess power in the residual compared to the detector noise. 

We follow the method used in the previous analyses performed in GWTC-1~\cite{LIGOScientific:2019fpa} and GWTC-2~\cite{LIGOScientific:2020tif}. However,  we use the new phenomenological waveform model PhenomXPHM \cite{Pratten:2020fqn,Garcia-Quiros:2020qpx} as the GR waveform model. 
For each gravitational wave event, in addition to calculating $\rhores$, additional \texttt{BAYESWAVE} runs are done on two hundred randomly selected time segments on a time window of  4096\,s symmetric around the event time. This allows us to calculate $p$-values of residual SNRs for individual events, which is equal to the probability of obtaining a background value of $\rhores$ higher than that of the event.
We perform the analysis on all the events listed in Table \ref{tab:events}.

\begin{figure}
	\centering
	\includegraphics[width=\columnwidth]{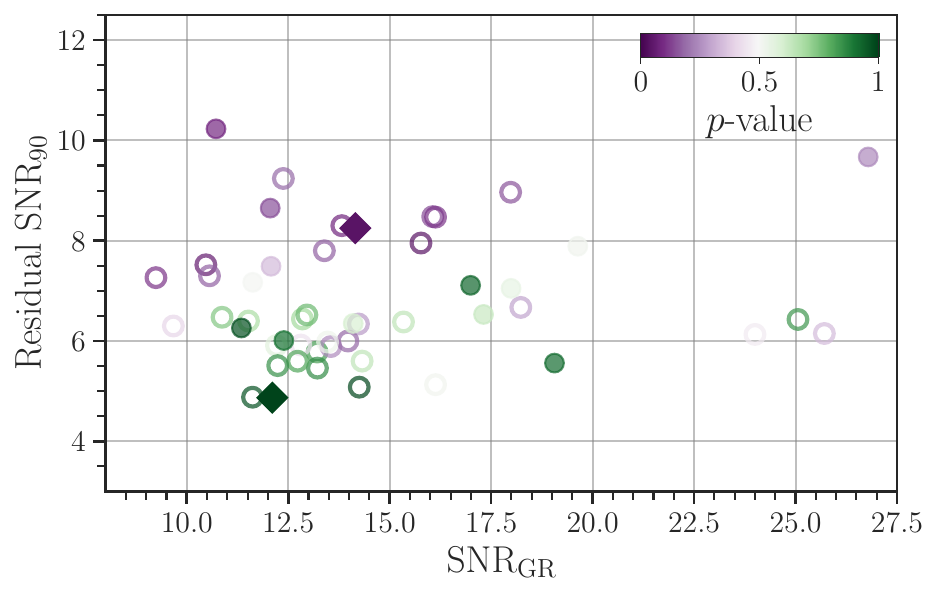}
	\caption{ Results of the residuals analysis (Sec.~\ref{sec:res}). Scatter plot of the maximum-likelihood template (${\rm SNR}_{\rm GR}$) and the  upper limit on the residual network SNR (${\rm SNR}_{90}$) for each event. The colorbar denotes the  $p$-values of individual events. Solid (empty) circles represent the  O3b (pre-O3b) events. The O3b events with highest (lowest) $p$-values are highlighted by green (purple) diamonds.
	}
	\label{fig:res:snr}
\end{figure}

The results from the residual analysis are summarized in Table \ref{tab:residuals}. For each event,  we have presented the SNR of the best-fit waveform $\rhosig$,  $\rhores$, fitting factor $\ff= \rhosig\, / (\rhores^2 + \rhosig^2)^{1/2}$, and $p$-values calculated from the background analysis. To analyze the trends between $\rhores$ and $\rhosig$, in Fig.~\ref{fig:res:snr} we  present the  scatter  of $\rhores$ and $\rhosig$. The absence of  correlation between $\rhores$ and $\rhosig$ in the figure indicates that data is consistent with GR templates and the values of $\rhores$ depend purely on the noise levels in the detectors at the detection of individual events. \NNAME{GW191222A}~shows the highest $p$-value~$= 1.0$ with  $\rhores= 4.87$ and  $\ff = 0.93$. Even though, \NNAME{GW200219D}~has the lowest fitting factor $\ff = 0.74$ with  $\rhores= 10.23$, its  $p$-value~$= 0.1$ is slightly above the lowest $p$-value~$= 0.05$ which corresponds to the event  \NNAME{GW200225B}. 

\begin{figure}
	\centering
	\includegraphics[width=\columnwidth]{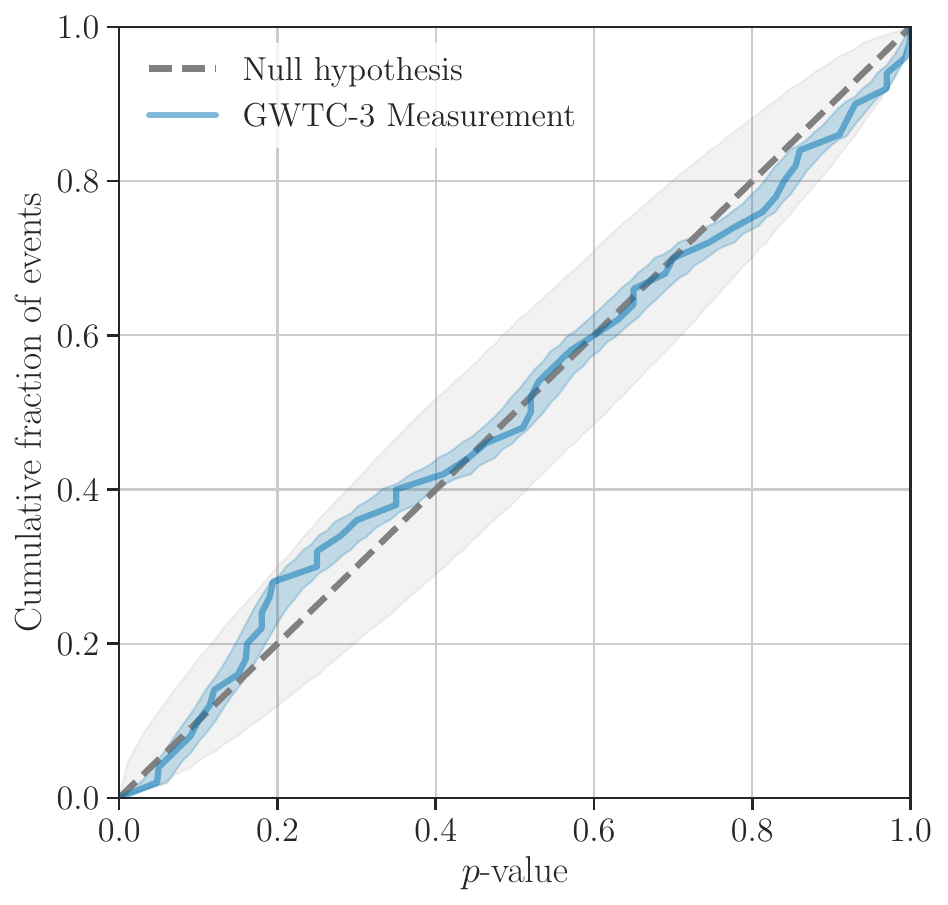}
	\caption{ Results of the residuals analysis (Sec.~\ref{sec:res}). The blue curve shows the fraction of events with $p$-values of the residual SNR  less than or equal to the abscissa (PP plot).  
		The light-blue  band  represents the $90\%$ credible interval of the observed $p$-values. The diagonal dashed line denotes the null hypothesis with the surrounding light-grey area denoting the $90\%$ uncertainty region of
		the null hypothesis due to the finite number of events.  
	}
	\label{fig:res:pp}
\end{figure}

If the left-over coherent network SNR were  purely from detector noise, we should expect the ${\rm SNR}_{90}$ $p$-values to be uniformly distributed within $[0, 1]$. 
To demonstrate the consistency of the observed  $p$-values with the noise (null) hypothesis, in Fig.~\ref{fig:res:pp}, we  present a probability--probability (PP) plot of the $p$-values.\footnote{However, the equivalent plot in \cite{LIGOScientific:2020tif} was between the observed p-values and the predicted p-values. See Appendix A of \cite{LIGOScientific:2020tif} for details.} To produce the PP plot, we have considered all the events in GWTC-3 that pass the FAR threshold. The measurement of $p$-values is subjected to uncertainty due to the  finite  size of background runs. If $N$ is the total number of  background trials around an event, and if $n$ of them produce $\rhores$ greater than that of the event, then the likelihood of the estimated $p$-value $\hat{p} = n/N$ is a binomial function,
\begin{equation}
	\mathcal{L}\left(\hat{p}\right) =  \binom{N}{n}~ p^n~ (1-p)^{N-n},
\end{equation}
where $p$ is the true $p$-value~\cite{LIGOScientific:2020tif}. Assuming uniform prior, we can obtain posterior distribution of $p$-value as a Beta distribution,
\begin{equation}
\text{P}(p|N,n) =  \text{Beta}(n+1,N-n+1)~.
\end{equation}

In Fig.~\ref{fig:res:pp}, the light-blue band around the PP curve represents the $90\%$ uncertainty region of the  $p$-value posteriors. The diagonal dashed line denotes  the prior hypothesis with surrounding light-gray band representing $90\%$ uncertainty region of the null hypothesis due to the finite number of events~\cite{Miller:2001ck,Wilks1948}. 

The PP plot is well with in the $90\%$ credible region of the null hypothesis indicating no significant deviation in the residual data from the expected incoherent noise distribution  in the individual instruments.

\begin{table}
	\caption{ Results of the residuals analysis (Sec.~\ref{sec:res}). For individual events we list the SNR of the best-fit waveform ($\text{SNR}_{\text{GR}}$), $90\%$ credible upper limit on the remnant coherent network SNR ($\text{SNR}_{90}$), fitting factor $\text{FF}_{90}$, and $p$-values calculate from the background analysis.}
	\label{tab:residuals}
	\centering
	\begin{tabular}{lrccc}
\toprule
           Events & ${\rm SNR}_{\rm GR}$ & Residual ${\rm SNR}_{90}$ & ${\rm FF}_{90}$ & $p$-value \\
 \midrule
 \NNAME{GW191109A} &                17.99 &                      7.05 &            0.93 &      0.55 \\
 \NNAME{GW191129G} &                14.10 &                      6.35 &            0.91 &      0.60 \\
 \NNAME{GW191204G} &                17.31 &                      6.53 &            0.94 &      0.63 \\
 \NNAME{GW191215G} &                12.39 &                      6.01 &            0.90 &      0.91 \\
 \NNAME{GW191216G} &                19.06 &                      5.56 &            0.96 &      0.92 \\
 \NNAME{GW191222A} &                12.11 &                      4.87 &            0.93 &      1.00 \\
 \NNAME{GW200115A} &                12.06 &                      8.65 &            0.82 &      0.16 \\
 \NNAME{GW200129D} &                26.79 &                      9.67 &            0.94 &      0.25 \\
 \NNAME{GW200202F} &                12.08 &                      7.49 &            0.85 &      0.35 \\
 \NNAME{GW200208G} &                11.35 &                      6.26 &            0.88 &      0.97 \\
 \NNAME{GW200219D} &                10.72 &                     10.23 &            0.74 &      0.10 \\
 \NNAME{GW200224H} &                19.63 &                      7.89 &            0.93 &      0.52 \\
 \NNAME{GW200225B} &                14.15 &                      8.25 &            0.86 &      0.05 \\
 \NNAME{GW200311L} &                16.99 &                      7.11 &            0.92 &      0.93 \\
 \NNAME{GW200316I} &                11.63 &                      7.17 &            0.85 &      0.51 \\
\bottomrule
\end{tabular}

\end{table}

    \subsection{Inspiral--merger--ringdown consistency test}
    \label{sec:imr}
    The IMR consistency test checks the consistency of the mass and spin of the remnant black hole inferred from the low- and high-frequency parts of the signal. 
To achieve this, we divide the GW signal into two parts in the frequency domain at the cutoff frequency $f_\text{c}^{\rm IMR}$ which is the dominant mode GW frequency of the innermost stable circular orbit (ISCO) of the remnant Kerr black hole~\cite{Ghosh:2016qgn, Ghosh:2017gfp}. The mass and spin of the remnant black hole are estimated by applying NR-calibrated fits~\cite{Healy:2016lce, Hofmann:2016yih, Jimenez-Forteza:2016oae,spinfit-T1600168,GW170104} to the median values of the redshifted component masses, dimensionless spins, and spin angles obtained using the full IMR signal and the waveform model \IMRPHM{}.
The low- and high-frequency regimes roughly correspond to the inspiral and postinspiral, respectively, of the dominant mode of the waveform. 
To make sure that the two regimes of the signal have enough information, we calculate the SNR of the inspiral and the postinspiral parts of the waveform for each event using their maximum \textit{a posteriori} parameter values obtained from the full IMR signal.

We analyze only those signals which have SNRs greater than 6 in both the inspiral and the postinspiral parts. 
This constraint was also used in previous studies~\cite{LIGOScientific:2019fpa,LIGOScientific:2020tif}.
We also impose an extra mass constraint $(1+z) M < 100\, M_{\odot}$ as in our previous analysis of GWTC-2 events~\cite{LIGOScientific:2020tif} to ensure enough inspiral signal for heavier BBHs.
The SNRs for the inspiral and the postinspiral regimes of the events analyzed are given in Table~\ref{tab:imrct_params}.

We independently estimate the posterior distributions of the mass $M_{\rm f}$ and the dimensionless spin $\chi_{\rm f}$ of the remnant black hole from both the inspiral and the postinspiral parts of the signal.
To constrain possible deviations from GR, two fractional deviation parameters $\Delta M_{\rm f}/ \bar{M}_{\rm f}$ and $\Delta \chi_{\rm f}/\bar{\chi}_{\rm f}$ are defined, where

\begin{align}
\dfrac{\Delta M_{\rm f}}{\bar{M}_{\rm f}} &= 2 \dfrac{M_{\rm f}^{\rm insp}-M_{\rm f}^{\rm postinsp}}{M_{\rm f}^{\rm insp}+M_{\rm f}^{\rm postinsp}} \,, \quad \dfrac{\Delta \chi_{\rm f}}{\bar{\chi}_{\rm f}} = 2 \dfrac{\chi_{\rm f}^{\rm insp} - \chi_{\rm f}^{\rm postinsp}}{\chi_{\rm f}^{\rm insp} + \chi_{\rm f}^{\rm postinsp}},
\end{align}
and $\bar{M}_{\rm f}$ and $\bar{\chi}_{\rm f}$ denote the mean values of final mass and final spin obtained from analyzing the inspiral and postinspiral parts of the signal, respectively. 
Here the superscripts denote the inspiral (insp) and the postinspiral (postinsp) portions of the signal. 
The two-dimensional posterior distribution of these fractional deviation parameters should peak around $(0,0)$ when the test is applied to a signal from a quasi-circular BBH coalescence in GR, given that we use a waveform model for such signals to analyze the data.
\begin{figure}
	\begin{center}
	\includegraphics[width=3.5in]{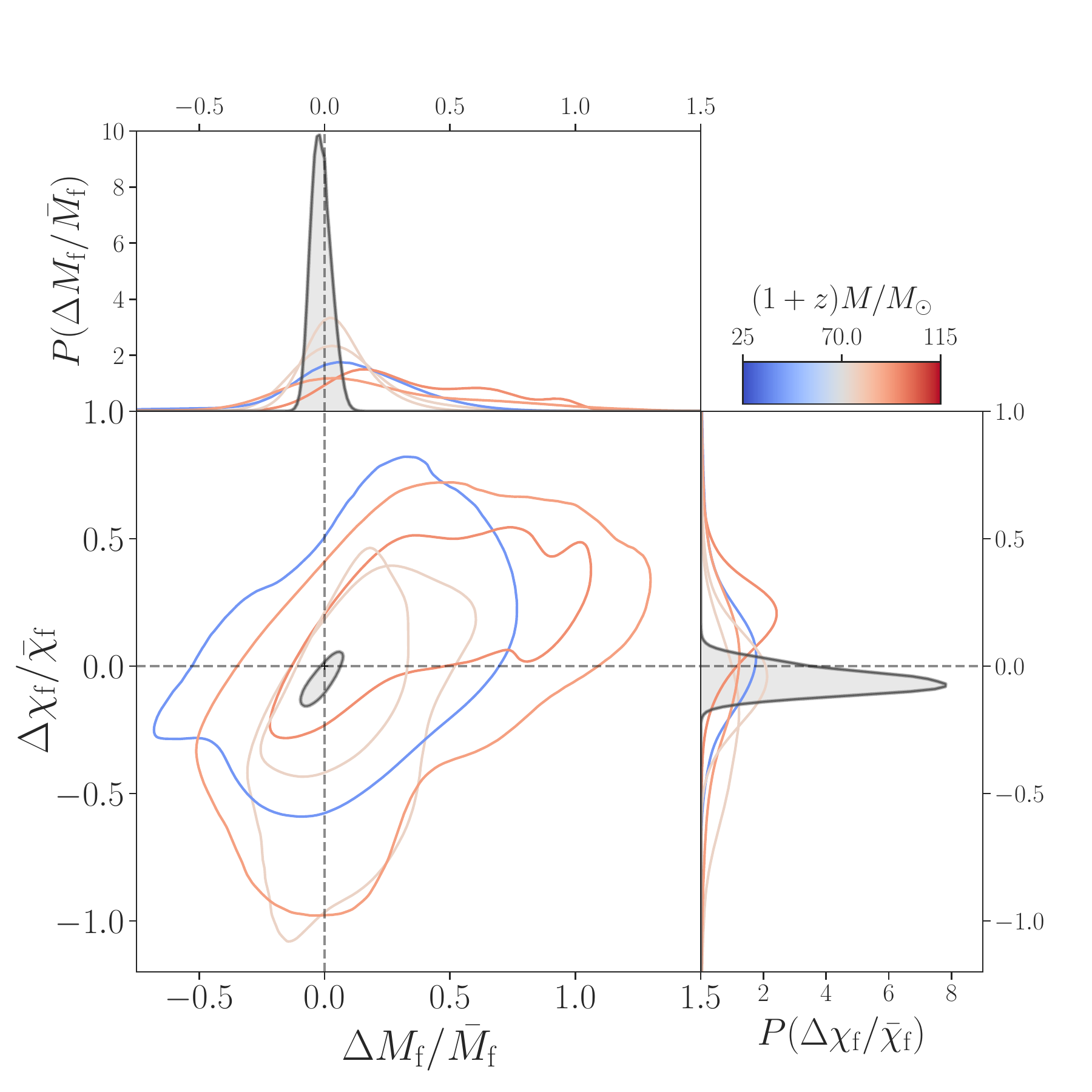}
	\end{center}
	\caption{
	Combined results of the IMR consistency test for BBH events which satisfy the selection criteria (see Table~\ref{tab:imrct_params} and Appendix~\ref{app:imr}). The combined bounds are obtained assuming the same deviation for all events.
	The main panel shows the 90\% credible regions of the two-dimensional posteriors on $(\dMf, \dchif)$ assuming a uniform prior, with $(0,0)$ being the expected value for GR. 
	The side panels show the marginalized posterior on $\dMf$ and $\dchif$. 
	The gray distributions correspond to posteriors obtained by combining individual results. 
	The other colored traces correspond to the O3b events given in Table~\ref{tab:imrct_params} where the color encodes the median redshifted total mass.   
  }
	\label{fig:imrct_posteriors}
\end{figure}
\begin{table}
\caption{
    Results from the IMR consistency test (Sec.~\ref{sec:imr}). 
	$f_\text{c}^{\rm IMR}$ denotes the cutoff frequency between the inspiral and postinspiral regimes; $\rho_\mathrm{IMR}$, $\rho_\mathrm{insp}$, and $\rho_\mathrm{postinsp}$ are the SNR in the full signal, the inspiral part, and the postinspiral part respectively; and the GR quantile $\QGR$ denotes the fraction of the reweighted posterior enclosed by the isoprobability contour that passes through the GR value, with smaller values indicating better consistency with GR. 
	The results are given only for O3b events which satisfy the selection criteria. 
	See Appendix~\ref{app:imr} for the updated results on GWTC-2 events.
}
\label{tab:imrct_params}
\centering
\begin{tabular}{l@{~} c@{\quad}r@{\quad}r@{\quad} r@{\quad} r}
\toprule
Event 		& $f_\text{c}^{\rm IMR}$ [Hz]  & $\rho_\mathrm{IMR}$ & $\rho_\mathrm{insp}$ & $\rho_\mathrm{postinsp}$ & $\QGR$ [\%] \\
\midrule
\NNAME{GW200129D} 	&  $\EVENTSELECTION{S200129mFCIMR}$	  &  \EVENTSELECTION{S200129mOPTSNR} 	&  \EVENTSELECTION{S200129mOPTSNRPREIMR} & \EVENTSELECTION{S200129mOPTSNRPOSTIMR} 	& \ImrEVENTSTATS{S200129mGRQUANTGWTC3}\phantom{${}^*$} \\
\NNAME{GW200208G} 	&  $\EVENTSELECTION{S200208qFCIMR}$	  &  \EVENTSELECTION{S200208qOPTSNR} 	&  \EVENTSELECTION{S200208qOPTSNRPREIMR} & \EVENTSELECTION{S200208qOPTSNRPOSTIMR} 	& \ImrEVENTSTATS{S200208qGRQUANTGWTC3}\phantom{${}^*$} \\
\NNAME{GW200224H} 	&  $\EVENTSELECTION{S200224caFCIMR}$	  &  \EVENTSELECTION{S200224caOPTSNR} 	&  \EVENTSELECTION{S200224caOPTSNRPREIMR} & \EVENTSELECTION{S200224caOPTSNRPOSTIMR} 	& \ImrEVENTSTATS{S200224caGRQUANTGWTC3}\phantom{${}^*$} \\
\NNAME{GW200225B} 	&  $\EVENTSELECTION{S200225qFCIMR}$	  &  \EVENTSELECTION{S200225qOPTSNR} 	&  \EVENTSELECTION{S200225qOPTSNRPREIMR} & \EVENTSELECTION{S200225qOPTSNRPOSTIMR} 	& \ImrEVENTSTATS{S200225qGRQUANTGWTC3}\phantom{${}^*$} \\
\NNAME{GW200311L} 	&  $\EVENTSELECTION{S200311bgFCIMR}$	  &  \EVENTSELECTION{S200311bgOPTSNR} 	&  \EVENTSELECTION{S200311bgOPTSNRPREIMR} & \EVENTSELECTION{S200311bgOPTSNRPOSTIMR} 	& \ImrEVENTSTATS{S200311bgGRQUANTGWTC3}\phantom{${}^*$} \\
\bottomrule
\end{tabular}
\end{table}

\begin{figure}[tb] 
	\begin{center}
	\includegraphics[width=\columnwidth]{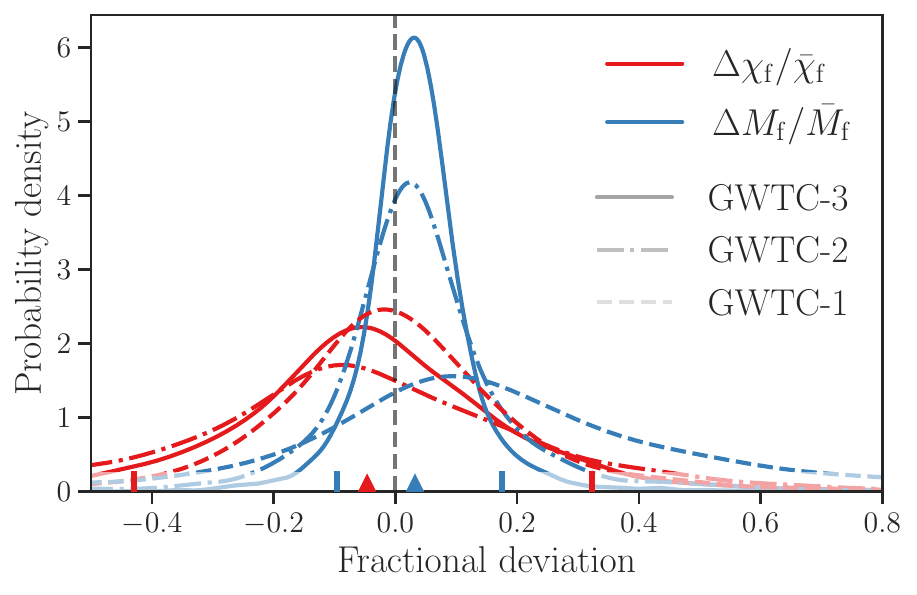}
	\end{center}
	\caption{
	Distributions on the remnant mass (blue) and spin (red) fractional deviation parameters obtained by hierarchically combining the GWTC-3 events (solid trace). 
	For comparison, we also show the results obtained using GWTC-2 (dot dashed traces) and GWTC-1 (dashed) events. 
	The vertical dashed line shows the GR prediction. 
	Triangles mark the GWTC-3 medians, and vertical bars the symmetric 90\%-credible intervals.
  }
	\label{fig:imrct_hier}
\end{figure}
%

\begin{figure}
	\begin{center}
	\includegraphics[width=3.5in]{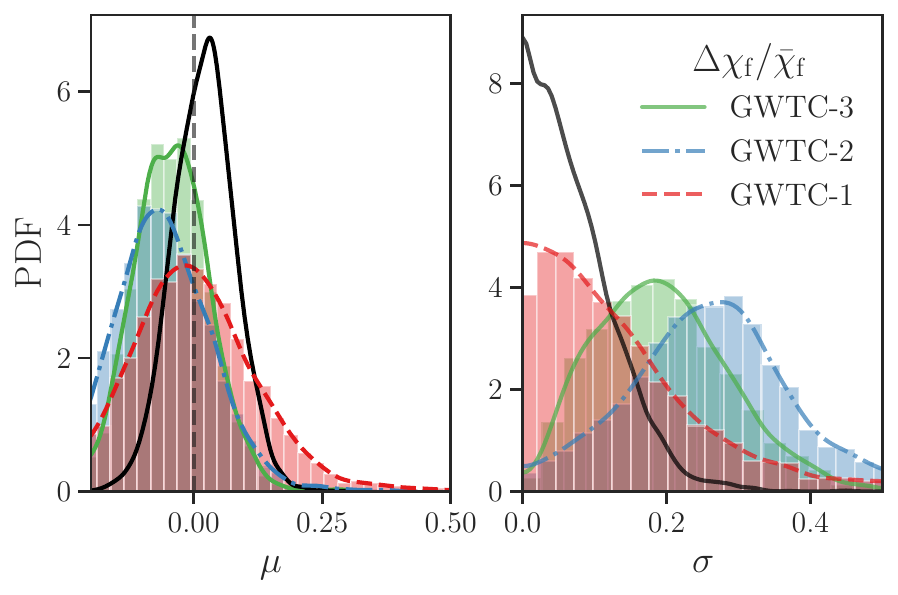}
	\end{center}
	\caption{Posteriors on the hyperparameters  $\mu$ and $\sigma$ of $\dchif$ distribution. The GWTC-2 and GWTC-3 posteriors on $\sigma$ show a marked deviation from zero primarily due to \NNAME{GW190814H}{} posterior on  $\dchif$ peaking away from zero. The black trace shows the posteriors for GWTC-3 events excluding \NNAME{GW190814H}{}. See Appendix \ref{app:imr} for more information about these deviations. The corresponding hyperparameters of the $\dMf$ distribution do not show any such deviation.}
	\label{fig:imrct_hier_spin_hyperposteriors}
\end{figure}



The parameter estimation runs employed the \IMRPHM{} waveform with uniform priors on the redshifted component masses and spins. 
These priors translate into nontrivial priors on $\dMf$ and $\dchif$. Thus, as in the previous analysis~\cite{LIGOScientific:2020tif}, we reweight the posteriors to obtain uniform priors on the deviation parameters. 
We provide our results in Fig.~\ref{fig:imrct_posteriors}, where we show the 90\% credible regions of the two-dimensional posteriors on the fractional deviation parameters for the O3b events which satisfy our selection criteria.

The reweighted posteriors on the fractional deviation parameters $\dMf$ and $\dchif$ of individual events are interpolated on a grid with bounds $[-2,2]$ for both the parameters, and the interpolated posteriors are then multiplied to obtain the combined posteriors. 
Here we assume the same deviation for all events to obtain the combined results. 
As shown in gray in Fig.~\ref{fig:imrct_posteriors}, the combined posteriors on the fractional deviation parameters of GWTC-3 events are consistent with the GR prediction with $\dMf = \ImrGWTCTHREE{DMFGWTC3PHENOM}$ and $\dchif = \ImrGWTCTHREE{DCHIFGWTC3PHENOM}$. 
The two-dimensional GR quantile values $\QGR$  for the events are given in Table~\ref{tab:imrct_params}. 
Here $\QGR$ is defined as the fraction of the posterior enclosed by the isoprobabilty contour that passes through $(0,0)$, the GR value. 
Smaller values indicate better consistency with GR.
The GR quantile of the combined distribution is $\ImrGWTCTHREE{GRQUANTGWTC3}\%$ which is similar to the value obtained for GWTC-2 ($\ImrGWTCTWO{GRQUANTGWTC2}\%$).
Among the O3b events, \NNAME{GW200225B}{} has the lowest $\QGR$ value of $\ImrEVENTSTATS{S200225qGRQUANTGWTC3}\%$ and \NNAME{GW200224H}{} has the highest value of $\ImrEVENTSTATS{S200224caGRQUANTGWTC3}\%$. 

We can also combine the results hierarchically, as discussed in Sec.~III B of our previous analysis~\cite{LIGOScientific:2020tif}.
Fig.~\ref{fig:imrct_hier} presents the results where the fractional mass (blue) and spin (red) deviation parameters for events from multiple observing runs are plotted with $\dMf = \ImrMfHierPop{GWTC3}$ and $\dchif = \ImrChifHierPop{GWTC3}$, which are consistent with the expected values in GR.
Treating $\dMf$ and $\dchif$ independently, we find that the Gaussian model parameters are constrained to $(\mu,\, \sigma) = (\ImrMfHierMu{GWTC3},\, \ImrMfHierSigma{GWTC3})$ and $(\ImrChifHierMu{GWTC3},\, \ImrChifHierSigma{GWTC3})$ for $\dMf$ and $\dchif$ respectively, with 90\% credibility. 
These bounds are not significantly different from the ones reported in GWTC-2~\cite{LIGOScientific:2020tif}, except for that of $\sigma$ for $\dchif$.
It peaks significantly away from zero as shown in Fig.~\ref{fig:imrct_hier_spin_hyperposteriors} due to \NNAME{GW190814H}{} whose updated posteriors (see Appendix~\ref{app:imr} for more details related to the updated GWTC-2 results) show marked deviation from GR.
We also show the posteriors excluding \NNAME{GW190814H}{} which peak at $\sigma = 0$.

\section{Tests of gravitational-wave generation}
\label{sec:gen}

    \subsection{Generic modifications}
    \label{sec:par}
    Deviations from GR, such as additional fields or higher-curvature corrections, may alter the binary's binding energy and angular momentum, and its energy and angular momentum flux \cite{Shiralilou:2021mfl,Bernard:2018hta,Bernard:2018ivi,Julie:2018lfp,Sotiriou:2006pq,Yagi:2011xp,Lang:2013fna,Lang:2014osa,Mirshekari:2013vb}. This in turn would result in modifications to the binary motion and, hence, to the gravitational-wave signal emitted by the system. A practical approach to quantifying such effects entails introducing a finite number of parameters that encapsulate possible deviations of a waveform from its GR prediction. We will focus here on parametrizations of the frequency-domain gravitational-wave phase evolution since observations are in general most sensitive to it (as opposed to changes in the amplitude). 

Small modifications to the gravitational-wave phase could accumulate for events with many detectable gravitational-wave cycles and thus parametrized tests initially focussed on the inspiral part of the waveform, whose duration in the detector band grows for low mass binaries. The inspiral can be treated perturbatively within the post-Newtonian framework~\cite{Blanchet:1995ez,Kidder:1995zr,Damour:2001bu,Blanchet:2005tk,Blanchet:2006gy,Arun:2008kb,Bohe:2012mr,Marsat:2012fn,Bohe:2013cla,Blanchet:2013haa,Damour:2014jta,Bohe:2015ana}, which expands observables in powers of $v/c$, with each $O([v/c]^{2n})$ being referred to as of $n$PN order. With the intrinsic parameters of the binary given, the coefficients at different orders of $v/c$ in the PN series are uniquely determined, and so is the perturbative expansion of the early-inspiral phasing within GR. Treating such PN coefficients as measurable parameters of the waveform is therefore a sensible consistency test of GR ~\cite{Blanchet:1994ez,Blanchet:1994ex,Arun:2006hn,Arun:2006yw,Yunes:2009ke,Mishra:2010tp,Li:2011cg,Li:2011vx}. While these parameterized waveforms could capture a wide variety of beyond-GR effects, the abrupt onset of waveform modifications, possible when nonperturbative phenomena such as dynamical scalarization are at play, may not be fully captured by them \cite{Sampson:2013jpa,Khalil:2019wyy}. However, in the spirit of null tests, these differences may still appear as apparent violations of GR. 

This approach can be applied by directly modifying coefficients in a specific waveform model that encodes PN information~\cite{Agathos:2013upa} or by adding corrections that correspond to deformations of a given inspiral PN coefficient at low frequencies and tapering the corrections to zero at a specific cutoff frequency~\cite{Abbott:2018lct,Mehta:2022pcn}. Corrections are applied in both cases at the level of the aligned-spin phasing; however, the first method can be leveraged to perform parametrized tests with precessing phenomenological templates, as these automatically inherit non-GR corrections introduced in the aligned-spin phase by virtue of the twisting-up construction \cite{LIGOScientific:2019fpa,LIGOScientific:2020tif}.

Here we present results obtained with the second method, which we apply to the frequency domain model \SEOBROM{} \cite{Bohe:2016gbl,Cotesta:2020qhw}, a reduced-order model of the time-domain aligned-spin approximant \SEOBNR{}. We do not include results obtained with the first method, for which an upgrade to the precessing \IMRP{} model is under development. Due to time constraints, these results will be presented elsewhere. Past analyses \cite{LIGOScientific:2019fpa,LIGOScientific:2020tif} showed good consistency between the two approaches, despite the differences in the waveform models being used and the physics content included. Dedicated studies would be needed to thoroughly assess the effect of waveform systematics on parametrized tests across parameter space and quantify the effects of specific approximations, such as the omission of precession and higher-order multipole moments. 

Fractional deviations are applied to the full phase as corrections scaling with $f^{\left(-5+n\right)/3}$ at each $n/2$-th PN order. Following previous works \cite{LIGOScientific:2019fpa,LIGOScientific:2020tif}, we reweight  the posteriors to reparametrize the results as fractional deviations applied to the nonspinning terms of a 3.5PN TaylorF2 phase \cite{Buonanno:2009zt}, which is obtained by applying the stationary phase approximation \cite{Cutler:1994} to time-domain post-Newtonian waveforms:
\begin{align}
\varphi_\mathrm{PN} (f)&= 2 \pi\,f\,t_\mathrm{c} - \varphi_\mathrm{c} - \frac{\pi}{4} \nonumber \\ 
&\quad+\frac{3}{128 \eta} \left(\pi \tilde{f} \right)^{-5/3} \sum^{7}_{i = 0} \left[\varphi_i+\varphi_{i\,l}\log(\pi \tilde{f} )\right] \left( \pi \tilde{f}  \right)^{i/3} .
\label{eq:par:TF2_phase}
\end{align}
Here, $\tilde{f} =G M (1+z) f/c^3$, with $M(1+z)$ being the redshifted total mass of the binary, $ \varphi_c ,t_c$ are the coalescence phase and time, and $\eta$ the symmetric mass ratio. This parametrization has the advantage of avoiding potentially singular behavior of the deviation coefficients, which might occur as a result of cancellations between the nonspinning and spin-dependent phasing coefficients. 

Phasing corrections to the inspiral phase are tapered off at the same cutoff frequency used in previous analyses, i.e., $f_{\mathrm{c}}^{\mathrm{PAR}}=0.35\,f_{\mathrm{peak}}^{22}$, where $f_{\mathrm{peak}}^{22}$ is the gravitational-wave frequency of the (2,2)-mode at the peak of the amplitude as defined in the model \SEOBNR{} \cite{Mehta:2022pcn}. 

We introduce the following parametric deviations to gravitational-wave inspiral phasing:
\begin{align}
\label{par:deviation_coefficients}
\left\lbrace \delta \hat{\varphi}_{-2} , \delta \hat{\varphi}_{0} , \delta \hat{\varphi}_{1}, \delta \hat{\varphi}_{2}, \delta \hat{\varphi}_{3}, \delta \hat{\varphi}_{4}, \delta \hat{\varphi}_{5l}, \delta \hat{\varphi}_{6},  \delta \hat{\varphi}_{6l}, \delta \hat{\varphi}_{7} \right\rbrace ,
\end{align}
\newline 
where each $\delta\hat{\varphi}_{i}$ represents the fractional deviation from the GR PN coefficient at the $i/2$-th PN order, following the parametrization adopted in previous analyses \cite{TheLIGOScientific:2016src,TheLIGOScientific:2016pea,Abbott:2017oio,Abbott:2018lct,LIGOScientific:2019fpa}. The subscript $l$ is used to denote coefficients of logarithmic-in-$f$ terms. We do not present bounds for the $2.5$PN non-logarithmic term, as this is degenerate with the coalescence phase, as can be seen from Eq.\,(\ref{eq:par:TF2_phase}). As predicted in GR, the coefficients corresponding to $-1$PN and $0.5$PN are identically zero, so we parametrize $\delta \hat{\varphi}_{-2}$ and $\delta \hat{\varphi}_{1}$ as \textit{absolute} deviations, with a prefactor equal to the $0$PN coefficient ($3/128\eta$); all other coefficients represent \emph{fractional} deviations from the GR value. 

As detailed in Sec.~\ref{sec:intro}, we consider all binaries that meet the significance threshold of FAR $< 10^{-3}~\mathrm{yr}^{-1}$ and impose the additional requirement that $\rm{SNR} \geq 6$ in the inspiral regime, as defined with respect to $f_{\mathrm{c}}^{\mathrm{PAR}}$. Eligible events are summarized in Table \ref{tab:par_events}. The parametrization presented above recovers GR in the limit $\delta \hat{\varphi}_i \rightarrow 0$, thus consistency with GR can be claimed if $0$ is included within a given confidence interval and, in what follows, we will report $90\%$ credible intervals for the posteriors of $\delta \hat{\varphi}_i$. We adopt uniform priors on $\delta \hat{\varphi}_i$ that are symmetric about zero and compute their posterior distributions using \linf{}. 

\begin{table}
\caption{Parametrized test event selection for all binaries meeting the FAR $< 10^{-3}~\mathrm{yr}^{-1}$ threshold. Here $f_{\text{c}}^{\rm PAR}$ denotes the cutoff frequency at which non-GR corrections to the inspiral phase are tapered away (see main text for details); $\rho_{\rm IMR}$ and $\rho_{\rm insp}$ are the \textit{optimal} SNRs computed on the full signal or on the region $f \leq f_{\text{c}}^{\rm PAR}$ respectively. The last column denotes if the event is included in parametrized tests on the PN deviation coefficients.
}
\label{tab:par_events}
\centering
\begin{tabular}{lc@{\quad}rr@{\quad}r}
\toprule
\rm{Event} &  $f_{\text c}^{\rm PAR} \; [ \rm{Hz} ]$ & $\rho_{\rm IMR}$ &  $\rho_{\rm insp}$  & Insp   \\
\midrule
\NNAME{GW191109A} &  \EVENTSELECTION{S191109dFCFTA}	&	\EVENTSELECTION{S191109dOPTSNR}	&	\EVENTSELECTION{S191109dOPTSNRPREFTA}		&	$-$	 \\
\NNAME{GW191129G} &  \EVENTSELECTION{S191129uFCFTA}	&	\EVENTSELECTION{S191129uOPTSNR}	&	\EVENTSELECTION{S191129uOPTSNRPREFTA}		&	\cmark	\\
\NNAME{GW191204G} &  \EVENTSELECTION{S191204rFCFTA}	&	\EVENTSELECTION{S191204rOPTSNR}	&	\EVENTSELECTION{S191204rOPTSNRPREFTA}	&	\cmark	 \\
\NNAME{GW191215G} &  \EVENTSELECTION{S191215wFCFTA}	&	\EVENTSELECTION{S191215wOPTSNR}	&	\EVENTSELECTION{S191215wOPTSNRPREFTA}		&	$-$ \\
\NNAME{GW191216G} &  \EVENTSELECTION{S191216apFCFTA}	&	\EVENTSELECTION{S191216apOPTSNR}	&	\EVENTSELECTION{S191216apOPTSNRPREFTA}	&	\cmark	\\
\NNAME{GW191222A} &  \EVENTSELECTION{S191222nFCFTA}	&	\EVENTSELECTION{S191222nOPTSNR}	&	\EVENTSELECTION{S191222nOPTSNRPREFTA}	&	$-$	 \\
\NNAME{GW200115A} &  \EVENTSELECTION{S200115jFCFTA}	&	\EVENTSELECTION{S200115jOPTSNR}	&	\EVENTSELECTION{S200115jOPTSNRPREFTA}	&	\cmark	 \\
\NNAME{GW200129D} &  \EVENTSELECTION{S200129mFCFTA}	&	\EVENTSELECTION{S200129mOPTSNR}	&	\EVENTSELECTION{S200129mOPTSNRPREFTA}		&	\cmark	 \\
\NNAME{GW200202F} &  \EVENTSELECTION{S200202acFCFTA}	&	\EVENTSELECTION{S200202acOPTSNR}	&	\EVENTSELECTION{S200202acOPTSNRPREFTA}		&	\cmark	 \\
\NNAME{GW200208G} &  \EVENTSELECTION{S200208qFCFTA}	&	\EVENTSELECTION{S200208qOPTSNR}	&	\EVENTSELECTION{S200208qOPTSNRPREFTA}		&	$-$ \\
\NNAME{GW200219D} &  \EVENTSELECTION{S200219acFCFTA}	&	\EVENTSELECTION{S200219acOPTSNR}	&	\EVENTSELECTION{S200219acOPTSNRPREFTA}		&	$-$	 \\
\NNAME{GW200224H} &  \EVENTSELECTION{S200224caFCFTA}	&	\EVENTSELECTION{S200224caOPTSNR}	&	\EVENTSELECTION{S200224caOPTSNRPREFTA}		&	$-$ \\
\NNAME{GW200225B} &  \EVENTSELECTION{S200225qFCFTA}	&	\EVENTSELECTION{S200225qOPTSNR}	&	\EVENTSELECTION{S200225qOPTSNRPREFTA}	&	\cmark	\\
\NNAME{GW200311L} &  \EVENTSELECTION{S200311bgFCFTA}	&	\EVENTSELECTION{S200311bgOPTSNR}	&	\EVENTSELECTION{S200311bgOPTSNRPREFTA}&	\cmark	 \\
\NNAME{GW200316I} &  \EVENTSELECTION{S200316bjFCFTA}	&	\EVENTSELECTION{S200316bjOPTSNR}	&	\EVENTSELECTION{S200316bjOPTSNRPREFTA}		&	\cmark	 \\
\bottomrule
\end{tabular}

\end{table}

As in previous analyses, we only allow the coefficients $\delta \hat{\varphi}_i$ to vary one at a time. It has been shown that this procedure is effective at picking up the deviations from GR that modify more than one PN coefficient~\cite{Sampson:2013lpa,Meidam:2017dgf,Haster:2020nxf}. Allowing multiple deviations in the same template is bound to produce less informative posteriors, due to correlations among different parameters and also with GR coefficients. Such correlations can be effectively reduced with alternative choices of the deformation parameters, which can be guided by a principal component analysis \cite{Pai:2012mv,Shoom:2021mdj}, thus increasing the effectiveness of multiparameter tests. It was also shown that synergies between third-generation detectors and space-based interferometry might dramatically improve the precision of multiparameter tests \cite{Gupta:2020lxa}.

\begin{figure*}[h!tp]
\centering
\includegraphics[width=0.8\textwidth]{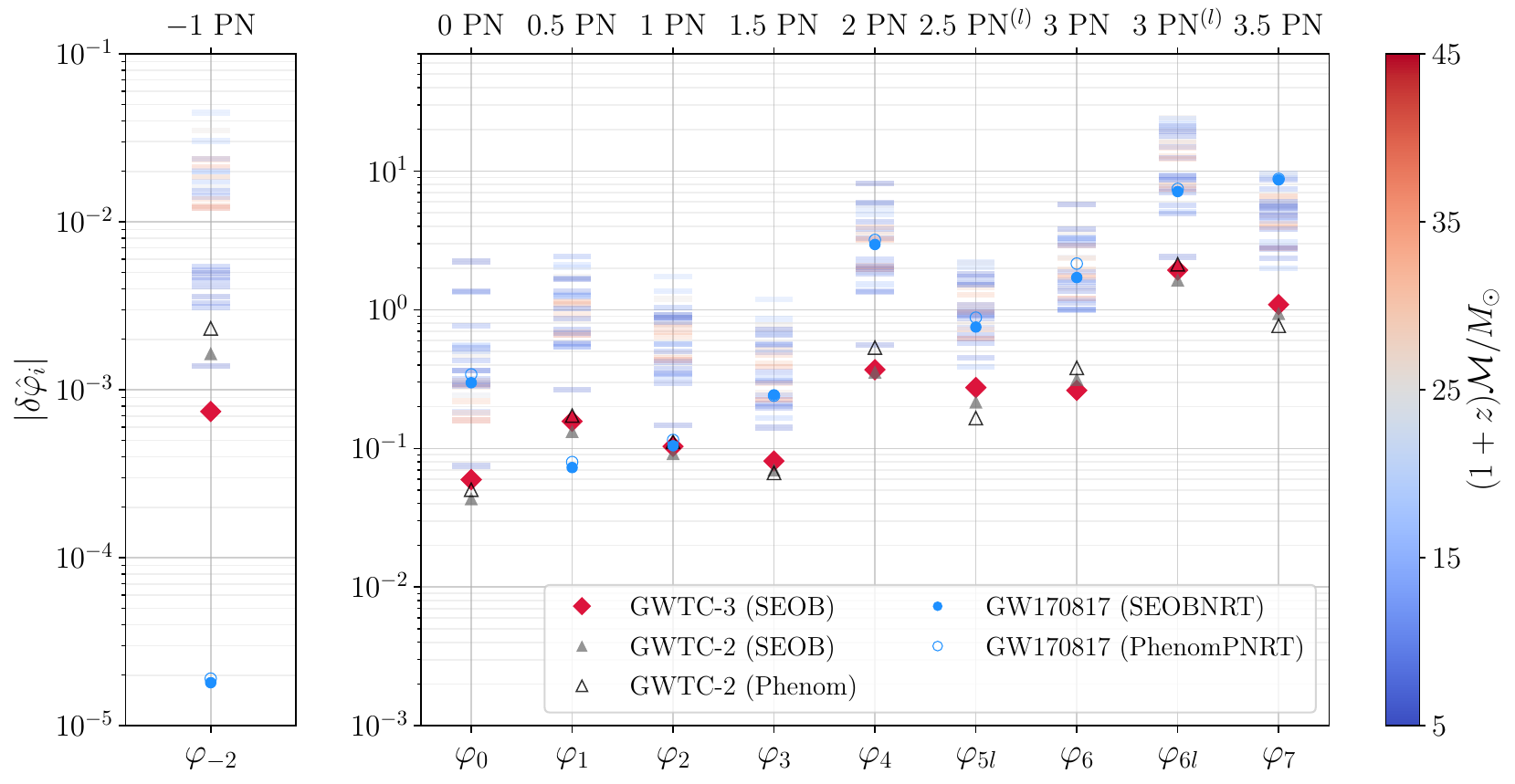}
\caption{$90\%$ upper bounds on the magnitude of the parametrized test coefficients discussed in Sec.\,\ref{sec:par}. Bounds marked by red diamonds were obtained with a pipeline based on the model \SEOBROM{}, combining all eligible GWTC-3 events, under the assumption that deviations take the same value for all the events. Filled (unfilled) gray triangles mark analogous results obtained with GWTC-2 data \cite{LIGOScientific:2020tif} using \SEOBROM{} (\textsc{IMRPhenomPv2}). We also show upper bounds obtained through the observation of the binary neutron star signal GW170817, using the \textsc{NRTidal} extensions of the two aforementioned models \cite{Dietrich:2017aum} as filled (unfilled) blue circles. Horizontal stripes indicate constraints obtained with individual events, with cold (warm) colors representing low (high) total mass events. The left and right panel show constraints on PN deformation coefficients, from $-1$PN to $3.5$PN order. The best improvement with respect to the GWTC-2 bounds is achieved for the $-1$PN term, thanks ot the inclusion of the NSBH candidate \NNAME{GW200115A}.}
\label{fig:par:seob_bounds}
\end{figure*}

In Fig.\,\ref{fig:par:seob_bounds}, we present the $90\%$ upper bounds on the deviation coefficients obtained from the combined distribution of events from GWTC-3, under the assumption that deviations take the same value for all the events. While the combined bounds are fully consistent with GR, we do note that for a number of events $\delta\hat{\varphi}_i = 0$ falls outside the $90\%$ credible interval of the deviation coefficients distributions. 
We find that the addition of an extra degree of freedom can enhance the weight of secondary modes observed in GR parameter estimation runs, as can be seen for \NNAME{GW191216G}{} and \NNAME{GW200316I}{}, where a secondary mode in mass ratio present in the GR parameter estimation runs dominates the results when inspiral deviation parameters are included. This is partly due to the singular behavior of the parametrization employed here, and indeed the application of reweighting alleviates the problem. Still, the presence of a secondary mode remains and so do the broad posteriors for the deviation coefficients. Appendix \,\ref{app:sys} contains a more extended discussion on the impact of noise properties and waveform systematics on our bounds. 

We find that there is no uniform improvement over the GWTC-2 results \cite{LIGOScientific:2020tif} across all deviation coefficients. This is consistent with the modest improvement factor due to the increased number of events, which can be estimated to be $\sim 1.2$. We find that this factor is comparable to fluctuations in the final bounds determined by individual events. The most striking difference with respect to the GWTC-2 analysis is the new constraint obtained for $\delta \hat{\varphi}_{-2}$, for which we obtain an upper bound $|\delta \hat{\varphi}_{-2}|\leq7.5\times10^{-4}$ at the $90\%$ credible level. This result improves upon the GWTC-2 bound by a factor $\sim 2$. The improvement is driven by the inclusion of \NNAME{GW200115A}, due to its long duration. The presence of a non-zero $-1$PN term can be associated to the emission of dipolar radiation, which is forbidden in GR but can be excited in alternative theories of gravity and is related to energy and momentum being transferred from the binary to additional fields. This result is less stringent than the one obtained with combined measurements of binary pulsars \cite{Nair:2020ggs} or with the observation of GW170817 \cite{Abbott:2018lct}. As can be seen in Fig.\,\ref{fig:par:seob_bounds}, a single observation of a low-mass binary, such as GW170817, allows to place tighter constraints than the ones obtained here for some of the lowest PN orders, due to the large number of observable inspiral cycles. 

\begin{figure*}[h!tp]
\centering
\includegraphics[width=\textwidth]{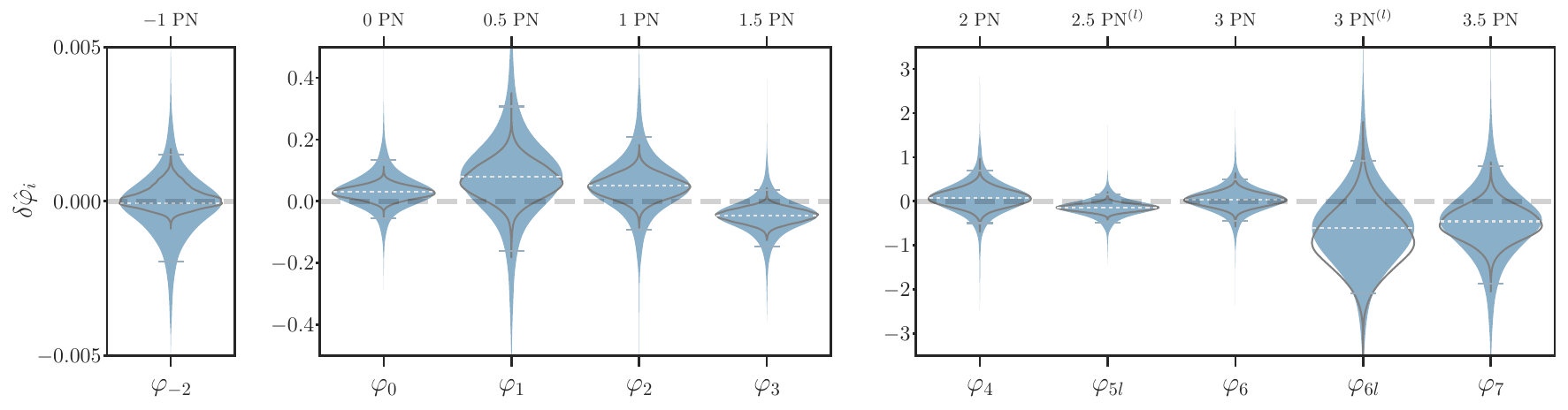}
\caption{Combined GWTC-3 results for the parametrized deviation coefficients of Sec.\,\ref{sec:par}. Filled distributions represent the results obtained hierarchically combining all events. This method allows the deviation coefficients to assume different values for different events. Unfilled black curves represent the distributions obtained in Fig.\,\ref{fig:par:seob_bounds}, by assuming the same value of the deviation parameters across all events. Horizontal ticks and dashed white lines mark the $90\%$ credible intervals and median values obtained with the hierarchical analysis.}
\label{fig:par:hier_bounds}
\end{figure*}

\begin{figure}
\raggedright
\includegraphics[width=0.95\columnwidth]{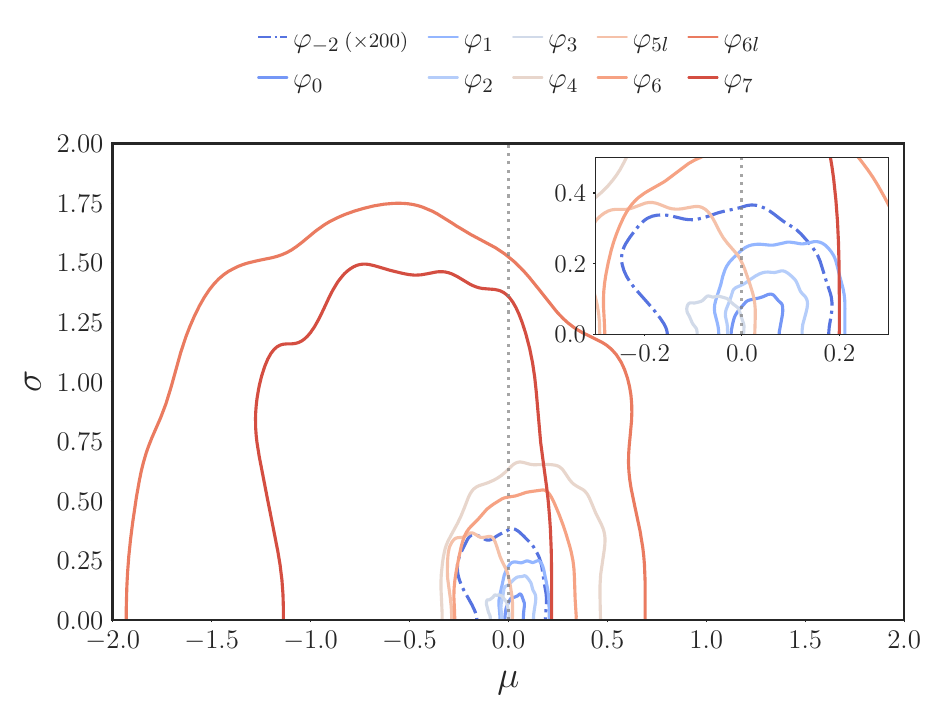}
\caption{Joint distribution for the hyperparameters $\mu$ and $\sigma$ of the GR deviation coefficients for the parametrized tests on generic gravitational-wave modifications of Sec.\,\ref{sec:par}. Contours mark $90\%$ credible regions and they all include $\mu=\sigma=0$, which corresponds to the GR prediction. These results were obtained with a pipeline based on \SEOBROM{}. The
$\phi_{-2}$-contour has been rescaled by a factor of $200$ to improve visibility, as deviations for this term are the most tightly constrained. Warm (cold) colors refer to deviations occurring at higher (lower) PN orders.} 
\label{fig:par:contour}
\end{figure}

We also computed hierarchically combined posteriors for all the deviation coefficients introduced above, where we allow the deviation coefficients to take independent values for each event. These are shown in Fig.\,\ref{fig:par:hier_bounds}, plotted against the combined posteriors employed to obtain the upper bounds of Fig.\,\ref{fig:par:seob_bounds}.  All results are consistent with the GR prediction with at least $90\%$ credibility. In Fig.\,\ref{fig:par:contour}, we show the joint distribution of the mean $\mu$ and standard deviation $\sigma$ of the population-marginalized posterior for the deviation coefficients of Eq.~\eqref{par:deviation_coefficients}. All distributions are consistent with the GR prediction for which $\mu=\sigma=0$, with deviations occurring at $-1$PN being the most tightly constrained.

In Table \ref{tab:par}, we report the medians, $90\%$ credible intervals, and GR quantiles
$Q_{\rm GR} = P(\delta\hat{\varphi}_i < 0)$ for both the hierarchical and the joint-likelihood approaches. Values of $Q_{\rm GR}$ significantly different from $50\%$ indicate that the null hypothesis falls in the tails of the combined distribution. 
In the hierarchical analysis the most constrained parameter is $\ParBestHierPopS{NAME} = \ParBestHierPopS{VALUE}$ at the $90\%$ credible level, while deviations in the 3.5PN order coefficient are the least constrained, with $\ParWorstHierPopS{NAME} = \ParWorstHierPopS{VALUE}$. For the majority of the PN coefficients, the hierarchical analysis of GWTC-3 data obtains tighter constraints than the ones obtained with GWTC-2 events \cite{LIGOScientific:2020tif}.   

\begin{table*}[t!]
\resizebox{\columnwidth}{!}{
\begin{tabular}{crrrrrrrr}
\toprule
               ${\varphi}_i$ & \phantom{X} & \multicolumn{4}{c}{General} & \phantom{X} & \multicolumn{2}{c}{Restricted} \\
\cline{3-6} \cline{8-9}
                &   &                       \multicolumn{1}{c}{$\mu$} &    \multicolumn{1}{c}{$\sigma$} &                    \multicolumn{1}{c}{$\delta\hat{\varphi}_i$} &       $Q_{\rm GR}$ & \phantom{X} &                    \multicolumn{1}{c}{$\delta\hat{\varphi}_i$} &       $Q_{\rm GR}$ \\
\midrule
${\varphi}_{-2}^{[\times 200]}$ &             &  $-0.02^{+0.16}_{-0.19}$ &  $<0.30$ &  $-0.01^{+0.32}_{-0.38}$ &  $54\%$ &             &   $0.02^{+0.12}_{-0.08}$ &  $40\%$ \\[4pt]
${\varphi}_{0}$ &             &   $0.03^{+0.05}_{-0.04}$ &  $<0.08$ &   $0.03^{+0.10}_{-0.09}$ &  $23\%$ &             &   $0.03^{+0.03}_{-0.03}$ &   $7\%$ \\[4pt]
\midrule
${\varphi}_{1}$ &             &   $0.08^{+0.11}_{-0.11}$ &  $<0.21$ &   $0.08^{+0.23}_{-0.24}$ &  $22\%$ &             &   $0.07^{+0.08}_{-0.08}$ &   $9\%$ \\[4pt]
${\varphi}_{2}$ &             &   $0.05^{+0.07}_{-0.07}$ &  $<0.14$ &   $0.05^{+0.16}_{-0.14}$ &  $21\%$ &             &   $0.04^{+0.06}_{-0.05}$ &   $8\%$ \\[4pt]
${\varphi}_{3}$ &             &  $-0.05^{+0.04}_{-0.05}$ &  $<0.08$ &  $-0.05^{+0.08}_{-0.10}$ &  $86\%$ &             &  $-0.04^{+0.02}_{-0.04}$ &  $99\%$ \\[4pt]
${\varphi}_{4}$ &             &   $0.08^{+0.33}_{-0.32}$ &  $<0.54$ &   $0.08^{+0.62}_{-0.58}$ &  $38\%$ &             &   $0.08^{+0.30}_{-0.27}$ &  $35\%$ \\[4pt]
${\varphi}_{5l}$ &             &  $-0.15^{+0.13}_{-0.14}$ &  $<0.29$ &  $-0.15^{+0.31}_{-0.34}$ &  $82\%$ &             &  $-0.14^{+0.12}_{-0.15}$ &  $97\%$ \\[4pt]
${\varphi}_{6}$ &             &   $0.04^{+0.25}_{-0.25}$ &  $<0.43$ &   $0.03^{+0.46}_{-0.47}$ &  $43\%$ &             &   $0.05^{+0.21}_{-0.24}$ &  $40\%$ \\[4pt]
${\varphi}_{6l}$ &             &  $-0.61^{+0.94}_{-0.92}$ &  $<1.25$ &  $-0.61^{+1.52}_{-1.48}$ &  $78\%$ &             &  $-0.89^{+1.05}_{-1.02}$ &  $91\%$ \\[4pt]
${\varphi}_{7}$ &             &  $-0.47^{+0.55}_{-0.63}$ &  $<1.20$ &  $-0.46^{+1.25}_{-1.41}$ &  $77\%$ &             &  $-0.53^{+0.57}_{-0.57}$ &  $93\%$ \\[4pt]
\bottomrule
\end{tabular}
}
\caption{Results from parametrized tests of gravitational-wave generation (Sec.~\ref{sec:par}).
Combined constraints on the deviation parameters $\delta\hat{\varphi}_i$ from the full set of GWTC-3 data using the the \SEOBROM{} waveform model. General (restricted) constraints are obtained under the assumption that deviation coefficients can (cannot) vary across the observed events. $Q_{\rm GR}$ indicates the quantile corresponding to the GR value for the distributions plotted in Fig.~\ref{fig:par:seob_bounds}. For general constraints, we also provide the mean $\mu$ and standard deviation $\sigma$ of the inferred hyperdistribution. For $\delta\hat{\varphi}_i$ and $\mu$, we report the median as well as the 90\%-credible intervals, while for $\sigma$ we only present upper bounds. For the restricted method, the null hypothesis can occasionally fall in the tail of the distribution, while the hierarchical analysis places it in the bulk of the inferred distribution. 
}

\label{tab:par}

\end{table*}

Results for the shifts to inspiral phase can also be mapped onto constraints on specific theories \cite{Mehta:2022pcn}, in particular via the parametrized post-Einstein (ppE) framework~\cite{Yunes:2009ke,Yunes:2016jcc}. For instance, bounds on the coupling constant of Einstein--dilaton--Gauss--Bonnet and dynamical Chern--Simons gravity were obtained using GWTC-2 events \cite{Nair:2019iur,Perkins:2021mhb, Wang:2021yll}. However, the upper bounds reported here depend on the parametrization being used and on specific details of the analysis, such as the frequency at which non-GR corrections are being tapered off \cite{Mehta:2022pcn}. The priors we impose on the deviation coefficients are not designed to suit any specific theory and, depending on the theory that is being considered, different sampling parameters and prior bounds might be preferable to the ones adopted here. Furthermore, in alternative theories of gravity multiple inspiral coefficients will be subject to deviations from GR; our bounds refer to single-coefficient deviations, that might capture at once deviations at several PN orders, so the mapping would be ambiguous. Likewise, one would also need a robust estimate of the error caused by neglecting currently unknown higher-order PN corrections and deviations in the merger--ringdown phase, which will also differ from the GR one \cite{Maselli:2019mjd, Carullo:2021dui}. Finally, it is not clear whether some of these theories, such as Lovelock, Chern--Simons or Horndeski gravity, of which Einstein--dilaton--Gauss--Bonnet gravity is a special case, would admit at all a well-posed initial value problem in their most general form \cite{Papallo:2017qvl}, although well-posed formulations are possible in the weak-coupling limit \cite{Okounkova:2017yby,Kovacs:2020pns,Kovacs:2020ywu}.

    \subsection{Spin-induced quadrupole moment}
    \label{sec:sim}
    \def \dks{\texttt{$\delta\kappa_{s}$}}
\def \pv2{\textsc{IMRPhenomPv2}}

Spinning objects have quadrupole and higher contributions to the multipole decomposition of their gravitational field due to their rotational deformations. Following the no-hair conjecture, the spin-induced multipole moments take unique values for black holes given their mass and spin~\cite{Hansen:1974zz, Carter}. Gravitational waveforms describing spinning compact binary systems encode information about these spin-induced multipole moment effects. The leading order term can be schematically represented as,
\begin{equation}\label{eq:sim} 
Q= -\, \kappa\, \chi^2 m^3.
\end{equation}
Here $Q$ is the quadrupole moment scalar and is the leading order term in the gravitational-wave phase at 2PN order. $m$ and $\chi$ are the mass and the dimensionless spin of the compact object. Along with this leading-order effect, we have included higher-order PN terms that appear through the inspiral phase~\cite{Arun:2008kb, Mishra:2016whh} of gravitational waveform. 

While  Kerr black holes have $\kappa=1$~\cite{Hansen:1974zz, Carter},  compact stars have a  value of $\kappa$ that differs from the black hole value, determined by the star's mass and internal composition. 
Numerical simulations of spinning  neutron stars show that the value of $\kappa$ can vary between ${\sim}2$ and ${\sim}14$ for these systems~\cite{Pappas:2012qg,Pappas:2012ns,Harry:2018hke}. Moreover, for currently available models of spinning boson stars,  $\kappa$ can have values ${\sim}10\textendash150$~\cite{FDRyan1997,Herdeiro:2014goa,Baumann:2018vus,Chia:2020psj}. More exotic stars like gravastars can even take negative values for $\kappa$~\cite{Uchikata:2015yma}. Hence, an independent measurement of $\kappa$ from gravitational-wave observations can be used to distinguish black holes from other exotic objects~\cite{Krishnendu:2017shb, Poisson:1997ha, Laarakkers:1997hb, Ryan:1995wh}. However, to fully understand the nature of compact objects, one may also include effects such as the tidal deformations that arise due to the external gravitational field \cite{Sennett:2017etc, Johnson-McDaniel:2018uvs, Pacilio:2020jza,Narikawa:2021pak} and tidal heating \cite{Cardoso:2017cfl,Maselli:2017cmm,Datta:2019epe,Datta:2019euh,Datta:2020gem, Datta:2020rvo} along with the spin-induced deformations, an extensive study of these effects is not in the scope of this paper.

For a spinning compact binary system, the coefficients $\kappa_i$, $i=1, 2$ represent the primary and secondary components' spin-induced quadrupole moment parameters. The correlation of $\kappa_i$ with the masses and spin parameters of the binary are evident from Eq.~(\ref{eq:sim}), which makes the simultaneous estimation of $\kappa_1$ and $\kappa_2$ hard. The higher-order terms present at the 3PN order help break this degeneracy, but are not enough to give reasonable constraints with our current detector sensitivities. However, a combination of these parameters can be measured~\cite{Krishnendu:2017shb, Krishnendu:2018nqa, Krishnendu:2019tjp,LIGOScientific:2020tif}. 
For this reason, we introduce the symmetric and anti-symmetric combinations of $\kappa_i$,
\begin{align}
\kappa_s=(\kappa_1+\kappa_2)/2,\\
\kappa_a=(\kappa_1-\kappa_2)/2. 
\end{align}

For binary black holes, $\kappa_s=1$ and $\kappa_a=0$. We measure the parameterised deviations $\kappa_s=1+\delta\kappa_s$ assuming $\kappa_a=0$. This assumption restricts our study to binaries consisting of compact stars with identical spin-induced deformations. {\blue If the data supports $\kappa_1\neq\kappa_2$,  expect a significant offset away from zero in $\delta\kappa_s$ measurements. Further studies are required for such situations}. 

The method employed here is the same as in previous analysis~\cite{LIGOScientific:2020tif}. We perform a Bayesian analysis with LALInference using a nested sampling algorithm to estimate the posteriors on $\delta\kappa_s$. The parametrized deviations $\delta\kappa_s$ are introduced in the inspiral phase of the \pv2 waveform model. 

We analyse the events listed in Table~\ref{tab:events} passing the selection criterion. Along with the FAR $< 10^{-3}~\mathrm{yr}^{-1}$ criteria, we consider two additional conditions for selecting the events for this test. First, we select inspiral-dominated events having an inspiral network SNR $\geq 6$. Since the test relies on at least one of the binary’s components having nonzero spin, we also drop events whose effective inspiral spin parameter measurements include zero at the 68\% credible level. Given the component masses $m_i$ and the dimensionless spins $\chi_{i}= \vec{S_{i}} \cdot \hat{L} / m_{i}^2$ pointing parallel to the orbital angular angular momentum axis, the effective inspiral spin parameter $\chi_{\rm{eff}}$ is defined as, $\chi_{\rm{eff}} = (m_{1}\chi_{1}+m_{2}\chi_{2})/(m_1+m_2)$~\cite{Ajith:2009bn}.  To compute the combined bounds we include events reported in~\cite{LIGOScientific:2020tif} satisfying both the above selection criteria. This gives us a total of 13 events considering the entire GWTC-3.

\begin{figure}
 \centering
 \includegraphics[width=3.in]{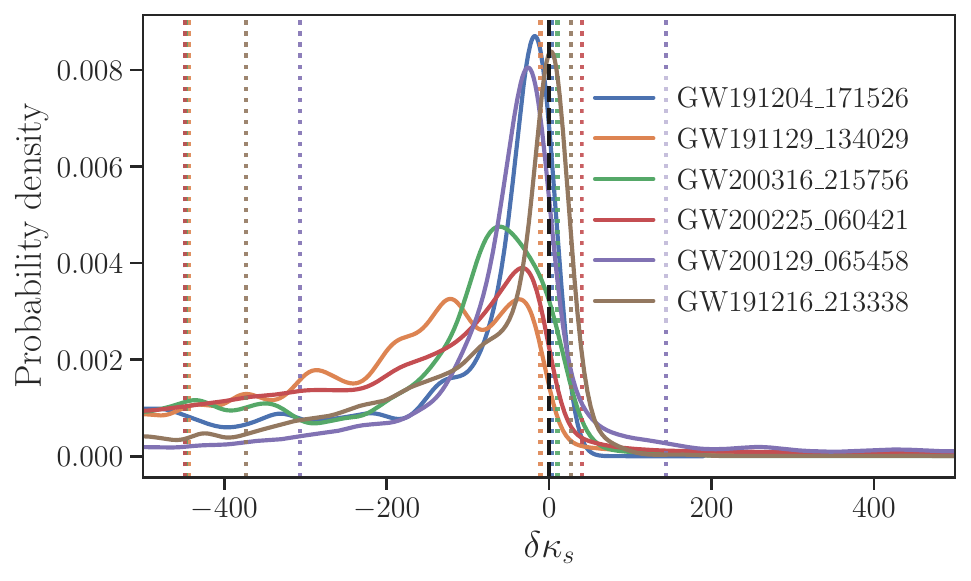}
 \caption{The posterior probability distribution on the spin-induced quadrupole moment parameter, $\delta\kappa_s$ from the events listed in the SIM column of Table~\ref{tab:events}, passing the selection criteria described in~Section~\ref{sec:sim}. The black dashed vertical line indicates the BBH value ($\delta\kappa_s=0$). The colored vertical lines show the 90\% symmetric bounds on $\delta\kappa_s$ calculated from the individual events assuming a uniform prior ranging between $[-500, 500]$ on  $\delta\kappa_s$. 
}
 \label{fig:sim}
\end{figure}

\begin{figure}
 \centering
 \includegraphics[width=3.0in]{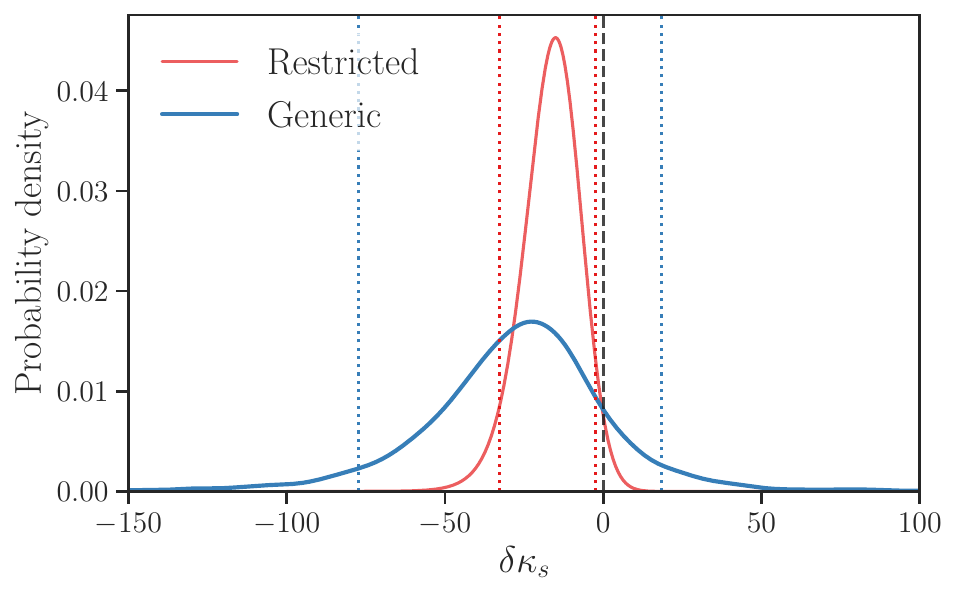}
 \caption{ Joint posterior probability distribution on the spin-induced quadrupole moment parameter $\delta\kappa_s$ from the GWTC-3 events. Bounds obtained by multiplying the likelihoods (restricted) and by hierarchically combining events (generic) are shown. The analysis is performed assuming uniform prior ranging between [-500, 500] on $\delta\kappa_s$.  }
 \label{fig:sim_combined}
\end{figure}

In Fig.~\ref{fig:sim}, the posterior distributions on $\delta\kappa_s$ for the events listed in Table~\ref{tab:events} are derived assuming a uniform prior on $\delta\kappa_s$ ranging between $[-500, 500]$. 
Individual events constrain positive values of  $\delta\kappa_s$ more strongly than negative ones. This is primarily because of how these parameters are correlated with the effective inpsiral spin parameter of the binary system~\cite{Krishnendu:2019tjp}. As most of the events we observe have small but positive $\chi_{\rm{eff}}$, the combined posterior and the 90\% bounds are expected to show this feature. 

We also consider a case where the analysis is restricted to only positive $\delta\kappa_s$ as is well motivated in the case of neutron stars \cite{Laarakkers:1997hb,Pappas:2012qg,Pappas:2012ns} and boson stars \cite{FDRyan1997}, in this case the event provides the tightest upper limits is GW191216\_213338, with 90\% credible bounds of $\delta\kappa_s < \SimBestPosUL{VALUE0}{}$.

We show the combined posterior distribution on $\delta\kappa_s$ from all the GW events passing the selection criteria in Fig.~\ref{fig:sim_combined}. The red curve draws the posterior distribution obtained by multiplying the likelihoods of each individual signal. In contrast, the population-marginalized posterior from the hierarchical analysis is shown in the blue curve. Dotted lines show the 90\% symmetric credible intervals, and a dashed line marks the BBH value ($\delta\kappa_s=0$). We estimate the combined symmetric 90\% bound on $\delta\kappa_s$ considering GWTC-3 events to be $\delta\kappa_s = \SimCombinedCI{SIMPLE_POP}$ and, conditional on positive values, $\delta\kappa_s < \SimCombinedCI{SIMPLE_POP_POS}$ from the joint likelihood analysis. With 90\% credibility, we find $\delta\kappa_s = \SimCombinedCI{HIER_POP}$ from the hierarchical analysis. The generic population results constrain $\delta\kappa_s < \SimCombinedCI{HIER_POP_POS}$ when we restrict to positive prior region. Also, we find the hyperparameters to be consistent with the Kerr BBH hypothesis with 90\% credible bounds with $\mu = \SimCombinedCI{HIER_MU}$ and $\sigma < \SimCombinedCI{HIER_SIGMA}$. Compared to the previous bounds reported in~\cite{LIGOScientific:2020tif}, $\mu = -24.6^{+30.7}_{-35.3}$ and $\sigma < 52.7$, the $\sigma$ estimate improves, meaning tighter constraints on $\delta\kappa_s$, while the peak of the distribution is shifted more towards the negative prior region. The shift in the peak or $\mu$ omits the BBH value with the 90\% credibility and can be associated to the poor $\delta\kappa_s$ constraints on the negative side of the prior region from the individual events, emerging from waveform degeneracies at $\delta\kappa_s <0$ with a certain region of the spin parameter space. A future study employing waveform models including higher harmonics may help break those degeneracies and hence to improve our overall parameter estimation~\cite{Krishnendu:2019tjp,siqm_pop}. Moreover, a more generic approach has been recently proposed~\cite{siqm_pop} that uses a hierarchical mixture-likelihood formalism to estimate the fraction of events in the population that deviated from BBH nature. With the increased number of detections in the future, it would be more natural to employ generic approaches that considers the population to be comprised of BBH and non-BBH subpopulations. 

The combined  log Bayes factor of $\log_{10} \mathcal{B}^{\rm Kerr}_{\delta\kappa_s\,\neq\,0} = \SimBF{SYM}{}$ is obtained supporting the BBH hypothesis over the hypothesis of all events being non-BBH. This changes to log Bayes factor of $\log_{10} \mathcal{B}^{\rm Kerr}_{\delta\kappa_s\,>\,0} = \SimBF{POS}{}$ if we only allow $\delta\kappa_s \geq 0$. The findings here are all consistent with the results reported in GWTC-2~\cite{LIGOScientific:2020tif} although the combined constraints are not directly compatible due to the different selection of events.

\section{Tests of gravitational-wave propagation}
\label{sec:liv}
GR predicts that gravitational waves  propagate nondispersively and  hence they are described by the dispersion relation  
$E^2=p^2c^2$, where $E$ and $p$ are the energy and  momentum of the wave. Detection of dispersion of gravitational waves can be seen as a signature of modifications to GR. For example, some of the Lorentz violating theories of gravity predict a modified dispersion relation~\cite{Calcagni:2009kc,AmelinoCamelia:2002wr, Horava:2009uw,Sefiedgar:2010we,Kostelecky:2016kfm}. We use a parameterized model~\cite{Will:1997bb, Mirshekari:2011yq} for dispersion of gravitational waves that helps search for the presence of dispersion using the data without referring to the details of the modified theory.

Our parameterized dispersion relation reads~\cite{ Mirshekari:2011yq}
\begin{equation}
\label{eq:liv:dispersion}
E^2 = p^2c^2 + A_\alpha p^\alpha c^\alpha\, ,
\end{equation}
where $A_\alpha$ and $\alpha$ are two phenomenological parameters characterizing dispersion. 
The modified dispersion relation causes  frequency modes of gravitational waves to propagate at different speeds,  changing the overall phase morphology of the gravitational wave that are observed with respect to the GR predictions. This can be incorporated in the waveform as  frequency-dependent corrections to its phase evolution~\cite{Mirshekari:2011yq,LIGOScientific:2019fpa}. Here we assume that the waveform obtained in the local wave zone~\cite{1980RvMP...52..299T}  of the system is consistent with GR~\cite{LIGOScientific:2019fpa}. 

For different  choices of $\alpha$, the modified dispersion leads to a deviation in the GR phasing formula. For example, $\alpha=0$ with $A_\alpha>0$ corresponds to the dispersion effect of a massive graviton with mass $m_g c^2= \sqrt{A_0}$~\cite{Will:1997bb}. We choose to test the dispersion relation for a set of eight discrete values of $\alpha$ between $0$ and $4$ with a step of $0.5$ excluding  $\alpha=2$. When $\alpha = 2$, the speeds of all the frequency components are modified in the same way; therefore, the gravitational-wave signal remains unchanged from the GR prediction except for an overall change in the time of arrival of the signal.

Our method is identical to the previous analyses performed in GWTC-1 and GWTC-2~\cite{LIGOScientific:2019fpa,LIGOScientific:2020tif}, except for the use of a more up-to-date IMRPhenomXP~\cite{Garcia-Quiros:2020qpx} waveform model as opposed to the PhenomPv2~\cite{Khan:2018fmp} waveform employed in GWTC-1 and GWTC-2~\cite{LIGOScientific:2019fpa,LIGOScientific:2020tif}. We perform parameter estimation using the nested sampling algorithm~\cite{Skilling:2004ns} as implemented in the \texttt{LALInference} package~\cite{Veitch:2014wba} and obtain bounds on the phenomenological parameters $A_\alpha$ for each event. As in the case of preceding analyses, we perform the sampling  for $A_\alpha<0$ and $A_\alpha>0$ separately~\cite{LIGOScientific:2019fpa,LIGOScientific:2020tif}, and then combine the posterior to produce the joint $A_\alpha$ posterior.  We choose uniform priors for 
the phenomenological parameters $A_\alpha$. However, while computing the bound on the graviton mass $m_g$, which is derived from $A_0$, we re-weight the posteriors such that the prior on $m_g$ is uniform.

Propagation effects are independent of the source properties. Therefore we can combine the results from individual events to compute overall constraints over the phenomenological dispersion parameters.   We obtain the combined posterior distributions of $A_\alpha$ by multiplying the likelihoods from individual events and weighting the product with the prior.  

We perform the analysis on the 12 BBH candidate events in the catalog that are listed in Table~\ref{tab:events}. Though we analyzed  \NNAME{GW191109A}, the posteriors obtained were too wide to be informative, and following the study regarding this event reported in  Appendix \ref{app:sys}, which  finds that nonstationarities in the detector noise could dominate over the signal, we exclude this event from further analysis.  Analysis of another BBH event,  \NNAME{GW200316I}~has sampling issues and is thus excluded from the analysis.
Further, we do not include NSBH event \NNAME{GW200115A}~in this analysis due to the computational constraints. Nonetheless, this is among the closest events in the catalog and would have a negligible impact on the joint bounds.

\begin{figure}
	\includegraphics[width=0.5\textwidth]{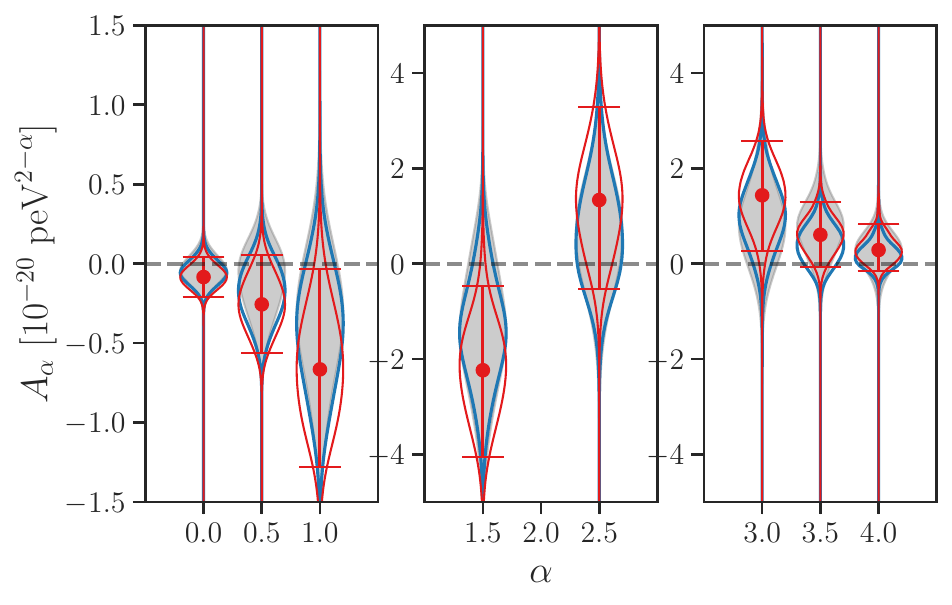}
	\caption{ Results for the modified dispersion analysis (Sec.~\ref{sec:liv}). The red violin plots show the combined posteriors of the parameter $A_\alpha$ calculated from the GWTC-3 events with the error bars denoting the $90\%$ credible intervals.
		For comparison,   we also present the combined posteriors after excluding the events \NNAME{GW200219D}~and \NNAME{GW200225B}~using blue violin plots.  The gray plots in the background are the combined posteriors corresponding to  GWTC-2~\cite{LIGOScientific:2020tif}. }
	\label{fig:liv:results_violin}
\end{figure}

Fig.~\ref{fig:liv:results_violin} shows the violin plots of joint posteriors on the phenomenological parameters $A_\alpha$ for various values of $\alpha$, which are obtained by combining posteriors from analysis of individual events.  The red violin plots represent the posteriors obtained from all 43 selected events (31 events from pre-O3b and 12 O3b events).  For some of the $\alpha$ values, the posteriors show biases with respect to the previous results~\cite{LIGOScientific:2020tif} due to the inclusion of O3b events. We have identified the events \NNAME{GW200219D}~and \NNAME{GW200225B}~as having the strongest impact in biasing the combined posterior.   These are the events with the lowest residual SNR $p$-values  among all the O3b events (see Table \ref{tab:residuals}). \NNAME{GW200225B}~shows $p$-value of  $0.05$ with fitting factor $\ff = 0.86$ and  \NNAME{GW200219D}~has $p$-value $=0.1$ with $\ff = 0.74$. These events require detailed analysis to understand the reasons for the observed deviations, which we leave for follow-up work.  
 For comparison, in Fig.~\ref{fig:liv:results_violin}, we also plot the combined posteriors from all the events excluding \NNAME{GW200219D}~and \NNAME{GW200225B}~(blue violin plots).  These are consistent with the   GWTC-2~\cite{LIGOScientific:2020tif} results (gray plots in the background)  and show  an average improvement of $\LivImprovFoutyoneEvents{AMP}{}$ over the previous results,  which is in agreement with the Gaussian expectation for improvement from $\LivEvents{GWTC-3 (41 events)}$ events  compared to $\LivEvents{GWTC-2}$ of GWTC-2 \cite{LIGOScientific:2020tif}. 

\begin{figure}
	\centering
	\includegraphics[width=0.5\textwidth]{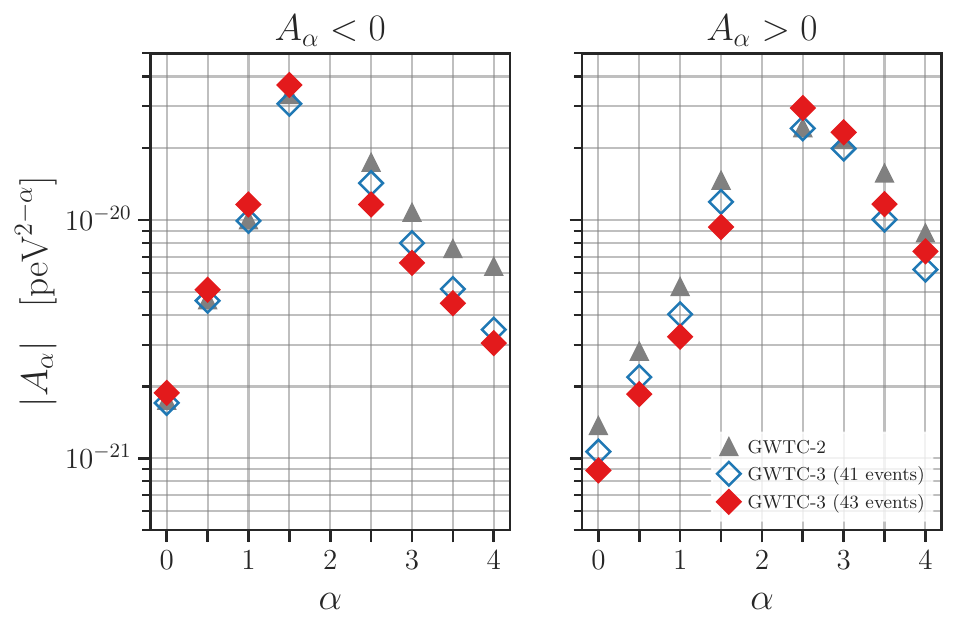}
	\caption{ Results for the modified dispersion analysis (Sec.~\ref{sec:liv}). The scatter plot of 90\% credible upper bounds on the modulus of deviation parameters $A_\alpha$. The one-sided bounds  are computed for positive and negative values of the parameters separately. Filled (open) diamond markers  represent the GWTC-3 bounds including (excluding) the events \NNAME{GW200219D}~and \NNAME{GW200225B}. The gray markers in the background denoted the numbers obtained from the previous analysis~\cite{LIGOScientific:2020tif}. }
	\label{fig:liv:results_summary}
\end{figure}

In Fig.~\ref{fig:liv:results_summary}, we present the scatter plot of 90\% credible upper bounds on  $|A_\alpha|$, for $A_\alpha>0$ and $A_\alpha<0$ separately. In the figure, red-filled  diamond markers  represent the GWTC-3 bounds. We also show the bounds from the analysis excluding the events  \NNAME{GW200219D}~and \NNAME{GW200225B}~in the blue diamond markers. For quantitative comparison, we list $|A_\alpha|$ bounds, including bounds on the graviton mass  $m_g$,  in Table~\ref{tab:liv:results_summary}. 
To demonstrate the level of bias in the posteriors with respect to the GR hypothesis, we  included the GR quantiles $Q_{\rm GR} = P(A_\alpha < 0)$ in Table~\ref{tab:liv:results_summary}.

The updated 90\% credible bound on the graviton mass obtained by combining posteriors of 43 GWTC-3 events is $m_g \leq \LivMgUL \mathrm{eV}/c^2$,  which is  $1.31$~
 times better than the Solar System bound of $3.16\times10^{-23}~\mathrm{eV}/c^2$~\cite{Bernus:2020szc}. Compared to the  previous GWTC-2 bound $3.09 \times 10^{-23}~\mathrm{eV}/c^2$~\cite{LIGOScientific:2020tif}, the improvement is a factor of $1.28$. 
\begin{table*}
	\caption{\label{tab:liv:results_summary}
	 Results for the modified dispersion analysis (Sec.~\ref{sec:liv}).
	The table shows 90\%-credible upper bounds on the graviton mass $m_g$ and the absolute value of the dimensionless phenomenological parameter $\bar{A}_\alpha = A_\alpha/\mathrm{eV}^{2-\alpha}$.  $Q_{\rm GR} = P(A_\alpha < 0)$ denotes the quantiles  corresponding to GR hypothesis. The $<$ and $>$ labels denote the bounds on $|\bar{A}_\alpha|$ for $A_\alpha>0$ and $A_\alpha<0$  respectively. We  also included bounds  computed from GWTC-2~\cite{LIGOScientific:2019fpa,LIGOScientific:2020tif} for comparison.}
	\centering
	\resizebox{\textwidth}{!}{\begin{tabular}{ccc*{7}{cccc}ccc}
 \toprule
 & $m_g$ & & \threec{$|\bar{A}_0|$} & & \threec{$|\bar{A}_{0.5}|$} & & \threec{$|\bar{A}_1|$} & & \threec{$|\bar{A}_{1.5}|$} & & \threec{$|\bar{A}_{2.5}|$} & & \threec{$|\bar{A}_3|$} & & \threec{$|\bar{A}_{3.5}|$} & & \threec{$|\bar{A}_4|$}\\
 \cline{4-6}
 \cline{8-10}
 \cline{12-14}
 \cline{16-18}
 \cline{20-22}
 \cline{24-26}
 \cline{28-30}
 \cline{32-34}
  & [$10^{-23}$ & & $<$ & $>$ & $Q_\text{GR}$ & & $<$ & $>$ & $Q_\text{GR}$ & & $<$ & $>$ & $Q_\text{GR}$ & & $<$ & $>$ & $Q_\text{GR}$ & & $<$ & $>$ & $Q_\text{GR}$ & & $<$ & $>$ & $Q_\text{GR}$ & & $<$ & $>$ & $Q_\text{GR}$ & & $<$ & $>$ & $Q_\text{GR}$\\
 &  eV$/c^2$] & & \twoc{[$10^{-45}$]} & [\%] & & \twoc{[$10^{-38}$]} & [\%] & & \twoc{[$10^{-32}$]} & [\%] & & \twoc{[$10^{-26}$]} & [\%] & & \twoc{[$10^{-14}$]} & [\%] & & \twoc{[$10^{-8}$]} & [\%] & & \twoc{[$10^{-2}$]} & [\%] & & \twoc{[$10^{4}$]} & [\%]\\
 \midrule
GWTC-2 & 3.09 &  & 1.75 & 1.37 & 66 &  & 0.46 & 0.28 & 66 &  & 1.00 & 0.52 & 79 &  & 3.35 & 1.47 & 83 &  & 1.74 & 2.43 & 31 &  & 1.08 & 2.17 & 17 &  & 0.76 & 1.57 & 12 &  & 0.64 & 0.88 & 25 \\
GWTC-3 (43 events) & 2.42 &  & 1.88 & 0.89 & 86 &  & 0.51 & 0.19 & 91 &  & 1.16 & 0.32 & 96 &  & 3.69 & 0.93 & 98 &  & 1.16 & 2.95 & 13 &  & 0.66 & 2.33 & 2 &  & 0.45 & 1.16 & 7 &  & 0.30 & 0.74 & 15 \\
\bottomrule
\end{tabular}
}
\end{table*}

\section{Polarizations}
\label{sec:pol}
Measuring the polarization content of gravitational waves is a way of constraining possible deviations from GR, as the theory allows only two of the six polarization states predicted by generic metric theories of gravity \cite{Eardley:1973br,Eardley:1974nw}. Assuming $M$ generic polarization modes, the frequency-domain strain data $\tilde{\boldsymbol{d}}(f)$ measured by a network of $D$ detectors can be written as the combination of a signal $\tilde{\boldsymbol{s}}(f)$ and noise $\tilde{\boldsymbol{n}}(f)$, or alternatively, as 
\begin{equation}
	\label{eq:pol_signal}
	\tilde{\boldsymbol{d}}(f)=\boldsymbol{F}\tilde{\boldsymbol{h}}(f)+\tilde{\boldsymbol{n}}(f),
\end{equation}
where $\tilde{\boldsymbol{s}}(f)=\boldsymbol{F}\tilde{\boldsymbol{h}}(f)$, $\boldsymbol{F}\in\mathbb{R}^{D\times M}$ are the beam pattern functions of the detectors and $\tilde{\boldsymbol{h}}(f)\in\mathbb{C}^{M}$ are the signal's polarization modes. We could interpret the gravitational-wave signal as a geometric projection on the subspace spanned by the basis vectors of $\boldsymbol{F}$. By projecting the data on the subspace orthogonal to these vectors, one can then construct null streams, i.e., linear combinations of the data containing no information on the signal \cite{GuerselTinto1989,ShourovLazzariniSteinSuttonSearleTinto2006}. Given $D$ detectors, it is possible to construct at most $D-M$ null streams. The projection operation can be formalized through the introduction of a null operator $\boldsymbol{P}$ \cite{Sutton:2009gi}
\begin{equation}
	\boldsymbol{P} = \boldsymbol{I} - \boldsymbol{F}(\boldsymbol{F}^{\dagger}\boldsymbol{F})^{-1}\boldsymbol{F}^{\dagger},
\end{equation}
where $\boldsymbol{I}$ is the identity matrix and $\dagger$ denotes conjugate transpose. The quantities $\boldsymbol{F}$ depend on the sky location of the signal, as well on the polarization angle and event time and, by construction, $\boldsymbol{P}\tilde{\boldsymbol{s}}(f) = \boldsymbol{0}$.

At least $M+1$ detectors are needed to apply the null stream method in the most generic case, although for specific sky locations less detectors will suffice to test certain polarization hypotheses \cite{Hagihara:2018azu,Hagihara:2019ihn}.   
The beam pattern functions of the breathing and longitudinal scalar modes are not linearly independent, and thus the maximum number of independent polarization modes is five \cite{PhysRevD.59.102002,Chatziioannou:2012rf}. Consequently, past analyses \cite{Abbott:2017oio,Abbott:2018lct,LIGOScientific:2020tif,Takeda:2020tjj} tested only pure polarization hypotheses, as these are fully characterized by two polarisation modes at most, and in this case it is possible to construct a null stream with the strain measured by three detectors. 

In this work, we use a method that allows tests of mixed polarization states with 2 and 3 detectors \cite{Wong:2021cmp}. This enables all our events to be used to compute combined Bayes factors, while the previous analysis \cite{LIGOScientific:2020tif} was restricted to 3-detector events. The method builds upon an \textit{effective antenna pattern function} $\bar{\boldsymbol{F}}\in\mathbb{C}^{D\times L}$ that is constructed from a subset of $L<M$ polarization modes. For each hypothesis to be tested, the relevant polarization state is projected into the chosen basis: thus, one orthogonalizes the data with respect to a smaller subspace spanned by the basis modes, rather than the assumed polarization modes. 
Each polarization mode $\tilde{h}_{m}$ can be rewritten as a linear combination of the basis modes, plus an additional orthogonal component
\begin{equation}
\tilde{h}_{m}(f)=\sum_{k=1}^{L}C_{m}^{\parallel}\tilde{h}_{\parallel,k}(f)+C_{m}^{\perp}\tilde{h}_{\perp}(f),
\end{equation}
with $C_{m}^{\parallel},C_{m}^{\perp}\in\mathbb{C}$.
We perform the null projection with respect to the subspace spanned by the component of the beam pattern vectors parallel to the basis mode(s). Therefore, the method is sensitive to any component of a given polarization hypothesis that is parallel to the chosen basis modes(s). The subspace removed by the null projection does not need to coincide with the polarization subspace of the hypothesis being tested.

We will conduct analyses employing either one ($L=1$) or two ($L=2$)  basis modes. The $L=2$ parameterization allows more freedom in the choice of the basis modes, but at the cost of a weaker distinguishability between different polarization hypotheses. The subspaces spanned by the beam pattern function vectors for different hypotheses, in fact, will generally have a larger overlap in the $L=2$ than in $L=1$ case. The polarization content is constrained to be a linear combination of the basis modes and, therefore, the $L=1$ analysis is expected to produce more stringent results, due to the strongest constraints imposed on the signal. On the other hand, the $L=2$ analysis will be able to capture orthogonal components missed by the $L=1$ analysis. 

Right ascension, declination and polarization angle are free parameters shared by both analyses. Additionally, we marginalize over the relative amplitude and phase with respect to the basis mode(s). While this implies the $L=2$ analysis has more degrees of freedom than the $L=1$ one, the polarization content of GR can be represented by both methods. In fact, in the quadrupolar approximation, $h_{+}$ and $h_{\times}$ only differ in terms of a relative amplitude and phase, which are marginalized over in the computation of the evidence, and therefore only one basis mode is sufficient to capture both. 

In the $L=2$ analysis, we choose one tensorial mode and one non-tensorial mode as the basis, to better capture possible extra degrees of freedom predicted by alternative theories of gravity. All events are analyzed with both models, to check consistency between the two formulations. 
The nature of null projection prohibits the comparison of model evidences with different number of basis modes \cite{Wong:2021cmp}; therefore, we conduct the analyses with one and two basis modes independently. 

We reanalyze here all of the events from the first three observing runs with a FAR below our $10^{-3} \,\mathrm{yr}^{-1}$ threshold. The pipeline employed here has been optimized to analyze the polarization content based on excess power in the data. Some events from the third observing run, namely \FULLNAME{GW190425B}, \FULLNAME{GW190720A}, \FULLNAME{GW190828B}, \FULLNAME{GW191129G}, \FULLNAME{GW200115A}, \FULLNAME{GW200202F}, and \FULLNAME{GW200316I}, do not have a strong enough time--frequency track for the pipeline to capture them and we exclude them from our analysis.

We test whether the data support a number of non-GR polarization hypotheses, which we denote by T (tensorial), V (vector) and S (scalar) or combinations of these terms. 
In the $L=1$ case, we choose the tensorial plus mode $h_{+}$ as the basis for all hypotheses involving tensor modes, the vector mode \textit{x} as the basis for the pure vector and vector--scalar hypotheses, and the scalar breathing mode as a basis for the scalar hypothesis. In the $L=2$ analyses, following the notation introduced above, we project mixed polarizations of the type T$\xi$ on a basis including $h_{+}$ and the relevant non-tensorial $\xi$ mode, VS polarizations on a vector--scalar basis and TVS polarizations on the $h_{+}$ and scalar--breathing mode. The basis modes chosen for each polarization hypothesis and the corresponding free parameters are summarized in Table \ref{tab:pol_table_3}.

\begin{table*}[!t]
\caption{The table summarizes the choices of basis used in the polarization test. \textit{+}, $\times$, \textit{b}, \textit{l}, \textit{x}, and \textit{y} represent the plus mode, cross mode, scalar breathing mode, scalar longitudinal mode, vector x mode, and vector y mode respectively. The first column shows the polarization hypothesis being tested, the third column reports the number of basis modes, and the last column reports the number of free parameters that are marginalized over in the computation of the evidence. }
\label{tab:pol_table_3}
\begin{tabular}{ c  c  c  c  c  c}
		\toprule
		Hypothesis & Description & \# of basis modes & Mode(s) & Basis mode(s) & Free parameters \\ [0.5ex] 
		\midrule
		$\mathcal{H}_{\textrm{T},1}$ & Pure tensorial & 1 & $+$, $\times$ & $+$ & 5 \\
		$\mathcal{H}_{\textrm{V},1}$ & Pure vectorial & 1 & $x$, $y$ & $x$ & 5 \\
		$\mathcal{H}_{\textrm{S},1}$ & Pure scalar & 1 & $b$ & $b$ & 2 \\
		$\mathcal{H}_{\textrm{TS},1}$ & Tensor--scalar & 1 & $+$, $\times$, $b$, $l$ & $+$ & 9  \\
		$\mathcal{H}_{\textrm{TV},1}$ & Tensor--vector & 1 & $+$, $\times$, $x$, $y$ & $+$ & 9  \\
		$\mathcal{H}_{\textrm{VS},1}$ & Vector--scalar & 1 & $x$, $y$, $b$, $l$ & $x$ & 9   \\
		$\mathcal{H}_{\textrm{TVS},1}$ & Tensor--vector--scalar & 1 & $+$, $\times$, $b$, $l$, $x$, $y$ & $+$ & 13   \\
		$\mathcal{H}_{\textrm{T},2}$ & Pure tensorial & 2 & $+$, $\times$ & $+$, $\times$& 2  \\
		$\mathcal{H}_{\textrm{V},2}$ & Pure vectorial & 2 & $x$, $y$ & $x$, $y$ & 2  \\
		$\mathcal{H}_{\textrm{TS},2}$ & Tensor--scalar & 2 & $+$, $\times$, $b$, $l$ & $+$, $b$ & 11\\
		$\mathcal{H}_{\textrm{TV},2}$ & Tensor--vector & 2 & $+$, $\times$, $x$, $y$ & $+$, $x$ & 11 \\
		$\mathcal{H}_{\textrm{VS},2}$ & Vector--scalar & 2 & $x$, $y$, $b$, $l$ & $x$, $b$ & 11 \\
		$\mathcal{H}_{\textrm{TVS},2}$ & Tensor--vector--scalar & 2 & $+$, $\times$, $b$, $l$, $x$, $y$ & $+$, $b$ & 19 \\
		\bottomrule
	\end{tabular}

\end{table*}

We present combined $\log_{10}$ Bayes factors assuming the events are independent from each other. Table~\ref{tab:pol_1} shows the combined $\log_{10}$ Bayes factors of the 2-detector and 3-detector events for the $L = 1$ analysis, with the last row showing the combined $\log_{10}$ Bayes factor for all eligible GWTC-3 events. The pure--scalar, pure--vector and vector--scalar hypotheses are significantly disfavored, while any mixed hypothesis involving tensor modes (i.e., tensor--scalar, tensor--vector, and tensor--vector--scalar) cannot be ruled out conclusively. 

Table~\ref{tab:pol_2} shows the combined $\log_{10}$ Bayes factors of the 3-detector events for the $L = 2$ analysis. There are no available data for O1, due to the fact that only two interferometers were operational at the time (LIGO Hanford and LIGO Livingston). In this case, the last row shows the combined $\log_{10}$ Bayes factor for O2 and O3 events. Mixed hypotheses in this case are more strongly disfavored than the pure vector hypothesis (the pure scalar hypothesis cannot be tested here, due to the fact that the longitudinal and breathing modes for interferometers are not linearly independent). This is particularly evident for the TVS hypothesis, which has the largest number of free parameters. All mixed hypotheses will be penalized by a higher Occam factor, due to the increased prior volume in the Bayesian evidence integral, as each mode is characterized by a relative amplitude and phase with respect to the basis modes. The number of free parameters necessary to represent each hypothesis, depending on the number of basis modes employed, is reported in the last column of Table \ref{tab:pol_table_3}.

If the polarization content of the signals is purely tensorial, mixed-mode hypotheses including a tensorial component will be also able to represent the data, as the signal space of the pure tensorial hypothesis is a subspace of that of the mixed-mode hypotheses including a tensorial component. In this case, we expect to see a larger penalization of mixed hypotheses in the $L = 2$ analysis, as the same polarization hypothesis will correspond to a larger parameter space than in the $L = 1$ analysis, as can be seen from Table \ref{tab:pol_table_3}. This expectation is confirmed when comparing the results reported in Tables~\ref{tab:pol_1}  and~\ref{tab:pol_2}. On the other hand, if we consider the vector--scalar hypothesis, the additional parameters in the $L = 2$ analysis will help to fit the tensorial component of the data: this explains why we see a less negative $\log_{10}$ Bayes factor in the $L = 2$ analysis than in the $L=1$ one for this hypothesis. A similar argument applies to the purely scalar and vectorial hypotheses.
 
These results support the conclusion that the population of events observed is consistent with the pure tensorial hypothesis, as predicted by GR, in line with the conclusion of past analyses of GWTC-1 and GWTC-2 data \cite{LIGOScientific:2019fpa,LIGOScientific:2020tif}.

\begin{table*}[!t]
\caption{Combined $\log_{10}$ Bayes factors $\mathcal{B}$ for various polarization hypotheses against the tensor hypothesis, using both 2-detector and 3-detector events. Polarization states have been projected onto one basis-mode as detailed in Sec.~\ref{sec:pol}. Positive (negative) values indicate that the hypothesis indicated in the superscript is favored (disfavored) with respect to the tensorial hypothesis. Error bars refer to $90\%$ credible intervals.}
\label{tab:pol_1}
	\begin{tabular}{c  c  c  c  c  c  c}
\toprule
		Events & $\log_{10}\mathcal{B}_{\textrm{T}}^{\textrm{S}}$ & $\log_{10}\mathcal{B}_{\textrm{T}}^{\textrm{V}}$ & $\log_{10}\mathcal{B}_{\textrm{T}}^{\textrm{TS}}$ & $\log_{10}\mathcal{B}_{\textrm{T}}^{\textrm{TV}}$ & $\log_{10}\mathcal{B}_{\textrm{T}}^{\textrm{VS}}$ & $\log_{10}\mathcal{B}_{\textrm{T}}^{\textrm{TVS}}$\\
\midrule
		O1 & $-0.04 \pm 0.07$ & $0.09 \pm 0.07$ & $0.04 \pm 0.07$ & $0.09 \pm 0.07$ & $0.09 \pm 0.07$ & $0.07 \pm 0.07$ \\
		O2 & $-0.42 \pm 0.12$ & $0.04 \pm 0.12$ & $0.08 \pm 0.12$ & $0.22 \pm 0.12$ & $0.09 \pm 0.12$ & $0.35 \pm 0.12$ \\
		O3a & $-1.85 \pm 0.21$ & $-1.04 \pm 0.20$ & $0.25 \pm 0.20$ & $0.07 \pm 0.20$ & $-1.05 \pm 0.20$ & $-0.18 \pm 0.20$ \\
		O3b & $-1.93 \pm 0.17$ & $-0.79 \pm 0.17$ & $-0.17 \pm 0.17$ & $-0.07 \pm 0.17$ & $-0.86 \pm 0.17$ & $-0.32 \pm 0.17$ \\
\midrule
		Combined & $-4.24 \pm 0.30$ & $-1.70 \pm 0.30$ & $0.20 \pm 0.30$ & $0.31 \pm 0.30$ & $-1.73 \pm 0.30$ & $-0.08 \pm 0.30$ \\
\bottomrule
	\end{tabular}
	\end{table*}

	\begin{table*}[!t]	
	\caption{Combined $\log_{10}$ Bayes factor $\mathcal{B}$ for various polarization hypotheses against the tensor hypothesis, for 3-detector events. Polarization states been projected onto two basis-modes as explained in Sec.~\ref{sec:pol}. Positive (negative) values indicate that the hypothesis indicated in the superscript is favored (disfavored) with respect to the tensorial hypothesis. Error bars refer to $90\%$ credible intervals.}
	\label{tab:pol_2}
	\begin{tabular}{c  c  c  c  c  c}
		\toprule
		Events & $\log_{10}\mathcal{B}_{\textrm{T}}^{\textrm{V}}$ & $\log_{10}\mathcal{B}_{\textrm{T}}^{\textrm{TS}}$ & $\log_{10}\mathcal{B}_{\textrm{T}}^{\textrm{TV}}$ & $\log_{10}\mathcal{B}_{\textrm{T}}^{\textrm{VS}}$ & $\log_{10}\mathcal{B}_{\textrm{T}}^{\textrm{TVS}}$\\
		\midrule
		O1 & $-$ & $-$ & $-$ & $-$ & $-$ \\
		O2 & $0.05 \pm 0.03$ & $0.01 \pm 0.03$ & $-0.02 \pm 0.03$ & $0.06 \pm 0.03$ & $0.01 \pm 0.03$ \\
		O3a & $-0.37 \pm 0.12$ & $-0.77 \pm 0.12$ & $-0.72 \pm 0.12$ & $-0.73 \pm 0.12$ & $-0.91 \pm 0.12$ \\
		O3b & $-0.09 \pm 0.10$ & $-0.22 \pm 0.10$ & $-0.35 \pm 0.10$ & $-0.38 \pm 0.10$ & $-0.38 \pm 0.10$ \\
		\midrule
		Combined & $-0.41 \pm 0.16$ & $-0.98 \pm 0.16$ & $-1.09 \pm 0.16$ & $-1.05 \pm 0.16$ & $-1.29 \pm 0.16$ \\
		\bottomrule
	\end{tabular}

	\end{table*}

\section{Remnant properties}
\label{sec:rem}

    \subsection{Ringdown}
    \label{sec:rin}
    The highly distorted black hole remnant formed from the merger emits gravitational radiation which is  referred to as ringdown. The late-ringdown waveform can be expressed as a superposition of quasi-normal modes (QNM) with a complex frequency~\cite{Vishveshwara1970b,Cunningham:1978zfa}. The real part of the complex frequency is the oscillation frequency  and the imaginary part is the inverse damping time of the mode. According to GR, for astrophysical black holes, the frequency and damping times are completely determined by the mass and spin of the remnant black hole~\cite{Carter,Penrose:1969pc,Hansen:1974zz,Gurlebeck:2015xpa}. In fact, at the current sensitivity, electric-like black hole charges have been shown to not leave a detectable imprint on ringdown measurements~\cite{Carullo:2021oxn}.
The relationship between frequency and remnant parameters, detection of multi-mode ringdown signals offers a unique test of the black hole nature of the merger remnant~\cite{Dreyer:2003bv,LISA_spectroscopy} and could be used to distinguish among different classes of ECOs~\cite{Berti:2015itd}.

The study of the QNM spectrum, and the post-merger waveform in general, contains a wealth of information about the remnant black hole. The spectrum of radiation emitted during the ringdown is usually expressed in terms of spheroidal  harmonic basis functions with spin-weight $-2$ denoted by ${}_{-2}S_{\ell m n}$. The indices $(\ell,m)$ represent the angular decomposition of the modes, whereas the index $n$ denotes various \emph{tones} of the spectrum starting with $n=0$. A schematic decomposition of the post-merger signal reads~\cite{LIGOScientific:2020tif},

\begin{widetext}
\begin{equation}\label{eq:ring_wf}
\begin{aligned}
	h_{+}(t) - i h_{\times}(t) = \sum_{\ell = 2}^{+\infty} \sum_{m = - \ell}^{\ell} \sum_{n = 0}^{+\infty} \; & \; \mathcal{A}_{\ell m n} \; \exp \left[ -\frac{t-t_0}{(1+z)\tau_{\ell m n}} \right] \exp \left[ -\frac{2\pi i f_{\ell m n}(t-t_0)}{1+z} \right] {}_{-2}S_{\ell m n}(\theta, \phi, \chi_{\rm f}), \\
\end{aligned}
\end{equation}
\end{widetext}
where $\mathcal{A}_{\ell m n}$ denotes the amplitude of the mode, $t_0$ is the start time of the ringdown model, and $z$ is the redshift of the source. The frequency and the damping time of a mode characterized by the three indices are denoted by $\tau_{\ell m n}$ and $f_{\ell m n}$, respectively, while $\chi_{\rm f}$ is the final spin. The polar and azimuthal angles $(\theta,\phi)$, measured relative to the final spin axis, describe the direction to the observer. These coordinates assume the spin of the black hole to be along the $\theta=0$ direction.
The contribution of counter-rotating perturbations is ignored, since it is expected to be negligible in the post-merger regime of the signals under consideration.
We approximate the spheroidal harmonics with spherical harmonics, leading to a feature, called \emph{mode-mixing} of QNMs~\cite{buonanno:124018,Kelly:2012nd,Berti:2014fga}, which mixes modes that have the same $m$ index but different $\ell$ indices. Mode mixing, especially of $\ell=|m|=2$ mode, is expected to be negligible for the modes that we consider here~\cite{Isi:2021iql}.

We present two analyses of ringdown: the time-domain ringdown analysis {\tt pyRing}~\cite{Carullo:2019flw, Isi:2019aib}, based on damped sinusoids, and the parametrized ringdown analysis \textsc{pSEOB}, based on the \textsc{SEOBNRv4HM} waveform model~\cite{Ghosh:2021mrv}.
\subsubsection{The \textsc{pyRing} analysis} 
\textsc{pyRing} employs a time-domain approach. The three types of templates used in the analysis are $\mathrm{Kerr_{220}}$, $\mathrm{Kerr_{221}}$, and $\mathrm{Kerr}_\mathrm{HM}$. $\mathrm{Kerr_{220}}$ has only the fundamental modes, i.e., $(\ell, |m|, n) = (2,2,0)$, $\mathrm{Kerr_{221}}$ has the fundamental modes and its first overtones (both $\pm m$, with the same damping time and frequencies of opposite sign).
Instead, $\mathrm{Kerr}_\mathrm{HM}$ includes the fundamental mode and higher moments (HMs) with $m = \ell$ and $m = \ell - 1$ modes with $\ell \leq 4$ (and $n = 0$). The $\mathrm{Kerr}_\mathrm{HM}$ template also takes into account the effect of mode-mixing due to the use of spherical harmonics instead of spheroidal harmonics to describe the ringdown.

While in the $\mathrm{Kerr_{220}}$ and $\mathrm{Kerr_{221}}$ models the amplitudes and phases are left free to vary, in the $\mathrm{Kerr}_\mathrm{HM}$ model, quasi-circular, aligned-spin NR fits are used to compute the mode amplitudes, frequencies, and damping times. The model calibration assumes that the remnant black hole originated from the quasi-circular coalescence of progenitor black holes with spins aligned with the orbital angular momentum. As other templates account for precession, this assumption does not affect the overall efficiency of the analysis. The progenitors' masses and spins are sampled with uniform priors.
In contrast to the GWTC-2 results~\cite{LIGOScientific:2020tif}, where they were independently measured, the remnant mass and spin are now also obtained through NR fits from the progenitor parameters~\cite{Jimenez-Forteza:2016oae}. This increases the coherence of the measurement, leading to tighter constraints on the remnant parameters. The model amplitudes are calibrated against NR simulations at $20 M$ after the peak of the $\ell=|m|=2$ complex strain, where $M$ is the total mass of the progenitor binary. The other two template types remain unchanged from previous analysis~\cite{LIGOScientific:2020tif}.

\textsc{pyRing} uses a reference time $t_0$, which is computed from an estimate of the peak of the strain $(h_+^2 + h_{\times}^2)$ from the full IMR analyses (with the approximant \texttt{IMRPhenomXPHM}) assuming GR. The sky location is fixed to coincide with the maximum likelihood value inferred from the full IMR analysis. When assuming the $\mathrm{Kerr_{221}}$ template, we fit the data starting at $t_0$~\cite{Giesler:2019uxc, Isi:2019aib}. To take into account the NR calibration of the template, the start time of the analysis is instead set to $15 G\bar{M}_\mathrm{f} (1+z)/c^3$ after $t_0$ when employing the $\mathrm{Kerr}_\mathrm{HM}$ template, where $\bar{M}_\mathrm{f}$ is the median IMR value of the remnant mass. This choice extrapolates the model outside its nominal validity region, however we have verified that this does not induce appreciable biases~\cite{Carullo:2018sfu}. In the $\mathrm{Kerr_{220}}$ case, due to the greater flexibility of this template, the analysis starts $10 G \bar{M}_\mathrm{f} (1+z)/c^3$ after $t_0$, which is when a linearised ringdown description is expected to dominate the signal for our analysis~\cite{PhysRevD.90.124032, Carullo:2018sfu, Baibhav:2023clw}. Both sky locations and start times at each detector, together with all configuration data required to reproduce the analysis, are released in~\cite{GWTC3:TGR:release}.

We report results where: i) the remnant parameters are constrained relative to the prior, and ii) the Bayesian evidence favors the presence of a signal over pure Gaussian noise when using our most sensitive template ($\mathrm{Kerr_{221}}$). Estimates of the remnant parameters obtained for the five events from O3b that pass the above criteria, are reported in Table~\ref{tab:rin:final_mass_spin}.

\begin{table*}[!t]
\caption{\label{tab:rin:final_mass_spin} The median, and symmetric $90\%$-credible intervals, of the redshifted final mass and final spin, inferred from the full IMR analysis (IMR) and the \textsc{pyRing} analysis (Sec.~\ref{subsec:rin1}) with three different waveform models ($\mathrm{Kerr_{220}}$, $\mathrm{Kerr_{221}}$, and $\mathrm{Kerr}_\mathrm{HM}$). 
A positive value of $\log_{10} \mathcal{B}^{\rm HM}_{\rm 220}$ indicates support for HM in the data, and a positive value of $\log_{10} \mathcal{B}^{\rm 221}_{\rm 220}$ shows support for the presence of the first overtone.
A positive value of $\log_{10} \mathcal{O}^{\rm modGR}_{\rm GR}$ quantifies the level of disagreement with GR.
The catalog-combined (including GWTC-2 events) log odds ratio is negative ($-0.88\pm 0.44$).}
\scalebox{0.95}{

\begin{tabular}{lllllllllllrrrr}
\toprule
Event & \multicolumn{4}{c}{Redshifted final mass} & \hphantom{X} & \multicolumn{4}{c}{Final spin} & \hphantom{X} & \multicolumn{1}{c}{Higher} & \hphantom{X} & \multicolumn{2}{c}{Overtones} \\
& \multicolumn{4}{c}{$(1+z)M_\mathrm{f} \; [M_{\odot}]$} & \hphantom{X} & \multicolumn{4}{c}{$\chi_{\mathrm{f}}$} & \hphantom{X} & \multicolumn{1}{c}{modes} & \hphantom{X} &  \multicolumn{2}{c}{} \\[0.075cm]
\cline{2-5}
\cline{7-10}
\cline{12-12}
\cline{14-15}
& IMR & $\mathrm{Kerr_{220}}$ & $\mathrm{Kerr_{221}}$ & $\mathrm{Kerr_{HM}}$ & \hphantom{X} & IMR & $\mathrm{Kerr_{220}}$ & $\mathrm{Kerr_{221}}$ & $\mathrm{Kerr_{HM}}$ & \hphantom{X} &  \multicolumn{1}{c}{$\log_{10} \mathcal{B}^{\rm HM}_{\rm 220}$} & \hphantom{X} & \multicolumn{1}{c}{$\log_{10} \mathcal{B}^{\rm 221}_{\rm 220}$} & \multicolumn{1}{c}{$\log_{10} \mathcal{O}^{\rm modGR}_{\rm GR}$} \\
\midrule

GW191109\_010717 &
$ 132.7^{+ 21.9 }_{- 13.8 } $ &
$ 187.1^{+ 21.1 }_{- 22.1 } $ &
$ 183.4^{+ 16.6 }_{- 16.1 } $ &
$ 180.3^{+ 21.6 }_{- 25.0 } $ &
\hphantom{X} &
$ 0.60^{+ 0.22 }_{- 0.19 } $ &
$ 0.86^{+ 0.06 }_{- 0.12 } $ &
$ 0.84^{+ 0.06 }_{- 0.09 } $ &
$ 0.80^{+ 0.07 }_{- 0.15 } $ &
\hphantom{X} & 
$ 0.00 $ &
\hphantom{X} &
$ 1.88 $ &
$ -0.33 $ \\[0.075cm]

GW191222\_033537 &
$ 114.2^{+ 14.3 }_{- 11.7 } $ &
$ 105.3^{+ 49.5 }_{- 27.1 } $ &
$ 105.2^{+ 40.0 }_{- 22.4 } $ &
$ 129.8^{+ 124.8 }_{- 65.8 } $ &
\hphantom{X} &
$ 0.67^{+ 0.08 }_{- 0.10 } $ &
$ 0.45^{+ 0.41 }_{- 0.40 } $ &
$ 0.47^{+ 0.35 }_{- 0.41 } $ &
$ 0.65^{+ 0.20 }_{- 0.55 } $ &
\hphantom{X} & 
$ 0.05 $ &
\hphantom{X} &
$ -0.89 $ &
$ -0.16 $ \\[0.075cm]

GW200129\_065458 &
$ 71.8^{+ 4.4 }_{- 3.9 } $ &
$ 58.0^{+ 15.4 }_{- 9.4 } $ &
$ 67.3^{+ 10.9 }_{- 10.3 } $ &
$ 130.9^{+ 159.9 }_{- 52.5 } $ &
\hphantom{X} &
$ 0.75^{+ 0.06 }_{- 0.06 } $ &
$ 0.38^{+ 0.38 }_{- 0.34 } $ &
$ 0.57^{+ 0.23 }_{- 0.41 } $ &
$ 0.66^{+ 0.20 }_{- 0.53 } $ &
\hphantom{X} & 
$ 0.15 $ &
\hphantom{X} &
$ 0.35 $ &
$ -0.18 $ \\[0.075cm]

GW200224\_222234 &
$ 90.3^{+ 6.4 }_{- 6.3 } $ &
$ 80.0^{+ 22.9 }_{- 15.4 } $ &
$ 86.8^{+ 17.9 }_{- 15.7 } $ &
$ 200.7^{+ 105.9 }_{- 98.7 } $ &
\hphantom{X} &
$ 0.73^{+ 0.06 }_{- 0.07 } $ &
$ 0.48^{+ 0.36 }_{- 0.42 } $ &
$ 0.57^{+ 0.26 }_{- 0.44 } $ &
$ 0.63^{+ 0.22 }_{- 0.40 } $ &
\hphantom{X} & 
$ 0.28 $ &
\hphantom{X} &
$ 0.15 $ &
$ -0.21 $ \\[0.075cm]

GW200311\_115853 &
$ 72.1^{+ 5.4 }_{- 4.7 } $ &
$ 70.7^{+ 23.4 }_{- 14.7 } $ &
$ 69.2^{+ 30.1 }_{- 16.0 } $ &
$ 277.4^{+ 128.5 }_{- 191.7 } $ &
\hphantom{X} &
$ 0.68^{+ 0.07 }_{- 0.08 } $ &
$ 0.31^{+ 0.43 }_{- 0.28 } $ &
$ 0.55^{+ 0.32 }_{- 0.45 } $ &
$ 0.56^{+ 0.25 }_{- 0.44 } $ &
\hphantom{X} & 
$ -0.02 $ &
\hphantom{X} &
$ -1.17 $ &
$ -0.08 $ \\[0.075cm]

\bottomrule
\end{tabular}
}
\end{table*}

There is agreement between the remnant properties obtained from the three waveform templates and the IMR analysis. 
The contribution of overtones or $\mathrm{HMs}$ during ringdown is quantified by log Bayes factors ($\log_{10} \mathcal{B}^{\rm HM}_{\rm 220}$, $\log_{10} \mathcal{B}^{\rm 221}_{\rm 220}$), which are reported in Table~\ref{tab:rin:final_mass_spin}. The Bayes factors depend sensitively upon the chosen priors \cite{Isi:2021iql}. When comparing a fit including all QNMs included in $\mathrm{Kerr}_\mathrm{HM}$, versus the same model including only the $\ell= |m| =2,\, n=0$ mode, we observe no strong evidence for the presence of $\mathrm{HMs}$.
Additionally, we search for evidence of $\mathrm{HMs}$ augmenting the $\mathrm{Kerr_{220}},\mathrm{Kerr_{221}}$ templates with each of the $(\ell, |m|, n) = (3,3,0), (3,2,0), (2,1,0), (2,0,0)$ modes separately. 
Following~\cite{Capano:2021etf}, to increase the sensitivity to these modes we assume the amplitudes of modes with opposite $m$ signs to be related by the non-precessing symmetry $A_{\ell,-|m|,n} = (-1)^{\ell} A^*_{\ell,|m|,n}$, while also imposing an amplitude hierarchy on the higher angular modes $\mathcal{A}_{\ell,m,n}/\mathcal{A}_{2,2,0} < 0.9$, as expected for quasi-circular binaries of moderate mass ratio. 
Under this set of assumptions, we find no significant evidence for the presence of higher modes in all events except for GW191109\_010717, which shows weak evidence for the presence of the $(3,2,0)$ and $(2,1,0)$ modes in addition to the $(2,2,0)$ one at times smaller than $t_0 + 5 \bar{M}_f$. However, as discussed below, non-stationary noise around the event time does not allow to draw reliable inference from this signal. Moreover for the parameter space under consideration, the templates signalling such evidence are composed only of fundamental $n=0$ modes and thus not reliable before $\sim t_0 + 10 \bar{M}_f$~\cite{PhysRevD.90.124032, Carullo:2018sfu, Baibhav:2023clw}. Comparing the $\mathrm{Kerr_{220}}$ and $\mathrm{Kerr_{221}}$ analyses, with both templates starting at the peak, we observe weak evidence for the presence of overtones only for the loudest among these signals (for example GW200129\_065458 shows a small such evidence). The estimates of the remnant parameters get closer to the full IMR waveform estimates when including overtones, in agreement with NR predictions.

The exception to the above discussion is GW191109\_010717, which shows an overestimation of the remnant mass and spin compared to IMR analyses. As discussed in Appendix~\ref{app:sys}, such discrepancies can be attributed to possible nonstationarities in the detector noise and we will exclude this event from any combined statement made below.

To investigate possible modifications to the ringdown spectrum of the remnant black hole, we add parametrised deviations to the $\mathrm{Kerr_{221}}$ template in the frequency and damping time with respect to their GR values for the $n=1$ mode. This parametrization can almost fully cover the two-tone parameter space and has a number of desirable features that facilitate both sampling and interpretation \cite{Isi:2021iql}.
Parameter estimation is performed over the same set of parameters appearing in the GR template, with the addition of the deviation parameters on which we impose uniform priors in the $[-1,1]$ range for the frequency $\delta \hat{f}_{221}$ and in the $[-0.9, 1]$ range for the damping time $\delta \hat{\tau}_{221}$. The lower bound on $\delta \hat{\tau}_{221}$ prevents issues due to the finite time resolution in the waveform sampling~\cite{LIGOScientific:2020tif}.
If GR provides an accurate description of the ringdown emission, we expect to observe posterior distributions of the deviation parameters to be centered around zero, together with a Bayesian evidence disfavouring the addition of non-GR parameters.

The inferred values of the frequency deviation parameters are consistent with GR for all events analysed, while weak constraints can be extracted on the damping times deviations from single events.
The damping time estimation of low-SNR events is more sensitive to violations of the Gaussianity and stationarity hypotheses compared to the frequency estimation~\cite{LIGOScientific:2020tif}. Additional studies investigating this behaviour will be required in the future to properly derive joint posteriors on this parameter when combining many weak events. The posterior distribution of $\delta \hat{\tau}_{221}$ often tends to rail towards the lower prior bound $-0.9$ for events with low SNR in the ringdown regime, as the data show little evidence for the first overtone.

\begin{figure}
\includegraphics[width=\columnwidth]{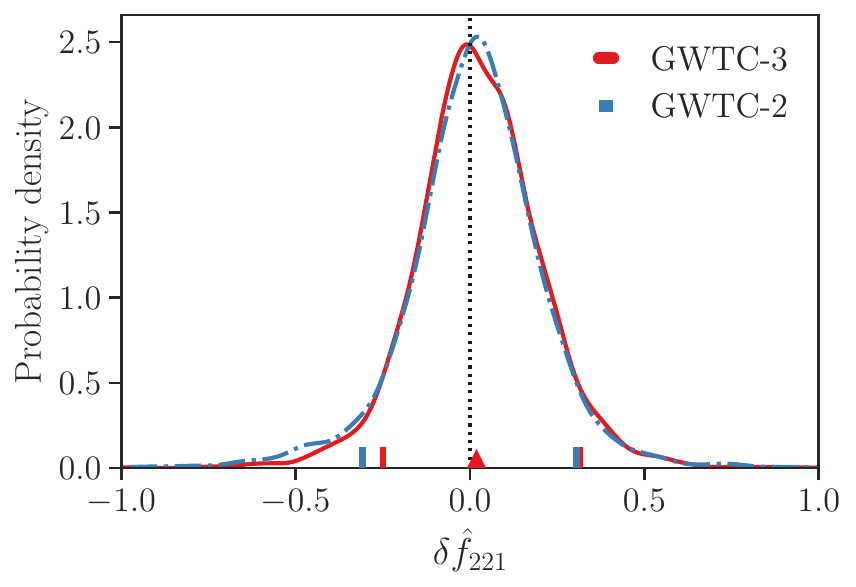}
\caption{
The posterior distribution of the fractional frequency deviation for the $\ell=|m|=2,n=1$ mode, $\delta \hat{f}_{221}$, from the \textsc{pyRing} joint hierarchical analysis (triangles and small vertical bars indicate respectively median and $90\%$ credible intervals). 
The measurements of $\delta \hat{f}_{221}$ from individual events, and its combined value using all available $21$ gravitational-wave events (red solid line), both show consistency with GR. 
Compared to the corresponding GWTC-2 constraint (dot--dashed blue line), the hierarchically combined posterior on the frequency deviation shows a $90\%$ CL shrinkage ratio of $\sim 8\%$. The median of the GWTC-2 posterior is not visible due to the overlap with the GWTC-3 one. See Sec.~\ref{subsec:rin1} for details.
}
\label{fig:rin:qnm_deviation_221_pyRing}
\end{figure}

To combine the set of measurements for all 21 available events we make use of a hierarchical analysis~\cite{LIGOScientific:2020tif}. 
The single events posteriors used to derive this joint bound are the marginalised $\delta \hat{f}_{221}$ posteriors obtained when allowing both the frequency and the damping time of the $221$ mode to deviate from the GR predictions.
We obtain a constraint on the frequency deviation equal to $\delta \hat{f}_{221} = 0.02^{+0.30}_{-0.27}$, overlapping with the GR predicted value for a Kerr black hole, 
and show its posterior probability distribution in Fig.~\ref{fig:rin:qnm_deviation_221_pyRing}. 
The corresponding hyperparameter values are: $\mu = 0.02^{+0.20}_{-0.18}, \, \sigma < 0.23$.
Although GW191109\_010717 is excluded from the combined analysis, we note that even though the mass and spin estimates coming from this event show some tension with the ones coming from an IMR analyses, the parametrised deviations do not indicate preference for additional parameters required to describe the ringdown emission. We do not obtain informative constraints on $\delta \hat{\tau}_{221}$.

The single event odds ratios $\log_{10} \mathcal{O}^{\rm modGR}_{\rm GR}$ values, computed following a procedure similar to our previous analysis~\cite{LIGOScientific:2020tif}, are reported in Table~\ref{tab:rin:final_mass_spin}. The highest $\log_{10} \mathcal{O}^{\rm modGR}_{\rm GR}$ value among O3b events, $-0.08$, corresponds to GW200311\_115853 and does not signal significant tension. By considering all the GWTC-3 events that passed our selection criteria (including previous GWTC-2 results), we find a combined log odds ratio of $-0.88 \pm 0.44$, at 90\% uncertainty, favouring the hypothesis that GR gives an accurate description of the observed ringdown signals. 

Finally, as an agnostic test of the consistency of the ringdown emission with GR predictions, a single damped sinusoid (DS) template is used to fit the data. In this case we are not assuming an underlying Kerr metric, nor that the object emitting the signal is a black hole, thus the frequency, damping time, and complex amplitude are considered as free parameters without imposing any predictions from GR. 
We adopt uniform priors on the frequency, damping time, log of the magnitude, and the phase of the complex amplitude. The fit starts at $10 G\bar{M}_\mathrm{f} (1+z)/c^3$ after $t_0$. The results on the frequency and damping time obtained in this case are reported in Table \ref{tab:rin:freq_tau_results}.

\begin{table}[b]
\caption{The median, and symmetric $90\%$ credible intervals of the frequency and damping time from a single damped-sinusoid (DS) analysis, compared to the IMR predictions for the fundamental $\ell=|m|=2,n=0$ mode from Sec.~\ref{subsec:rin1}}.
{{
\begin{tabular}{lccccc}
\toprule
Event & \multicolumn{2}{c}{Redshifted} & \hphantom{X} & \multicolumn{2}{c}{Redshifted} \\
& \multicolumn{2}{c}{frequency [Hz]} & \hphantom{X} & \multicolumn{2}{c}{damping time [ms]} \\[0.075cm]
\cline{2-3}
\cline{4-6}
& IMR & DS & \hphantom{X} & IMR & DS \\
\midrule

GW191109\_010717 &
$120^{+8}_{-6}$ &
$106^{+6}_{-5}$  &
\hphantom{X} &
$7.8^{+2.5}_{-1.0}$ &
$15.3^{+4.8}_{-3.5}$\\[0.075cm]

GW191222\_033537 &
$147^{+13}_{-14}$ &
$156^{+755}_{-55}$  &
\hphantom{X} &
$6.9^{+1.1}_{-0.8}$ &
$13.0^{+31.9}_{-10.3}$\\[0.075cm]

GW200129\_065458 &
$251^{+9}_{-11}$ &
$250^{+40}_{-29}$  &
\hphantom{X} &
$4.5^{+0.5}_{-0.4}$ &
$2.8^{+2.0}_{-1.5}$\\[0.075cm]

GW200224\_222234 &
$197^{+9}_{-8}$ &
$192^{+12}_{-21}$  &
\hphantom{X} &
$5.6^{+0.6}_{-0.5}$ &
$5.7^{+5.6}_{-2.6}$\\[0.075cm]

GW200311\_115853 &
$236^{+10}_{-13}$ &
$322^{+625}_{-246}$  &
\hphantom{X} &
$4.4^{+0.5}_{-0.4}$ &
$21.4^{+25.6}_{-19.2}$\\[0.075cm]

\bottomrule
\end{tabular}}}\label{tab:rin:freq_tau_results}
\end{table}

The resulting values agree with the measured values obtained from IMR analyses assuming GR and with those obtained from the additional \texttt{pSEOB} ringdown test discussed below, except for GW191109\_010717, where we find an overestimation of the frequency and damping time from the \textsc{pyRing} analysis with respect to the full IMR analyses, which is compatible with the overestimation of the remnant mass and spin.

\label{subsec:rin1}
\subsubsection{The \textsc{pSEOBNRv4HM} analysis}
\begin{figure*}
  \centering
\includegraphics[height=7cm]{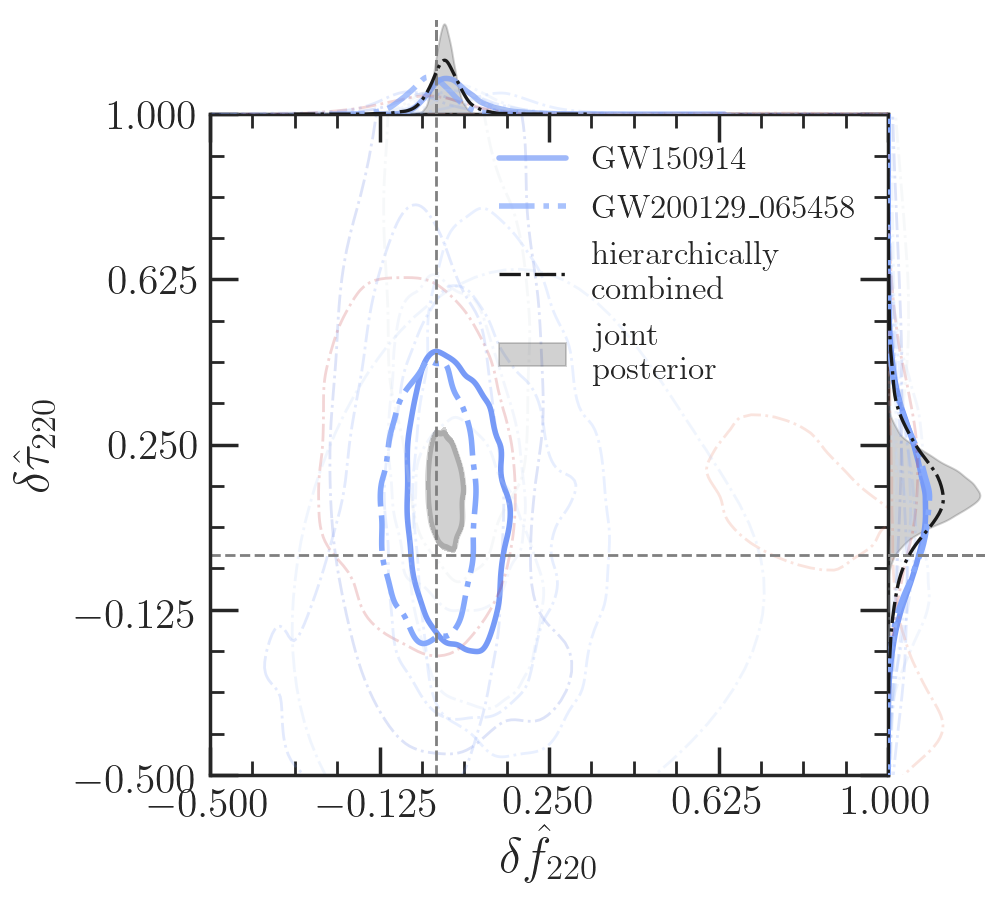}
\includegraphics[height=6.5cm]{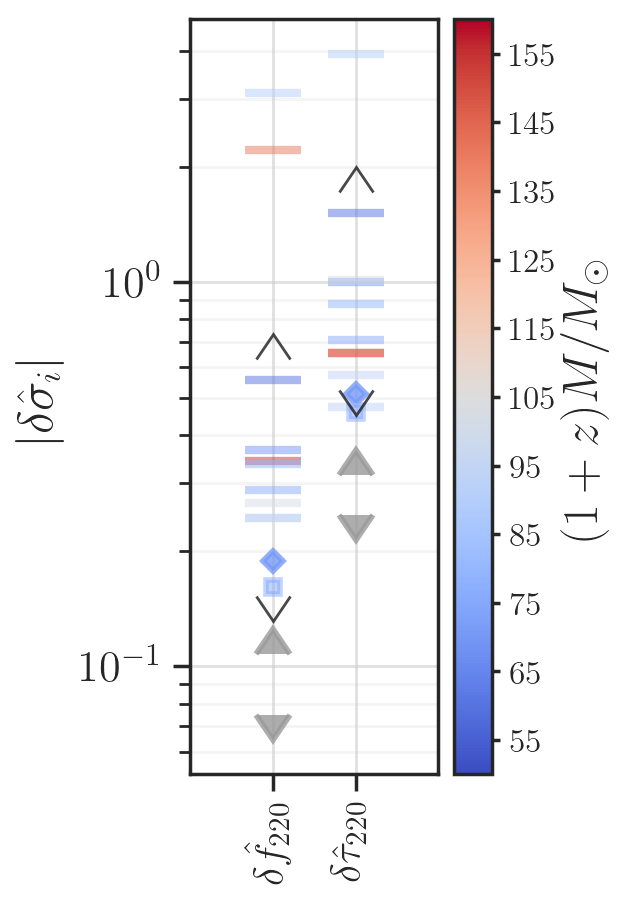}
  \caption{\emph{Left panel}: The 90\% credible levels of the posterior probability distribution of the fractional deviations in the frequency and damping time of the $(2, 2, 0)$ QNM, $(\delta f_{220},\delta \tau_{220})$ and their corresponding one-dimensional marginalized posterior distributions, for events from O1, O2 and O3 passing a SNR threshold of $8$ in both the pre- and post-merger signal. Posteriors for GW150914 and \NNAME{GW200129D}~are separately shown. The joint constraints on $(\delta f_{220},\delta \tau_{220})$ obtained multiplying the posteriors (given a flat prior) from individual events are given by the filled grey contours, while the hierarchical method of combination yields the black dot dashed curves  in the one-dimensional marginalized posteriors. \emph{Right panel}: 90\% credible interval on the one-dimensional marginalised posteriors on $\delta \sigma_i=(\delta f_{220},\delta \tau_{220})$, colored by the median redshifted total mass $(1 + z)M$, inferred assuming GR. Filled gray (unfilled black) downward triangles mark the constraints obtained when all the events are combined by multiplying posteriors (hierarchically). For comparison, we mark the previously published bounds from~\cite{LIGOScientific:2020tif} using filled/unfilled upward triangles. The bounds from  \NNAME{GW200129D}~(square) and GW150914 (diamond) are indicated by the separate markers. See Sec.~\ref{subsec:rin3} for details.}
  \label{fig:rin:qnm_deviation220pSEOBNR}
\end{figure*}

The \texttt{pSEOBNRv4HM} ringdown analysis~\cite{Ghosh:2021mrv,LIGOScientific:2020tif}, which uses parameterized spinning EOB waveforms with higher modes calibrated to non-precessing NR simulations~\cite{Cotesta:2018fcv,Brito:2018rfr}, measures QNM frequencies within the framework of a complete IMR BBH waveform model. It makes full use of gravitational-wave signal modelling and hence the complete SNR of the signal.  In this approach, the ringdown start-time is built into the model, based on calibrations to NR, and may hence be considered complementary to the post-merger time-domain ringdown analysis described in the previous section.

In the \texttt{SEOBNRv4HM} model, starting from estimates of the initial binary's masses and spins, NR fits~\cite{Taracchini:2013rva,Hofmann:2016yih} are used to predict the mass and spin of the remnant object, which are then used to predict the ringdown frequencies and damping times~\cite{Berti:2005ys,Berti:2009kk}. Thus, the frequency and damping time of the $(\ell,\pm m,0)$ QNM, $(f_{\ell m 0}, \tau _{\ell m 0})$ are functions of the initial masses and spins: 
\begin{eqnarray}
f_{\ell m 0}^{\text{GR}} = f_{\ell m 0}^{\text{GR}}(m_1,m_2,\chi_1, \chi_2)\,, \label{eq:pseob_f_tau_masses_spins_a} \\
\tau _{\ell m 0}^{\text{GR}} = \tau_{\ell m 0}^{\text{GR}}(m_1,m_2,\chi_1, \chi_2)\,. \label{eq:pseob_f_tau_masses_spins_b}
\end{eqnarray}

In the parameterized version of the \texttt{SEOBNRv4HM} model used here (\texttt{pSEOBNRv4HM}), deviations in the frequency and damping time are described through fractional deviations, $(\delta \hat{f}_{\ell m 0},\delta \hat{\tau}_{\ell m 0})$, from the corresponding GR predictions as:
\begin{eqnarray}
f_{\ell m 0} &=& f_{\ell m 0}^{\text{GR}}\, (1 + \delta \hat{f}_{\ell m 0})\,,\label{eq:nongr_freqs_a} \\ 
\tau _{\ell m 0} &=& \tau _{\ell m 0}^{\text{GR}}\, (1 + \delta \hat{\tau}_{\ell m 0})\,. \label{eq:nongr_freqs_b}
\end{eqnarray}

We use \texttt{LALInference}~\cite{Veitch:2014wba} to stochastically sample over the parameter space of $\{\delta \hat{f}_{\ell m 0}, \delta \hat{\tau}_{\ell m 0}\}$, along with the set of GR parameters, and finally reconstruct $(f_{\ell m 0}, \tau _{\ell m 0})$ using Eqs.~(\ref{eq:pseob_f_tau_masses_spins_a})--(\ref{eq:nongr_freqs_b}).  
 Here, we keep our analysis identical to the one presented in previous analysis~\cite{LIGOScientific:2020tif} and restrict ourselves to fractional deviations of the least-damped dominant QNM, i.e., $(\delta \hat{f}_{220},\delta \hat{\tau}_{220})$, keeping the other QNMs fixed at their nominal GR values. 

The analysis is performed on events from Table~\ref{tab:events} with an SNR $\geq 8$ in the pre- \emph{and} post-inspiral regimes, following  Table~\ref{tab:imrct_params}. A reasonable SNR in the inspiral is required to break the degeneracy between the fundamental ringdown frequency deviation parameter and the remnant mass. The detector-frame total mass threshold criterion used (due to computational limitations) in the previous analysis~\cite{LIGOScientific:2020tif} was relaxed in this analysis. In reporting joint constraints, we include events from past analyses. These include the 3 events from O3a~\cite{LIGOScientific:2020tif}, and five events from first two observing runs and O3a, reported for the first time in~\cite{Ghosh:2021mrv}. The final list of events that were used for this analysis is listed in Table~\ref{tab:rin:pseob_freq_tau_results}.

The results of the analysis is summarised in Fig.~\ref{fig:rin:qnm_deviation220pSEOBNR}. The left panel of Fig.~\ref{fig:rin:qnm_deviation220pSEOBNR} shows the two-dimensional posteriors (along with the marginalised one-dimensional posteriors) of the frequency and damping time for all the events listed in Table~\ref{tab:rin:pseob_freq_tau_results}. The contours are colored by the median redshifted total mass $(1 + z)M$ of the corresponding binary. We also show the 90\% credible bounds on the fractional deviations from GR in the right panel color coded by the median redshifted mass of the binary. We specifically highlight the bounds from two events, GW150914 and \NNAME{GW200129D}. 
GW150914 previously provided the strongest bounds with this method~\cite{Ghosh:2021mrv}. However, \NNAME{GW200129D}, similar in source properties to GW150914, but with an SNR of 26.5~\cite{GWTC3}, provides currently provides the strongest single event bound, $\delta \hat{f}_{220} = -0.01^{+0.08}_{-0.08};  \delta \hat{\tau}_{220}=0.11^{+0.23}_{-0.23}$.  The joint constraints are reported using two methods: multiplying posteriors (given a flat prior on the deviation parameters) and hierarchically combining events. The joint bounds from these two methods read
\begin{equation}
    \delta \hat{f}_{220} = 0.02^{+0.03}_{-0.03};  \delta \hat{\tau}_{220}=0.14^{+0.11}_{-0.11}
\end{equation}
by multiplying the posteriors and 
\begin{eqnarray}
    \delta \hat{f}_{220} = 0.02^{+0.07}_{-0.07}~~~[\mu = 0.02^{+0.04}_{-0.04}; \sigma < 0.06] \\
    \delta \hat{\tau}_{220}=0.13^{+0.21}_{-0.22}~~~[\mu = 0.13^{+0.13}_{-0.13}; \sigma < 0.19]
\end{eqnarray}
by combining hierarchically. The numbers in the square brackets are the hyperparameter estimates. 
There is a significant improvement from results previously published in~\cite{LIGOScientific:2020tif} (filled/unfilled upward/downward triangles in the right plot of Fig.~\ref{fig:rin:qnm_deviation220pSEOBNR}), which were: $\delta \hat{f}_{220} = 0.03^{+0.38}_{-0.35}$; $\delta \hat{\tau}_{220} = 0.16^{+0.98}_{-0.98}$ using the hierarchical method and $\delta \hat{f}_{220} = 0.03^{+0.07}_{-0.06}$; $\delta \hat{\tau}_{220} = 0.13^{+0.18}_{-0.18}$ using the restricted method.

We note a couple of points about the joint posteriors. First, though we perform single-event analyses over \NNAME{GW191109A}~and \NNAME{GW200208G}, we do not include them  in the computation of the joint bounds (hierarchical or simple combination).
The posterior on $\delta \hat{f}_{220}$ show multi-modalities for these two events. For \NNAME{GW200208G}, the secondary of two modes is consistent with the GR prediction $\delta \hat{f}_{220} = 0$, while for \NNAME{GW191109A}, none of the modes are. 
 Follow-up investigations with synthetic signals in segments of data immediately adjacent to the event suggests the possibility of noise systematics not accounted for.
The same study rules out, within our statistical uncertainties, any systematic bias due to missing physics in the \texttt{SEOBNRv4HM} waveform model.

We also note that the joint posterior distribution on $\delta \hat{\tau}_{220}$ in the left plot of Fig.~\ref{fig:rin:qnm_deviation220pSEOBNR} does not include the GR prediction at the 90\% credible level. Although insufficient to claim a violation of GR, this  apparent deviation definitely warrants further investigation. The trend of overestimating the combined damping time is consistent with what is observed on an event-by-event analysis, where the posterior on $\delta \hat{\tau}_{220}$, although consistent with 0 is biased towards positive values. Hence a combination of information across multiple events is expected to reduce statistical uncertainties and make this bias more prominent. One possible reason might be a prior on ($\delta \hat{f}_{220}, \delta \hat{\tau}_{220}$) which is asymmetric around $0$ with greater support for positive values. This is because, since ($f_{220}, \tau_{220}$) are strictly positive quantities, the priors on ($\delta \hat{f}_{220}, \delta \hat{\tau}_{220}$) are strictly greater than $-1$. However, the upper prior boundary is free to be as large as is required for the posterior to not rail against it and it usually greater than $1$. For events with moderately high SNRs analysed with this method, the effect of the prior on the final posterior can be non-negligible. We also note that while the posteriors on the fractional deviation show more support towards positive values, the frequency and damping time reconstructed using Eqs. (\ref{eq:nongr_freqs_a}) and (\ref{eq:nongr_freqs_b}) are consistent with those predicted using estimates of initial masses and spins from \cite{GWTC3} and NR fits~\cite{spinfit-T1600168}. This gives us more confidence in the measured QNMs,  while also pointing to the possibility that correlations among the remnant parameters may be responsible for the apparent deviation. Further, as has been argued in \cite{LIGOScientific:2020tif}, imperfect noise modelling can also lead to overestimation of damping time~\cite{Ghosh:2021mrv}. Finally, we can not rule out the statistical uncertainties of working with a sample of just 12 events.

\begin{table*}[t]

\begin{tabular}{llllccc}
\toprule
Event & $f_{220}$ (Hz) & $\tau_{220}$ (ms) & $(1+z)M_{\rm f}/M_{\odot}$ & $\chi_{\rm f}$ \\[0.075cm]
\midrule
\hline

GW150914 &
$254.6^{+16.1}_{-12.2}$ &
$4.51^{+1.10}_{-0.99}$ &
$71.6^{+8.6}_{-11.0}$ &
$0.76^{+0.10}_{-0.20}$
\\[0.075cm]

GW170104 &
$287.8^{+99.4}_{-36.1}$ &
$4.70^{+3.24}_{-2.24}$ &
$69.4^{+13.6}_{-28.1}$ &
$0.84^{+0.12}_{-0.57}$
\\[0.075cm]

GW190519\_153544 &
$123.6^{+11.9}_{-13.0}$ &
$10.33^{+3.56}_{-3.07}$ &
$155.5^{+24.0}_{-29.9}$ &
$0.81^{+0.10}_{-0.28}$
\\[0.075cm]

GW190521\_074359 &
$204.6^{+14.6}_{-11.7}$ &
$5.32^{+1.48}_{-1.21}$ &
$86.4^{+12.2}_{-14.3}$ &
$0.73^{+0.12}_{-0.26}$
\\[0.075cm]

GW190630\_185205 &
$247.0^{+29.0}_{-49.8}$ &
$3.86^{+2.25}_{-1.73}$ &
$65.7^{+18.3}_{-39.2}$ &
$0.62^{+0.26}_{-0.62}$
\\[0.075cm]

GW190828\_063405 &
$254.3^{+20.2}_{-17.7}$ &
$6.22^{+2.53}_{-2.34}$ &
$83.1^{+11.1}_{-18.2}$ &
$0.89^{+0.06}_{-0.25}$
\\[0.075cm]

GW190910 112807 &
$174.2^{+11.7}_{-7.5}$ &
$9.52^{+3.13}_{-2.68}$ &
$123.5^{+14.7}_{-18.1}$ &
$0.90^{+0.05}_{-0.11}$
\\[0.075cm]

\NNAME{GW191109A} &
$136.6^{+11.2}_{-18.3}$ &
$15.09^{+3.62}_{-2.74}$ &
$170.4^{+25.3}_{-15.1}$ &
$0.94^{+0.02}_{-0.04}$
\\[0.075cm]

\NNAME{GW200129D} &
$246.4^{+14.5}_{-18.1}$ &
$4.68^{+1.01}_{-0.97}$ &
$74.2^{+7.4}_{-10.0}$ &
$0.76^{+0.10}_{-0.22}$
\\[0.075cm]

\NNAME{GW200208G} &
$460.7^{+40.7}_{-271.7}$ &
$18.25^{+47.49}_{-14.10}$ &
$71.5^{+23.8}_{-11.1}$ &
$1.00^{+0.00}_{-0.45}$
\\[0.075cm]

\NNAME{GW200224H} &
$196.1^{+10.2}_{-9.6}$ &
$7.00^{+1.86}_{-1.71}$ &
$101.6^{+10.4}_{-14.0}$ &
$0.85^{+0.07}_{-0.16}$
\\[0.075cm]

\NNAME{GW200311L} &
$241.8^{+19.9}_{-20.0}$ &
$4.72^{+1.75}_{-1.45}$ &
$75.3^{+12.4}_{-17.4}$ &
$0.76^{+0.13}_{-0.39}$
\\[0.075cm]

\bottomrule

\end{tabular}

\caption{Redshifted damping times and frequencies of the 220 mode as well as final redshifted mass and spin as inferred from the analysis for different events that are analysed using the \textsc{pSEOBNRv4HM} method (Sec.~\ref{subsec:rin3}).}
\label{tab:rin:pseob_freq_tau_results}
\end{table*}

\label{subsec:rin3}

    \subsection{Echoes}
    \label{sec:ech}

Mergers of certain classes of exotic compact objects that do not have a horizon can cause in-going gravitational waves (e.g., resulting from merger) to reflect multiple times off effective radial potential barriers, with wave packets leaking  out to infinity at regular times; these are called echoes \cite{Cardoso:2016rao,Cardoso:2017cqb,Cardoso:2016oxy}.
Previous analysis~\cite{LIGOScientific:2020tif} employed a morphology-dependent method~\cite{Lo:2018sep} using the waveform template proposed in~\cite{Abedi:2016hgu} to search for GW echoes. Here, however, we employ a method that is independent of the morphology of the signal~\cite{Tsang:2018uie, Tsang:2019zra}.

We make use of the \texttt{BAYESWAVE} pipeline~\cite{bayeswave,Cornish:2014kda,Littenberg:2014oda} to search for echoes and compute the Bayes factors for the signal versus noise hypothesis. We further analyze the background around each event to quantify the significance of the Bayes factors.

Our method employs \emph{combs} of decaying identical sine--Gaussians as the basis functions or the generalized wavelets~\cite{Tsang:2019zra,Tsang:2018uie}. The sine--Gaussians are parameterized by an amplitude $A$, a central frequency $f_0$, a damping time $\tau$, and a reference phase $\phi_0$. In addition to these parameters, basis functions include three extra parameters which characterize the way in which the sine--Gaussians are arranged in the wavelet;  the central time of the first echo $t_0$, the time separation between sine--Gaussians $ \Delta t$,  a phase difference  between them $\Delta \phi$, an amplitude damping factor $\gamma$, and a widening factor $w$. 

  To perform the search, we use uniform priors for all the wavelet parameters except the damping time $\tau$ and the  amplitude $A$. The central frequency, $f_0$ is sampled uniformly from  the interval between the lower cut-off frequency and half the sampling rate of the analysis, $\interval{30}{1024}$ Hz. Also, $\Delta t$, $\phi_0$ , $\Delta \phi$ and  $\gamma$ are  sampled uniformly with respective ranges,  $\Delta t \in \interval{0}{0.7}$ s, $\phi_0 \in \interval{0}{2\pi}$, $\Delta \phi \in \interval{0}{2\pi}$, $\gamma \in \interval{0}{1}$,  and $w \in \interval{1}{2}$. The damping time $\tau$ is sampled such that the corresponding the  quality factor $Q= 2\pi f_0 \tau$ is distributed uniformly  in the interval $\interval{2}{40}$, which ensures  $\tau$ to be within the time  scales  set  by  masses of the events ($\sim \interval{3 \times 10^{-4}}{0.2}$ s). The wavelet amplitude $A$ is sampled based  on  signal-to-noise  ratio  as described in \cite{Cornish:2014kda}.

	To  construct  background  distributions  for  the  log Bayes factors  $\log_{10} \mathcal{B}^{\rm{S}}_{\rm{N}}$,  we  use  stretches of data randomly picked in an asymmetric time window around each coalescence event. Backgrounds are composed by 100 trials in $\interval{t_{\rm event} - 3072\,{\rm s}}{t_{\rm event}-300\,{\rm s}}$ and 100 trials in $\interval{t_{\rm event} +1024\,{\rm s}}{t_{\rm event}+2048\,{\rm s}}$, , where $t_{\rm event}$ is the coalescence time of the event, each trial analyzing an interval of $4\,{\rm s}$. The start of the time window after coalescence is chosen in order to avoid the presence of putative echoes signals in the background trials.

	The foreground run is instead performed on a $4$ s time interval starting at $t_{\rm event} + 3\tau_{220}$, since we  want to start  analyzing at  a  time  that  is  safely  beyond  the plausible  duration  of  the ringdown  of  the  remnant  object.  $\tau_{220}$ is  a  conservatively  long  estimate  for  the  decay time of the 220 mode in the ringdown, obtained from the fitting formula for $\tau_{220}(M_{\rm f},\chi_{\rm f})$ \cite{Berti:2005ys,Kamaretsos:2011um}. We take the upper bound of $ 90\%$ credible interval of $\tau_{220}$ distribution computed from final mass and spin samples for the event. This is typically of the order of a  few  milliseconds.

   The complementary empirical cumulative distribution function of background distributions of $\log_{10} \mathcal{B}^{\rm{S}}_{\rm{N}} $ are used to quantify the search outcome via $p$-values for each event. We compute the $p$-value for $\log_{10} \mathcal{B}^{\rm{S}}_{\rm{N}}$ for each event, which is the fraction of background $\log_{10} \mathcal{B}^{\rm{S}}_{\rm{N}}$ above the log Bayes factor of the event.

\begin{figure} 	
	\centering
	\includegraphics[width=\columnwidth]{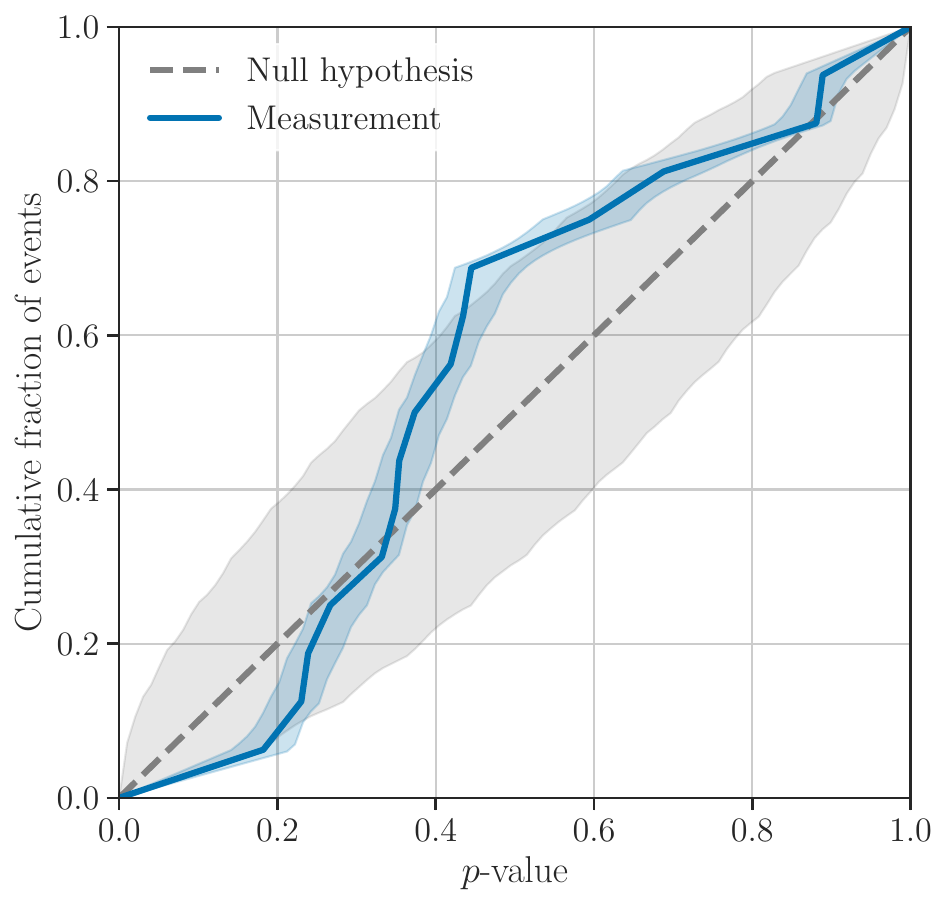}
	\caption{ Results of the echoes analysis (Sec.~\ref{sec:ech}). Plot of fraction of events for which the echoes signal-to-noise $p$-value is less than
		or equal to the abscissa. The light-blue  band  represents the $90\%$ credible interval of the observed $p$-values, while the diagonal dashed line is expectation from the null hypothesis. The light-gray band around the diagonal line represents the $90\%$ uncertainty band of the null hypothesis.} \label{fig:echoes_MI:pp}
\end{figure}

For all the events, the signal versus noise  Bayes factor $\mathcal{B}^{\rm{S}}_{\rm{N}}$  are within the corresponding background distributions, and the corresponding $p$-values are tabulated in Table \ref{tab:pvalues}. 
If  echoes are not present in the data, we expect the $p$-values to follow a uniform distribution between $[0,1]$. Fig.~\ref{fig:echoes_MI:pp} plots the $p$-value versus the cumulative fraction of events and the dotted dash line shows the prediction if no signal is present with corresponding $90\%$ uncertainty regions marked in light-color bands. We follows the recipe described in Sec.~\ref{sec:res} to make the PP plot. As can be seen, the measurement is consistent with the absence of echoes within 90\% credible region.
We conclude that we find no statistically significant evidence for echoes from the morphology-independent search we carried out.
 As the methodology employed here is different from that of our previous analysis~\cite{LIGOScientific:2020tif}, and relies of the $p$-values, one cannot have a fair comparison of the results between the two.

	\begin{table}[]
		\caption{ Results of the echoes analysis (Sec.~\ref{sec:ech}). List of $p$-values for signal to noise Bayes Factor $\mathcal{B}^{\rm{S}}_{\rm{N}}$ for the events that are analysed. In the absence of any echoes signal these should be uniformly distributed between $[0,1]$. Fig.~\ref{fig:echoes_MI:pp} shows the corresponding PP plot with 90\% credible intervals superimposed on it. There is no evidence for the presence of echoes.}
			\begin{tabular}{lrccc}
			\toprule
				
				Event     & $p$-value \\ 
				
				\midrule
				
				\NNAME{GW191109A} & 0.35   \\ 
				\NNAME{GW191129G}  & 0.35   \\ 
				\NNAME{GW191204G}  & 0.37   \\ 
				\NNAME{GW191215G}  & 0.23   \\ 
				\NNAME{GW191216G} & 0.88   \\ 
				\NNAME{GW191222A}  & 0.89   \\ 
				\NNAME{GW200115A}  &  0.44  \\ 
				\NNAME{GW200129D}  & 0.33   \\ 
				\NNAME{GW200202F} & 0.43  \\ 
				\NNAME{GW200208G}  & 0.24  \\ 
				\NNAME{GW200219D} & 0.18    \\ 
				\NNAME{GW200224H} & 0.59    \\ 
				\NNAME{GW200225B}  & 0.69   \\ 
				\NNAME{GW200311L} & 0.42  \\ 
				\NNAME{GW200316I} & 0.27    \\ 
				\bottomrule
			\end{tabular}
	
		\label{tab:pvalues}
		\centering
	\end{table}

\section{Conclusions and outlook}
\label{sec:conclusion}

Gravitational-wave observations provide a unique tool to test fundamental physics. The strongly gravitating, highly dynamical and radiative spacetime associated with the late inspiral, merger and ringdown of compact binaries facilitates tests of general relativity in a regime that is unaccessible otherwise. BBH and BNS mergers observed in the past observing runs already set  limits on possible deviations from GR~\cite{PhysRevLett.116.221101,TheLIGOScientific:2016wfe,TheLIGOScientific:2016agk,GW150914_paper,GW170104,Abbott:2017oio,Hagihara:2019ihn,Pardo:2018ipy,Abbott:2018wiz,Abbott:2018lct,GWTC3:TGR:release,GWTC1,LIGOScientific:2019fpa,LIGOScientific:2020tif}. Here we discuss a pool of tests aimed at unearthing deviations from GR using the events detected during the second part of the third observing run of Advanced LIGO and Advanced Virgo. We perform ten tests of GR on the 15 events that have a false alarm rate less than $10^{-3}\,\rm{yr}^{-1}$.  These tests are the same ones as in the previous analysis~\cite{LIGOScientific:2020tif}, except with the following updates. Our search for post-merger echoes is morphology-independent in this paper and the method to test for non-GR polarization modes is refined to address mixed polarizations as opposed to scalar-only, vector-only, and tensor-only hypotheses as was the case in~\cite{LIGOScientific:2020tif}. Furthermore, some of the tests rely on more up-to-date waveforms; in the residuals and inspiral-merger-consistency tests, we account for higher order multipole moments for all the events from the second part of the third observing run.

 We subtract the maximum-likelihood GR waveform from the data to verify the consistency of the residuals with detector noise, thereby showing the consistency of the signals in the data with GR. Independent estimates of the mass and spin of the merger remants, from the inspiral and postinspiral parts of the waveform for different events show mutual consistency.
The fractional changes in the final mass and spin from this test, assuming they take the same values for all the events and combining all the events analyzed so far, are constrained to $\dMf = \ImrGWTCTHREE{DMFGWTC3PHENOM}$ and $\dchif = \ImrGWTCTHREE{DCHIFGWTC3PHENOM}$ at 90\% credibility. 

Tests aimed at looking for parametrized departures from GR in the post-Newtonian phasing coefficients all find consistency with GR within the statistical uncertainties. The most well-constrained parameter is the absolute value of the $-1$PN coefficient,   which is bound to $\leq 7.3\times10^{-4}$ at 90\% credibility, assuming its value is the same for all the events.
As certain modified theories of gravity predict dispersion of gravitational waves, we searched for this effect and found no evidence for dispersion.  The bound on the graviton mass is updated to $m_g \leq \LivMgUL \mathrm{eV}/c^2$, at 90\% credibility. A general metric theory of gravity admits up to six modes of gravitational-wave polarization. We searched for non-GR polarization modes and found no signature of such modes. 

Analyses to measure the spin-induced quadrupole moments of the binary components found no signatures of exotic compact objects. Further, tests for deviations from GR in the ringdown of the remnant black hole were carried out using two independent methods and the frequency deviation parameters are constrained to $\delta \hat{f}_{221} = 0.01^{+0.27}_{-0.28}$ and $\delta \hat{f}_{220}=0.02^{+0.07}_{-0.07}$, at 90\% credibility,  by hierarchically combining the results from the events that are analyzed. We also found no evidence for post-merger echoes from the merger remnant from our morphology-independent search.

Future observing runs with improved detector sensitivities will provide a larger catalog of compact binary observations
and events with larger SNR. These observations will enable us to carry out more stringent tests of GR in parts of the parameter space that are not covered here. Developing accurate waveform models that cover the diverse physics that will be revealed by the sources is an important step towards this goal.
Finally, devising new tests or improving the existing ones, optimizing their sensitvity to predictions from specific modified theories of gravity, can play a very important role in constraining beyond-GR physics using future gravitational-wave observations.

\acknowledgments
Analyses in this paper made use of
\textsc{NumPy} \cite{Numpy2020},
\textsc{SciPy} \cite{2020SciPy-NMeth},
\textsc{Astropy} \cite{astropy:2013,astropy:2018},
\textsc{IPython} \cite{ipython},
\textsc{qnm} \cite{Stein:2019mop},
\textsc{PESummary} \cite{Hoy:2020vys},
and \textsc{GWpy} \cite{gwpy};
plots were produced with
\textsc{Matplotlib} \cite{matplotlib}, and
\textsc{Seaborn} \cite{seaborn}.
Posteriors were sampled with 
\textsc{Stan} \cite{JSSv076i01},
\textsc{CPNest} \cite{CPNest},
\textsc{PyMultinest} \cite{Feroz:2009,Feroz:2013},
\textsc{Bilby} \cite{Ashton:2018jfp,Romero-Shaw:2020owr},
and \linf{} \cite{Veitch:2014wba}.
Power spectral densities are generated through the software \texttt{BAYESWAVE} \cite{Cornish:2014kda,Littenberg:2014oda}.
%
\newif\ifcoreonly
\newif\ifkagra
\newif\ifheader
%
\coreonlyfalse \kagratrue  \headerfalse
\ifheader
\begin{center}{\bf\Large
\ifkagra
Conference proceedings acknowledgements for \\ the LIGO Scientific Collaboration, the Virgo Collaboration and the KAGRA Collaboration
\else
Conference proceedings acknowledgements for \\ the LIGO Scientific Collaboration and the Virgo Collaboration
\fi
}\end{center}
\fi
This material is based upon work supported by NSF’s LIGO Laboratory which is a major facility
fully funded by the National Science Foundation.
The authors also gratefully acknowledge the support of
the Science and Technology Facilities Council (STFC) of the
United Kingdom, the Max-Planck-Society (MPS), and the State of
Niedersachsen/Germany for support of the construction of Advanced LIGO 
and construction and operation of the GEO600 detector. 
Additional support for Advanced LIGO was provided by the Australian Research Council.
The authors gratefully acknowledge the Italian Istituto Nazionale di Fisica Nucleare (INFN),  
the French Centre National de la Recherche Scientifique (CNRS) and
the Netherlands Organization for Scientific Research (NWO), 
for the construction and operation of the Virgo detector
and the creation and support  of the EGO consortium. 
The authors also gratefully acknowledge research support from these agencies as well as by 
the Council of Scientific and Industrial Research of India, 
the Department of Science and Technology, India,
the Science \& Engineering Research Board (SERB), India,
the Ministry of Human Resource Development, India,
the Spanish Agencia Estatal de Investigaci\'on (AEI),
the Spanish Ministerio de Ciencia e Innovaci\'on and Ministerio de Universidades,
the Conselleria de Fons Europeus, Universitat i Cultura and the Direcci\'o General de Pol\'{\i}tica Universitaria i Recerca del Govern de les Illes Balears,
the Conselleria d'Innovaci\'o, Universitats, Ci\`encia i Societat Digital de la Generalitat Valenciana and
the CERCA Programme Generalitat de Catalunya, Spain,
the National Science Centre of Poland and the European Union – European Regional Development Fund; Foundation for Polish Science (FNP),
the Swiss National Science Foundation (SNSF),
the Russian Foundation for Basic Research, 
the Russian Science Foundation,
the European Commission,
the European Social Funds (ESF),
the European Regional Development Funds (ERDF),
the Royal Society, 
the Scottish Funding Council, 
the Scottish Universities Physics Alliance, 
the Hungarian Scientific Research Fund (OTKA),
the French Lyon Institute of Origins (LIO),
the Belgian Fonds de la Recherche Scientifique (FRS-FNRS), 
Actions de Recherche Concertées (ARC) and
Fonds Wetenschappelijk Onderzoek – Vlaanderen (FWO), Belgium,
the Paris \^{I}le-de-France Region, 
the National Research, Development and Innovation Office Hungary (NKFIH), 
the National Research Foundation of Korea,
the Natural Science and Engineering Research Council Canada,
Canadian Foundation for Innovation (CFI),
the Brazilian Ministry of Science, Technology, and Innovations,
the International Center for Theoretical Physics South American Institute for Fundamental Research (ICTP-SAIFR), 
the Research Grants Council of Hong Kong,
the National Natural Science Foundation of China (NSFC),
the Leverhulme Trust, 
the Research Corporation, 
the Ministry of Science and Technology (MOST), Taiwan,
the United States Department of Energy,
and
the Kavli Foundation.
The authors gratefully acknowledge the support of the NSF, STFC, INFN and CNRS for provision of computational resources.

\ifkagra
This work was supported by MEXT, JSPS Leading-edge Research Infrastructure Program, JSPS Grant-in-Aid for Specially Promoted Research 26000005, JSPS Grant-in-Aid for Scientific Research on Innovative Areas 2905: JP17H06358, JP17H06361 and JP17H06364, JSPS Core-to-Core Program A. Advanced Research Networks, JSPS Grant-in-Aid for Scientific Research (S) 17H06133 and 20H05639 , JSPS Grant-in-Aid for Transformative Research Areas (A) 20A203: JP20H05854, the joint research program of the Institute for Cosmic Ray Research, University of Tokyo, National Research Foundation (NRF) and Computing Infrastructure Project of KISTI-GSDC in Korea, Academia Sinica (AS), AS Grid Center (ASGC) and the Ministry of Science and Technology (MoST) in Taiwan under grants including AS-CDA-105-M06, Advanced Technology Center (ATC) of NAOJ, Mechanical Engineering Center of KEK.
\fi


{\it We would like to thank all of the essential workers who put their health at risk during the COVID-19 pandemic, without whom we would not have been able to complete this work.}

\vspace{5mm}

\appendix
\section{Effect of waveform systematics and noise artifacts on the tests}
\label{app:sys}
In the context of null tests of GR, the posterior distributions on deformation parameters are sensitive to every assumption that goes into the analysis. These include assumptions about the relevant physics included in the waveforms employed and detector noise model, among others. Hence, any deviation from GR seen in the posteriors, statistically significant or not, could be due to the breakdown of one or more of the above assumptions. We discuss below some of these features and discuss the specific case of \NNAME{GW191109A}{} as an example.

\subsection{Waveform systematics and parameter degeneracies}

The increasing number of events detected results in a large variety of systems observed, spanning regions of parameter space where different approximants lead to visibly different estimates of source parameters~\cite{GWTC3,Ossokine:2020kjp,Nitz:2020mga,Huang:2020pba}. Such discrepancies are expected for non-standard events, and are a direct consequence of specific approximations inherent to each waveform model (such as neglecting precessional effects, higher order multipole moments, etc.) or lack of a uniform coverage in the parameter space of NR waveforms, which are used to train or calibrate the waveform models. Extreme spins, highly unequal masses and nearly edge-on inclinations are all elements that can enhance such differences \cite{Biscoveanu:2021nvg,Varma:2016dnf,Colleoni:2020tgc}.

Another aspect to consider in this context is the presence of waveform degeneracies, where two or more templates provide a good fit to the signal: in this case, posteriors can show a complicate bi(multi)-modal structure, with different modes clearly separated in parameter space. The addition of non-GR parameters complicates the picture, increasing the dimensionality of the likelihood surfaces to be explored and possibly enhancing such multimodalities. Multi-modal features might appear also in the distributions for deviation parameters, with some of the modes having consistent support away from zero. It has been pointed out that this can be a direct consequence of a known degeneracy between source parameters and deviation coefficients~\cite{Ghosh:2021mrv}. In this sense, one needs to be extremely cautious in classifying support for values away from GR as violations from GR, without thorough cross-comparisons of multiple analyses ~\cite{Johnson-McDaniel:2021yge} which might be individually prone to systematic biases related to the partial inclusion of beyond-GR effects \cite{Vallisneri:2013rc}.
The \texttt{pSEOBNRv4HM} model, an IMR model where ringdown complex frequencies deviations are free to vary, is one of the analyses which commonly suffers from such degeneracies~\cite{LIGOScientific:2020tif, Ghosh:2021mrv}. One example of degeneracy is due to the strong correlation between the fundamental ringdown frequency and the mass of the remnant black hole. Such degeneracy is broken when the mass is constrained independently from other phases of the coalescence. This considerations led to the requirement of considering only events with $\rm {SNR} > 8$ in the pre-merger regime for this analysis.
While additional selection criteria can help in minimizing these effects, it is important to carefully interpret the results of each test keeping in mind the assumptions underlying the analysis of the signal.

\subsection{Non-stationarity and non-Gaussianity of the detector noise}

An additional important aspect of data analysis is the impact of transient non-Gaussian noise or glitches affecting the data around the time of the event analysed. Glitches can lead to systematic deviations in the null parameter that is measured and mimic false deviations from GR~\cite{LIGOScientific:2020tif, Kwok:2021zny}. These issues will affect both low and high mass systems in different ways, depending on the amplitude and duration of the noise transient \cite{Oliver:10.1103}, and might exacerbate waveform systematics effects, which tend to be more severe for short signals. Furthermore, since the parameterized models we employ invoke parameters over and above the GR source parameters, these additional parameters can capture any residual noise artifacts that are left after methods such as deglitching \cite{Cornish:2014kda,Cornish:2020dwh} are employed to mitigate non-stationarities or non-Gaussianities. As we detect more and more signals, including high-mass ones, these considerations become of increasing importance while testing for deviations from GR. Already in the first half of the third observing run, a few events have shown deviations which were understood to be an effect of incomplete models of the noise background (e.g., in the ringdown regime)~\cite{LIGOScientific:2020tif}. Gaussian noise fluctuations alone are expected to cause deviations from GR for $\sim 1$ in ten events with our choice of credible intervals, though the results of our hierarchical analyses should be robust against these effects.
As an example, we next discuss how statistically significant deviations seen in some of the analyses, as mentioned in the main text, may be understood.

\subsection{Anomalies observed in the case of \NNAME{GW191109A}}

Among the different events we analysed, \NNAME{GW191109A}{} is the only event which led to statistically significant deviations in three of our null tests (i.e., modified dispersion tests, Sec.\,\ref{sec:liv}, and two ringdown tests, Sec.\,\ref{sec:rin}).

\NNAME{GW191109A}{} is the heaviest system in our event list and hence has the shortest signal duration. It is also the only event considered in this work for which both Livingston and Hanford frames required mitigation, due to the presence of glitches overlapping with the inspiral track of the signal (see Table XIV of GWTC-3~\cite{GWTC3}). Such noise artifacts belong to the category of slow scattering glitches \cite{Soni:2021cjy}, which are due to light scattering and lead to long duration arches ($\sim 2$--$2.5  {\rm s}$) in time--frequency spectograms, as discussed in \cite{LIGO:2020zwl} (see also Sec. IIIB of GWTC-3 \cite{GWTC3} for a discussion of data quality of O3b events). Data below $30 (40)$ Hz were affected in Livingston (Hanford), with noise directly overlapping the time of the trigger, and glitches were subtracted with \texttt{BayesWave} \cite{Cornish:2014kda,Cornish:2020dwh}. The same glitch subtraction algorithm was applied only to another event in our selection list, \NNAME{GW200115A}{}. However, \NNAME{GW200115A}{} has a long inspiral compared to \NNAME{GW191109A}{}, which lasts for less than a second in detector band, meaning the glitch overlaps with a significant portion of the signal. GR parameter estimation runs \cite{GWTC3} already show evidence of a multimodal likelihood surface for this event.

Previous investigations have shown that \texttt{BayesWave} deglitching is not expected to affect parameter estimation~\cite{Cornish:2020dwh,Pankow:2018qpo} nor tests of GR~\cite{Kwok:2021zny}. However, these works did not consider situations in which two detectors are affected by the type of glitch that impacted \NNAME{GW191109A}{}. 

Injections performed around the trigger time in detector noise could partially reproduce the deviations seen in the actual analyses of the event, indicating noise properties might be the main explanation behind our results. 
Further studies would be required to assess the reliability of tests of GR in similar situations, as well as the sensitivity of different analyses to residual noise features. These investigations are outside the scope of this paper and, as a cautionary measure, we drop \NNAME{GW191109A}{} from combined statements in the context of ringdown and modified dispersion tests.

\section{Revisiting inspiral-merger-ringdown consistency test results of GWTC-2 events}
\label{app:imr}

\begin{figure}
	\begin{center}
	\includegraphics[width=3.5in]{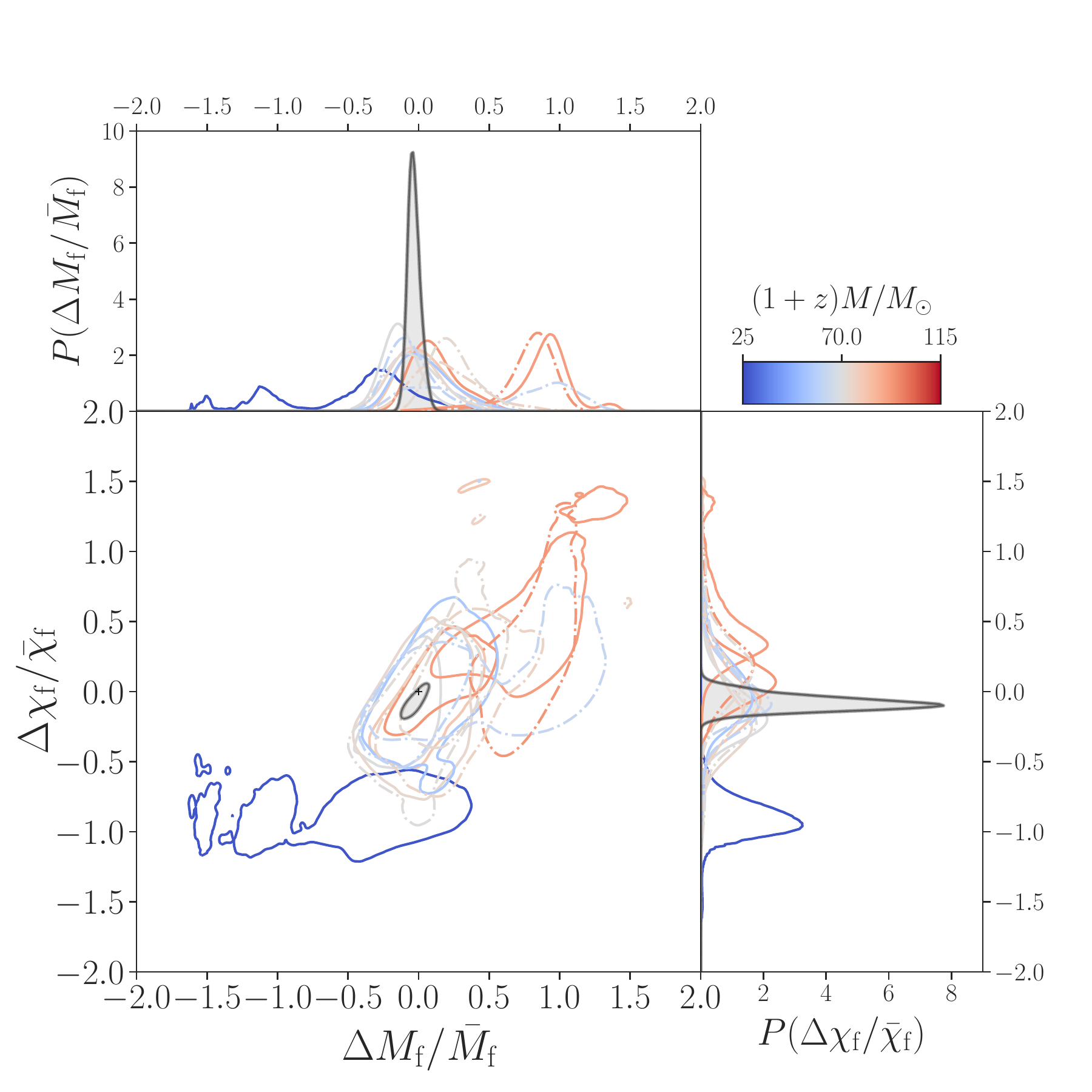}
	\end{center}
	\caption{
	Same as Fig.~\ref{fig:imrct_posteriors} but for GWTC-2 events (see Table~\ref{tab:imrct_params_gwtc2}). 
	The gray distribution corresponds to the joint posterior of GWTC-2 events. 
	O3a (O1 and O2) events are plotted with solid (dot--dashed) traces.
  }
	\label{fig:imrct_posteriors_gwtc2}
\end{figure}

\begin{table}
\caption{
    Same as Table~\ref{tab:imrct_params} but here we include the updated results of GWTC-2 events which satisfy the inspiral--merger--ringdown consistency test selection criteria. 
    Only the $\QGR$ values are updated compared to results given in Table~IV of our previous analysis~\cite{LIGOScientific:2020tif}.
}
\label{tab:imrct_params_gwtc2}
\centering
\begin{tabular}{l@{~} c@{\quad}r@{\quad}r@{\quad} r@{\quad} r}
\toprule
Event 		& $f_\text{c}^{\rm IMR}$ [Hz]  & $\rho_\mathrm{IMR}$ & $\rho_\mathrm{insp}$ & $\rho_\mathrm{postinsp}$ & $\QGR$ [\%] \\
\midrule
GW150914	& 132	  & 25.3	& 19.4	& 16.1	& \ImrEVENTSTATS{GW150914GRQUANTGWTC3}\phantom{${}^*$} \\	
GW170104	& 143	  & 13.7	& 10.9	& 8.5	& \ImrEVENTSTATS{GW170104GRQUANTGWTC3}\phantom{${}^*$} \\	
GW170809	& 136	  & 12.7	& 10.6	& 7.1	& \ImrEVENTSTATS{GW170809GRQUANTGWTC3}\phantom{${}^*$} \\
GW170814	& 161	  & 16.8	& 15.3	& 7.2	& \ImrEVENTSTATS{GW170814GRQUANTGWTC3}\phantom{${}^*$} \\
GW170818	& 128	  & 12.0	& 9.3	& 7.2	& \ImrEVENTSTATS{GW170818GRQUANTGWTC3}\phantom{${}^*$} \\
GW170823	& 102	  & 11.9	& 7.9	& 8.5	& \ImrEVENTSTATS{GW170823GRQUANTGWTC3}\phantom{${}^*$} \\ 
\midrule
\NNAME{GW190408H}	& 164	  &  15.0 	&  13.6  & 6.4 	& \ImrEVENTSTATS{S190408anGRQUANTGWTC3}\phantom{${}^*$} \\
\NNAME{GW190503E}	& 99	  &  13.7 	&  11.5  & 7.5 	& \ImrEVENTSTATS{S190503bfGRQUANTGWTC3}\phantom{${}^*$} \\
\NNAME{GW190513E}	& 125 	  &  13.3 	&  11.2  & 7.2  & \ImrEVENTSTATS{S190513bmGRQUANTGWTC3}\phantom{${}^*$} \\
\NNAME{GW190521E}	& 105	  &  25.4 	&  23.4  & 9.9  & \ImrEVENTSTATS{S190521rGRQUANTGWTC3}\phantom{${}^*$} \\
\NNAME{GW190630E}	& 135     &  16.3   &  14.0  & 8.2 	& \ImrEVENTSTATS{S190630agGRQUANTGWTC3}\phantom{${}^*$} \\
\NNAME{GW190814H}   & 207 	  &  24.8 	&  23.9  & 6.9 	& \ImrEVENTSTATS{S190814bvGRQUANTGWTC3}\phantom{${}^*$} \\
\NNAME{GW190828A} 	& 132	  &  16.2 	&  13.8  & 8.5 	& \ImrEVENTSTATS{S190828jGRQUANTGWTC3}\phantom{${}^*$} \\
\bottomrule
\end{tabular}
\end{table}

In this section, we revisit the IMR consistency test results of GWTC-2 events which are summarized in Table IV of our previous analysis~\cite{LIGOScientific:2020tif}. 
Here we describe the main reasons for this reanalysis.

First, for some events the parameter estimation analyses of the inspiral and the postinspiral parts used different prior bounds. 
This is not necessarily problematic but it can lead to the two-dimensional distribution of the prior on the fractional deviation parameters $(\dMf, \dchif)$ peaking away from $(0,0)$. 
Such priors are undesirable, since we do not want to prefer a GR deviation \emph{a priori}.
The prior distributions on the deviation parameters for GW170823 and \NNAME{GW190503E}{} peaked significantly away from $(0, 0)$, so we reanalyzed these events using the same priors for masses and spins in the inspiral and postinspiral analyses.
The GR quantile value for \NNAME{GW190503E}{} ($\ImrEVENTSTATS{S190503bfGRQUANTGWTC3}\%$) is significantly higher than its previous value ($53.2\%$). 
This can be attributed to the fact that the new prior for this event peaks at zero whereas the old prior peaked close to the peak of the posterior (well away from zero). 
The new and old posteriors peak at almost the same place, causing the reweighted posterior to shift further away from $(0,0)$.

Second, in  our GWTC-1 analysis~\cite{LIGOScientific:2019fpa}, the prior distributions on the fractional mass and spin parameters of O1 and O2 events were computed only using the prior range on component masses, and not accounting for the additional constraints on the mass priors. 
This was discontinued for O3a events in GWTC-2~\cite{LIGOScientific:2020tif}, where prior samples were generated considering the full set of priors. 
However, the GWTC-1 priors were used for O1/O2 events.
To maintain uniformity, we recomputed the priors for O1/O2 events which were then used to reweight the posteriors. 
The old prior for the event GW170814 favored fractional mass deviation parameters further away from zero compared to the new prior which pushed the portion of the posterior with less probability closer to the origin.
This is likely the reason why the GR quantile value of GW170814 in our previous analysis~\cite{LIGOScientific:2020tif} is significantly higher ($22.9\%$) than the current value ($\ImrEVENTSTATS{GW170814GRQUANTGWTC3}\%$).

Third, we change the limits of the fractional deviation parameters, $\dMf$ and $\dchif$. 
As can be seen from Fig.~3 of the GWTC-2 analysis~\cite{LIGOScientific:2020tif}, the $90\%$ credible regions of the posteriors on $(\dMf, \dchif)$ for a few of the GWTC-2 events such as \NNAME{GW190814H}{} were not closed within the range of the deviation parameters for which they were calculated. 
The ranges of deviation parameters were earlier set to $[-1.5,1.5]$ for $\dMf$ and $[-1,1]$ for $\dchif$. 
We now set the ranges of both deviation parameters to $[-2,2]$, which encloses all the 90\% credible regions we find.
This change in the ranges of deviation parameters has at most a small effect on the GR quantiles, with the largest absolute difference of 0.5 percentage points for \NNAME{GW190828A}{}.

The new results obtained with these three changes are given in Table~\ref{tab:imrct_params_gwtc2} and the posteriors are shown in Fig.~\ref{fig:imrct_posteriors_gwtc2}. 
The three events whose contours do not enclose the origin are GW170823 (orange dot--dashed), \NNAME{GW190503E}{} (orange solid), and \NNAME{GW190814H}{} (blue solid).
Additionally, GW170814, GW170818, and \NNAME{GW190828A}{} show some small multimodal structures. 
The possible reasons for the high $\QGR$ values for GW170823 and \NNAME{GW190814H}{} have already been discussed in our previous study~\cite{LIGOScientific:2020tif}.
Specifically, GW170823 is the event with the lowest SNR among the events in Table~\ref{tab:imrct_params_gwtc2} and has a relatively high redshifted mass, while \NNAME{GW190814H}{} has a low SNR in its postinspiral part due to its low redshifted mass leading to significant bias in its final spin measurements compared to the relatively high SNR inspiral regime.
This bias leads to the marginalized \NNAME{GW190814H}{} $\dchif$ posterior (blue solid) in the right side panel of Fig.~\ref{fig:imrct_posteriors_gwtc2} peaking significantly away from the GR value. 
\NNAME{GW190503E}{}, and \NNAME{GW190814H}{} also have prominent bimodal and multimodal posteriors which can be seen as additional peaks away from GR value in the marginalized $\dMf$ and $\dchif$ posteriors. 
This significant bias from the GR value for \NNAME{GW190814H}{} is the reason why we see the $\sigma$ posterior of $\dchif$ in Fig.~\ref{fig:imrct_hier_spin_hyperposteriors} peaking away from zero for GWTC-2 and, in turn, for GWTC-3. 
Considering the change in $\QGR$ values with respect to our previous analysis~\cite{LIGOScientific:2020tif}, we have already explained why we see significant differences for GW170814 and \NNAME{GW190503E}{} with absolute differences of $13.0$ and $41.1$ percentage points, respectively.
The next largest absolute change is $2.3$ percentage points for GW170818 and for most events the change is at most $0.5$ percentage points.

To summarize, we changed our calculation of the reweighted posteriors in three ways. We corrected the cases where the mismatch of priors used in the inspiral and postinspiral runs led to the prior on the deviation parameters peaking well away from zero, recomputed the prior distribution on the deviation parameters for the O1/O2 events, and made the limits on the deviation parameters uniform across all events. With these changes, we note that $\QGR$ changes appreciably only for two events: GW170814 and \NNAME{GW190503E}{}. We use the results from these new reweighted posteriors to obtain the combined results.

\bibliography{cbc-group,software}

\clearpage

\nolinenumbers

\iftoggle{endauthorlist}{

 \let\author\myauthor
 \let\affiliation\myaffiliation
  \let\maketitle\mymaketitle

 \pacs{}

 \newpage
\maketitle
}

\end{document}